\documentclass[fleqn,usenatbib]{mnras}

\usepackage[T1]{fontenc}
\usepackage{ae,aecompl}
\usepackage{xcolor}
\usepackage{orcidlink}

\usepackage{graphicx}	% Including figure files
\usepackage{amsmath}	% Advanced maths commands
\usepackage{amssymb}	% Extra maths symbols
\usepackage{bm}

\usepackage{newtxtext,newtxmath}
\newcommand{\fb}{FIREbox}
\newcommand{\hr}{FIREbox$^{\it HR}$}

\title[Star formation efficiency at Cosmic Dawn]{Elevated UV luminosity density at Cosmic Dawn explained by non-evolving, weakly mass-dependent star formation efficiency}

\author[R. Feldmann et al.]{Robert Feldmann\orcidlink{0000-0002-1109-1919}$^{1}$\thanks{E-mail: robert.feldmann@uzh.ch},
Michael Boylan-Kolchin\orcidlink{0000-0002-9604-343X}$^{2}$\thanks{Authors listed in alphabetical order},
James S. Bullock\orcidlink{0000-0003-4298-5082}$^{3}$,
Onur Çatmabacak\orcidlink{0000-0003-4067-1434}$^{1}$,
\newauthor
Claude-André Faucher-Giguère\orcidlink{0000-0002-4900-6628}$^{4}$,
Christopher C. Hayward\orcidlink{0000-0003-4073-3236}$^{5}$,
Dušan Kereš\orcidlink{0000-0002-1666-7067}$^{6}$,
Alexandres Lazar\orcidlink{0000-0002-6852-6282}$^{3}$,
\newauthor
Lichen Liang\orcidlink{0000-0001-9422-0095}$^{7}$,
Jorge Moreno\orcidlink{0000-0002-3430-3232}$^{8, 9}$,
Pascal A. Oesch\orcidlink{0000-0001-5851-6649}$^{10, 11}$,
Eliot Quataert\orcidlink{0000-0001-9185-5044}$^{12}$,
Xuejian Shen\orcidlink{0000-0002-6196-823X}$^{13}$, and
\newauthor
Guochao Sun\orcidlink{0000-0003-4070-497X}$^{4}$
\\ \\
$^{1}$Department of Astrophysics, University of Zurich, Zurich CH-8057, Switzerland\\
$^{2}$Department of Astronomy, The University of Texas at Austin, 2515 Speedway Stop C1400, Austin, TX 78712, USA\\
$^{3}$Department of Physics and Astronomy, University of California, Irvine, CA 92697, USA\\
$^{4}$CIERA and Department of Physics and Astronomy, Northwestern University, 1800 Sherman Ave, Evanston, IL 60201, USA\\
$^{5}$Center for Computational Astrophysics, Flatiron Institute, 162 5th Avenue, New York, NY 10010, USA\\
$^{6}$Center for Astrophysics and Space Sciences, University of California San Diego, San Diego, CA 92093, USA\\
$^{7}$Canadian Institute for Theoretical Astrophysics, University of Toronto, Toronto, ON. M5S 3H8, Canada\\
$^{8}$Department of Physics and Astronomy, Pomona College, Claremont, CA 91711, USA\\
$^{9}$The Observatories of the Carnegie Institution for Science, 813 Santa Barbara Street, Pasadena, CA 91101, USA\\
$^{10}$Department of Astronomy, University of Geneva, Chemin Pegasi 51, 1290 Versoix, Switzerland \\
$^{11}$Cosmic Dawn Center (DAWN), Niels Bohr Institute, University of Copenhagen, Jagtvej 128, K\o benhavn N, DK-2200, Denmark \\
$^{12}$Department of Astrophysical Sciences, Princeton University, Princeton, NJ 08544, USA\\
$^{13}$Department of Physics \& Kavli Institute for Astrophysics and Space Research, Massachusetts Institute of Technology, Cambridge, MA 02139, USA
}

\date{}

\pubyear{2024}

\begin{document}
\label{firstpage}
\pagerange{\pageref{firstpage}--\pageref{lastpage}}
\maketitle

% Abstract: It should be a single paragraph not more than 250 words (200 words for Letters)
\begin{abstract}
Recent observations with the James Webb Space Telescope (JWST) have uncovered unexpectedly high cosmic star formation activity in the early Universe, mere hundreds of millions of years after the Big Bang. These observations are often understood to reflect an evolutionary shift in star formation efficiency (SFE) caused by changing galactic conditions during these early epochs.
We present \hr{}, a high-resolution, cosmological hydrodynamical simulation from the \emph{Feedback in Realistic Environments} project, which offers insights into the SFE of galaxies during the first billion years of cosmic time. \hr{} re-simulates the cosmic volume ($L=22.1$ cMpc) of the original \fb{} run with eight times higher mass resolution ($m_{\rm b}\sim{}7800\,M_\odot$), but with identical physics, down to $z\sim{}6$.
\hr{} predicts ultraviolet (UV) luminosity functions in good agreement with available observational data. The simulation also successfully reproduces the observed cosmic UV luminosity density at $z\sim{}6-14$, demonstrating that relatively high star formation activity in the early Universe is a natural outcome of the baryonic processes encoded in the FIRE-2 model.
According to \hr{}, the SFE -- halo mass relation for intermediate mass halos ($M_{\rm halo}\sim{}10^9-10^{11}\,M_\odot$) does not significantly evolve with redshift and is only weakly mass-dependent. These properties of the SFE -- halo mass relation lead to a larger contribution from lower mass halos at higher $z$, driving the gradual evolution of the observed cosmic UV luminosity density.
A theoretical model based on the SFE -- halo mass relation inferred from \hr{} allows us to explore implications for galaxy evolution. 
Future observations of UV faint galaxies at $z>12$ will provide an opportunity to further test these predictions and deepen our understanding of star formation during Cosmic Dawn.
\end{abstract}


%%%%%%%%%%%%%%%%%%%%%%%%%%%%%%%%%%%%%%%%%%%%%%%%%%

%%%%%%%%%%%%%%%%% BODY OF PAPER %%%%%%%%%%%%%%%%%%

\section{Introduction}

The first billion years of cosmic time were a pivotal epoch in the history of our Universe that witnessed the formation of the first galaxies, the rapid growth of supermassive black holes, and the reionization of intergalactic hydrogen \citep{Stark2016, Dayal2019, Inayoshi2020, Robertson2022}. Before the launch of the James Webb Space Telescope (JWST), the ultraviolet (UV) luminosity and star formation rate (SFR) of galaxies at high redshifts were primarily constrained by observations from the Hubble Space Telescope (HST) and ground-based facilities. These observations provided crucial insights into the galaxy population up to $z\sim{}10$, revealing a steep faint-end slope of the UV luminosity function and a fast decline in their overall normalization with increasing redshift (e.g., \citealt{McLure2013, Bouwens2015, Finkelstein2015, Oesch2018, Bouwens2022}). However, the number densities and properties of galaxies at $z\gtrsim{}9$ were not well constrained with challenges arising from, e.g., low number statistics and lensing uncertainties \citep{Bouwens2017, Atek2018}.

The advent of JWST has transformed our understanding of high-redshift galaxy evolution thanks to its excellent sensitivity, resolution, and wavelength coverage. Among many other findings, JWST revealed a higher-than-expected density of UV bright galaxies at high $z$ \citep{Naidu2022, Finkelstein2022, Labbe2023a, Finkelstein2023, Casey2023}, confirmed spectroscopically the presence of galaxies up to $z\sim{}14$ \citep{Curtis-Lake2023, Robertson2023, Harikane2023a, Castellano2024, Carniani2024}, and uncovered the population of galaxies most likely responsible for reionization \citep{Atek2024}. Furthermore, JWST-based studies showed an elevated UV luminosity density $\rho_{\rm UV}$ at $z>10$ compared with both previous observational estimates and modeling predictions \citep{Donnan2023, Donnan2023a, Harikane2023a, Harikane2023, Finkelstein2023, Bouwens2023, Chemerynska2024, Conselice2024} suggesting that an important component of galaxy theory is either missing or not sufficiently understood. Solving this conundrum is critical, not only to gain a better understanding of the physical processes driving galaxy evolution at high $z$, but also to constrain the impact of galaxies on cosmic reionization.

The elevated UV luminosity density may arise from an increase in the star formation efficiency\footnote{The SFE, as used here, is a dimensionless quantity that measures the SFR in a galaxy relative to the growth rate of its host halo, see Section \ref{sect:SFE}.} (SFE) of galaxies at higher redshift (e.g., \citealt{Harikane2023a, Ceverino2024, Chakraborty2024}). Two main arguments support this statement. 

First, models with a non-evolving SFE -- halo mass relation match the observed UV luminosity density at $z\sim{}6-10$ but underpredict $\rho_{\rm UV}$ at $z>10$ (e.g., \citealt{Mason2015, Tacchella2018, Harikane2022}). This argument is further strengthened by many empirical or semi-analytical models appearing to favor significant evolution to match observations \citep{Sun2016, Behroozi2019, Sabti2022a, Sipple2023, Qin2023, Chakraborty2024, Wang2024a}. Cosmological hydrodynamical simulations targeting galaxies at $z>10$ have not fully resolved this issue as they yield significantly varying predictions for the UV luminosity density (e.g., \citealt{Wilkins2022, Kannan2023, Ceverino2024}), perhaps indicating gaps in our understanding of the physics during the Cosmic Dawn. Intriguingly, recent simulations analyzing $z=8-12$ suggest that this tension could be resolved with standard astrophysical modeling  \citep{Sun2023a}.

Secondly, the observation of massive, luminous galaxies suggests that at least some galaxies are efficient in converting accreted baryonic matter into stars \citep{Labbe2023a, Boylan-Kolchin2023, Casey2023}, potentially pointing to a failure of feedback processes limiting star formation \citep{Bassini2023, Dekel2023, Li2023a, Mirocha2023a}. We note, however, that some observational studies suggest a mild to absent evolution in the SFE of galaxies at high redshift \citep{Stefanon2021, Harikane2022}. Moreover, the presence of UV bright galaxies may also be explained in other ways, e.g., via a highly stochastic star formation activity \citep{Sun2023, Pallottini2023, Sun2023a, Shen2023b, Kravtsov2024}, low dust attenuation \citep{Ferrara2023}, AGN contamination \citep{Hegde2024}, or a top-heavy IMF that increases the UV luminosity per SFR \citep{Harikane2023}.

In this work, we analyze the SFE of UV faint galaxies and address the origin of the elevated cosmic UV luminosity density with the help of \hr{}, a novel hydrodynamical, cosmological simulation from the \emph{Feedback in Realistic Environments} (FIRE) project\footnote{\url{https://fire.nort hwestern.edu}} \citep{Hopkins2014}. This simulation is a re-run of the original \fb{} cosmological volume simulation \citep{Feldmann2023} to $z=6.3$ but with 8 times more particles ($2\times{}2048^3$) and at the numerical resolution of standard FIRE zoom-in runs. \hr{} is well suited to analyze the cosmic UV luminosity given its accurate modeling of baryonic processes with the FIRE-2 physics model \citep{Hopkins2018} and its high numerical resolution, which enables it to properly resolve UV faint galaxies and their interstellar medium (ISM). \hr{} also extends previous zoom-in studies with FIRE-2 physics focusing on the SFE and stellar mass -- halo mass relation of high-redshift galaxies \citep{Ma2018b, Ma2019, Sun2023a} as it covers a much larger volume thus improving statistics. Crucially, many properties of galaxies simulated with FIRE-2 physics, such as their masses, SFRs, morphologies, and ISM compositions, have already been validated against observations over a broad range in redshift (e.g., \citealt{Wetzel2016, Hopkins2018, Chan2018, Ma2019, Garrison-Kimmel2019, Feldmann2023, Liang2023, Gensior2023, Bassini2024, Marszewski2024}), i.e., the results of \hr{} are genuine predictions and not the result of tuning to high-redshift observations.

As we will demonstrate in this work, the above arguments in favor of a strongly evolving SFE -- halo mass relation are not substantiated by \hr{}. Instead, the simulation predicts a non-evolving SFE -- halo mass relation and, yet, well reproduces both the UV luminosity functions and the UV luminosity density evolution at $z\sim{}6-14$, suggesting that no fundamentally new physics is involved in setting the UV luminosities of galaxies of low to intermediate mass at these redshifts.

This paper is structured as follows. In Section \ref{sect:methodology} we introduce the \hr{} simulation and describe our data analysis. The UV luminosity functions and the UV luminosity density in the simulation are discussed in Section \ref{sect:UVLFdens}. We measure the SFE -- halo mass relation for different SFR tracers in Section  \ref{sect:SFE} and link these results to the stellar mass to halo mass relation (SHMR) in Section \ref{sect:SHMR}. A theoretical model based on the measured SFE -- halo mass relation is presented in Section \ref{sect:TheoreticalModel}. Its implications for the UV luminosity density evolution and the cosmic reionization history are explored in Sections \ref{sect:ModelImplicationsUV} and \ref{sect:ModelImplicationsReionization}. Caveats are addressed in Section \ref{sect:Caveats}. Finally, we summarize our main findings and conclude in Section \ref{sect:Summary}.

\section{Methodology}
\label{sect:methodology}

\subsection{Simulation set-up and included physics}
\label{sect:Sim}

The simulation introduced in this work, \hr{}, is a re-run of the original \fb{} simulation presented in \cite{Feldmann2023} at a higher numerical resolution but with otherwise identical physics.
\hr{} traces the evolution of gas, stars, and dark matter in a cubic volume with a $L=22.1$ cMpc side length and periodic boundary conditions from the initial redshift $z_{\rm init}=120$ down to $z=6.3$. Initial conditions for the simulation were created with the MUlti-Scale Initial Conditions (MUSIC\footnote{\url{https://www-n.oca.eu/ohahn/MUSIC}}) code \citep{Hahn2011} for Planck 2015 cosmological parameters ($h=0.6774$, $\Omega_{\rm m} = 0.3089$, $\Omega_{\rm b} = 0.0486$, $\Omega_\Lambda=0.6911$, $\sigma_8 = 0.8159$, $n_{\rm s} = 0.9667$; \citealt{PlanckCollaboration2015a}) and with a transfer function calculated by the Code for Anisotropies in the Microwave Background (CAMB\footnote{\url{https://www.camb.info}}; \citealt{Lewis2000}). These initial conditions are identical to those of the original \fb{} run except that they include additional high wavenumber modes. Gas particles are assigned an initial metallicity of $2\times{}10^{-6}$ and a temperature of 201.95 $K/\mu$, where $\mu$ is the mean molecular weight.

The simulated volume consists initially of 2048$^3$ gas particles with a mass of $7823\,M_\odot$ and the same number of dark matter particles with a  mass of $41899\,M_\odot$. The force softening lengths (Plummer equivalent) of gas particles are adaptive and tied to the gas interparticle separation, with a minimum value of 0.5 proper pc at $z\leq{}9$ and 5 comoving pc at $z\geq{}9$. Softening lengths of dark matter particles (star particles) are held fixed at 40 proper pc (4 proper pc) at $z\leq{}9$ and at 400 comoving pc (40 comoving pc) at $z\geq{}9$.

Starting from the initial conditions, the evolution of the matter components in the cosmic volume is followed numerically with GIZMO\footnote{\url{http://www.tapir.caltech.edu/~phopkins/Site/GIZMO.html}}. Gravitational forces are calculated with a tree gravity solver \citep{Springel2005a, Springel2008} while the equations of fluid dynamics are solved with the meshless-finite-mass method \citep{Hopkins2015a}. The FIRE-2 model, an updated version of FIRE, is used to account for baryonic processes and we refer the reader to \cite{Hopkins2018}  for details. Specifically, heating and cooling rates include terms arising from, e.g., free-free, photoelectric, photoionization/recombination, metal-line, fine-structure, and molecular processes \citep{Katz1996a, Verner1996, Wiersma2009b, Ferland1998}. Radiative effects from sources in the simulation volume are calculated in the LEBRON approximation \cite{Hopkins2018}. Photo-ionization and -heating from a spatially uniform, but time-dependent, ultraviolet (UV) background is also included starting at $z=10.6$ \citep{Faucher-Giguere2009}. A Sobolev-length like approximation is used to calculate the shield-shielding from local sources and the UV background \citep{Gnedin2009a, Faucher-Giguere2010, Rahmati2013a}. A total of 15 species (hydrogen, helium, nine metal species, and 4 tracer field for different $r$-process models) are followed in the baryonic components. Metal diffusion from unresolved turbulence is included following \cite{Su2017a}.
Stars may form only in dense ($>1000$ cm$^{-3}$), self-gravitating, Jeans unstable gas at a volumetric rate of $\rho_{\rm SFR}=f_{\rm H2}\,\rho_{\rm gas}/t_{\rm ff}$, where $t_{\rm ff}$ is the local free-fall time and $f_{\rm H2}$ is the molecular gas fraction. The latter is calculated as described in \cite{Krumholz2011c}. Star formation is implemented in a stochastic fashion, whereby individual gas particles transform into star particles, inheriting attributes such as position, mass, momentum, and metallicity in the process. Feedback from stellar sources include mass, momentum, metal, and energy (kinetic, thermal, and radiative) injections as a result of stellar winds and supernovae (type II and type Ia). Feedback from active galactic nuclei (AGNs) is not included in \hr{}.

A total of 97 simulation save-points are stored covering redshifts $z=99$ to $z=6.3$ with an average spacing of $\sim{}9$ Myr. These include 28 full data snapshots (level 0 and 1) as well as 69 `snipshots' (level 2) that store only the most relevant subset of the full data; see \cite{Feldmann2023} for details on the content of snapshot and snipshots. A snapshot takes up about 600-700 GB, while a snipshot requires about 230-260 GB, adding to a grand total of about 35 TB to store \hr{}. Advancing the simulation from the initial state to $z=6.3$ took 5.04 million core hours and a wall-clock time of 20.6 days.

\subsection{Post-processing}
\label{sect:postprocessing}

We use the AMIGA halo finder (AHF; \citealt{Gill2004, Knollmann2009}) to identify dark matter (sub-)halos and the galaxies at their centers. Halo masses and sizes are defined using the virial overdensity criterion by \cite{Bryan1998}. Halos with 300 or more dark matter particles ($M_{\rm halo}\gtrsim{}10^7\,M_\odot$) are analyzed, resulting in about $2.6\times{}10^4$ ($9.8\times{}10^4$, $2.1\times{}10^5$, $3.6\times{}10^5$, $4.4\times{}10^5$) halos at $z=15$ ($z=12$, $z=10$, $z=8$, $z=6.3$). Physical galaxy properties such as stellar masses and star formation rates are measured within spherical apertures with a 3 proper kpc radius from the halo center. This radius is reduced to the virial radius (for main halos) or the tidal radius (for proper sub-halos) if those are smaller than 3 kpc. The aperture size is motivated by JWST observations which show that 2-dimensional aperture diameters of 0.3 and 0.7 arcsec ($1.8$ kpc and $4$ kpc at $z=6$) typically contain 70-80\% and 85-90\% of the total flux \citep{Naidu2022, Donnan2023, Finkelstein2023, Tacchella2023, Adams2024}. However, our results are not very sensitive to the exact choice of this radius. Unless explicitly stated otherwise, satellite galaxies and sub-halos are included in the analysis, except for those objects that are within 3 proper kpc of their parent central galaxy and main halo to avoid double counting.

\begin{figure*}
\begin{tabular}{c}
\includegraphics[width=160mm]{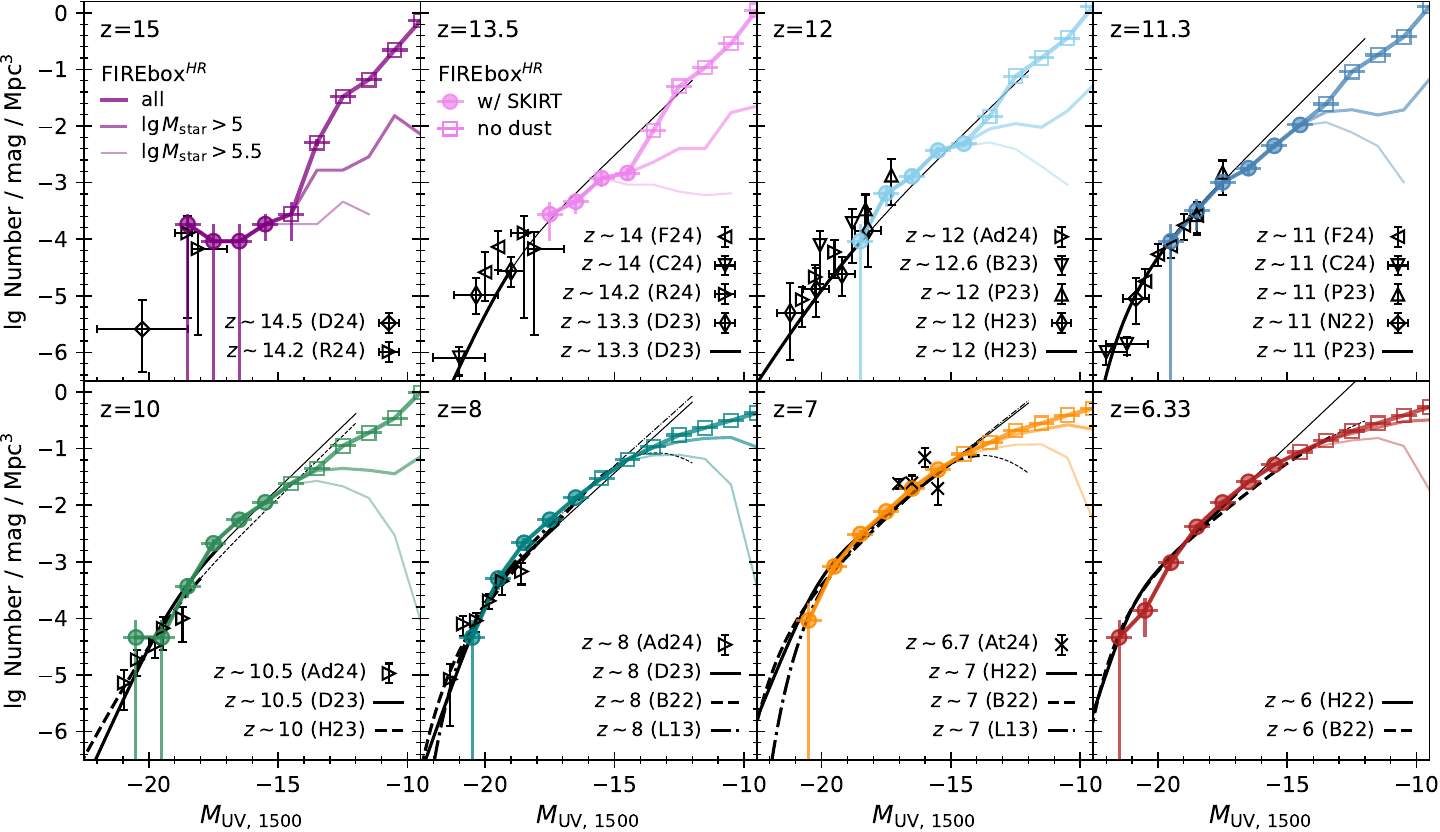}
\end{tabular}
\caption{Rest-frame ultraviolet luminosity functions (UV-LFs) of galaxies in \hr{} at $z\sim{}6-15$ (colored circles, squares, and lines) and comparison with observational data (black symbols and thick lines; thin black lines show extrapolations of the observed UV-LFs to faint magnitudes): \citealt{McLure2013} (L13), \citealt{Harikane2022} (H22), \citealt{Naidu2022} (N22), \citealt{Bouwens2022} (B22), \citealt{Bouwens2023} (B23), \citealt{Donnan2023} (D23), \citealt{Harikane2023} (H23), \citealt{Perez-Gonzalez2023} (P23), \citealt{Adams2024} (Ad24), \citealt{Atek2024} (At24), \citealt{Casey2023} (C24), \citealt{Donnan2024} (D24), \citealt{Finkelstein2024} (F24), and \citealt{Robertson2023b} (R24). The UV luminosities are calculated with the dust-radiative transfer code SKIRT for the more massive galaxies (filled circles) and in an optically thin approximation for the remaining, largely dust-free, galaxies (empty circles). Vertical error bars of the UV-LFs in \hr{} are obtained via bootstrapping, while horizontal error bars indicate bin-widths. Thinner colored lines indicate the UV-LFs when galaxies with stellar masses below $10^5\,M_\odot$ ($\sim{}12-15$ star particles) or below $3\times{}10^5\,M_\odot$ ($\sim{}35-50$ star particles) are excluded. \hr{} reproduces the observed UV LFs at moderate to low luminosities and their evolution over $z\sim{}6-14$.}
\label{fig:UVLF}
\end{figure*}

UV magnitudes of galaxies are calculated with the help of the radiative transfer code SKIRT version 9 \citep{Camps2020}, accounting for dust attenuation, or with a simpler approach described further below in the dust-free approximation. Either method employs the Binary Population and Spectral Synthesis (BPASS, \citealt{Eldridge2009}) library version 2.2.1 \citep{Stanway2018} that includes binary stellar systems and with a \cite{Chabrier2003} initial stellar mass function (IMF), exponential cutoff below 1\,$M_\odot$, high mass slope of -2.3, and a maximum mass of 300 $M_\odot$.
The input to SKIRT are the positions, velocities, initial masses (set to $m_{\rm b}$), metallicities, and ages of all star particles within the virial radius of a given halo. We also provide the distance to the 32th nearest star particle neighbor as a smoothing length, thus adaptively spreading out the light distribution of each star particle to reduce particle noise. Furthermore, we provide positions, velocities, masses, smoothing lengths, metallicities, and temperatures of all gas particles in the halo. The dust-to-metal ratio is set to 0.4 \citep{Dwek1998b, Draine2007j, Watson2011, Li2019} and a Milky-Way grain size distribution is adopted with $R_{\rm V}=3.1$ and $b_{\rm c}=6\times{}10^{-5}$ \citep{Weingartner2001b}. For this dust model FIRE galaxies reproduce the observed IRX-$\beta$ relation and the $L_{\rm IR}$-SFR relation \citep{Liang2021a, Liang2023}. However, as will be shown later, the main findings of this paper are not strongly affected by the presence of dust. We defer a more thorough exploration of dust models to future work.
To reduce computational cost we consider only 2 camera angles per galaxy, face-on and edge-on based on the angular momentum vector of star particles in the central 1 kpc. 
Each voxel in the data cube output by SKIRT stores $\nu{}f_\nu$, where $f_\nu(x, y, \nu)$ is the specific intensity (spectral radiance), for a projected 2-dimensional spatial position $(x, y)$ and frequency $\nu$. Subsequently, the rest-frame specific intensity is averaged over the spectral dimension using a boxcar filter $B$ centered on $0.15\,\mu{}{\rm m}$ with a width of $0.02\,\mu{}{\rm m}$, resulting in the map $f_{\rm UV}(x, y) = \int{}d\nu f_\nu(x, y, \nu) B(\nu) / \nu  / \int{} d\nu B(\nu) / \nu $. Next, the map $f_{\rm UV}$ is integrated over a circle with a 3 proper kpc radius centered on the galaxy, resulting in an estimate of the spectral flux density. The latter is converted into an absolute magnitude in the AB system \citep{Oke1983} via $M_{\rm UV}=-2.5\lg{}f_{\rm UV}-48.6$. While this first method accurately accounts for the effects of dust-radiative transfer, it also incurs a high computational cost.

An alternative, computationally cheaper approach is to calculate UV magnitudes of galaxies in the dust-free limit by summing the luminosities of their star particles. To this end, we bi-linearly interpolate the spectral luminosities (per unit stellar mass) provided by the BPASS library for the given ages and metallicities of all star particles within a 3 proper kpc ball from the center of the galaxy. After multiplying with the initial stellar mass we calculate the spectral UV luminosity of a given star particle by averaging of the 0.15 $\mu{}m$ UV filter band as described above for the SKIRT analysis. We finally sum the spectral UV luminosities and convert them into absolute UV magnitudes.

In this work, we combine both approaches to obtain accurate UV luminosities at a lower computational cost. Specifically, we employ the first method for the $n$-th most massive halos at a given snapshot and the second approach for the remaining halos. Here, $n$ is in the range of 51 to 1001 depending on snapshot. It is chosen to conservatively include all halos with $M_{\rm halo}\geq{}5\times{}10^8\,M_\odot$ at $z\geq{}13$, with $M_{\rm halo}\geq{}10^9\,M_\odot$ at $z\sim{}10-12$, and with $M_{\rm halo}\geq{}3\times{}10^9\,M_\odot$ at $z\leq{}9$. As demonstrated by Fig.~\ref{fig:MUV_SKIRT_vs_nodust} in the appendix, the importance of dust attenuation depends strongly on halo mass but not redshift. 
For halos less massive than $3\times{}10^9\,M_\odot$ the magnitude difference is lower than 0.05 mag on average, allowing us to replace the radiative transfer calculation with the simpler, dust-free estimate in the interest of reducing the computational cost.

Uncertainties for most estimates are calculated via bootstrapping with the help of Python's {\sc scipy.stats.bootstrap} routine. Parametrized fitting uses Python's {\sc scipy.optimize.curve\_fit} function with weights usually set to the 1-$\sigma$ uncertainties obtained via bootstrapping.

\section{Results}

\subsection{UV luminosity function and luminosity density in \hr{}}
\label{sect:UVLFdens}

Because of its high dynamic range, \hr{} includes galaxies with a wide range of properties, including low mass, UV faint galaxies below current observational limits as well as moderately luminous galaxies with significant levels of dust attenuation. For instance, the simulation volume contains about 2300 (540) galaxies with UV magnitudes brighter than $-14$ at $z=6.3$ ($z=10$) as well as a smaller number of luminous galaxies with UV magnitudes reaching up to -21 at $z\sim{}6-7$, -20 at $z\sim{}8-11$, and -18 at $z\sim{}12-15$. To put this in perspective, $z\sim{}8$ galaxies with a rest-frame, dust-attenuated UV magnitude of -14 (-20) have stellar masses of the order of $10^{6}\,M_\odot$ ($3\times{}10^8\,M_\odot$) and halo masses of $10^9\,M_\odot$ ($5\times{}10^{10}\,M_\odot$), see Fig.~\ref{fig:MUV_Mhalo_Mstar} in the appendix. Consequently, \hr{} allows us to both probe the shape of the UV luminosity function (LF) and to explore the evolution of the cosmic UV luminosity density from the end of Cosmic Dawn to the final stages of hydrogen reionization.

\begin{figure*}
\begin{tabular}{cc}
\includegraphics[width=85mm]{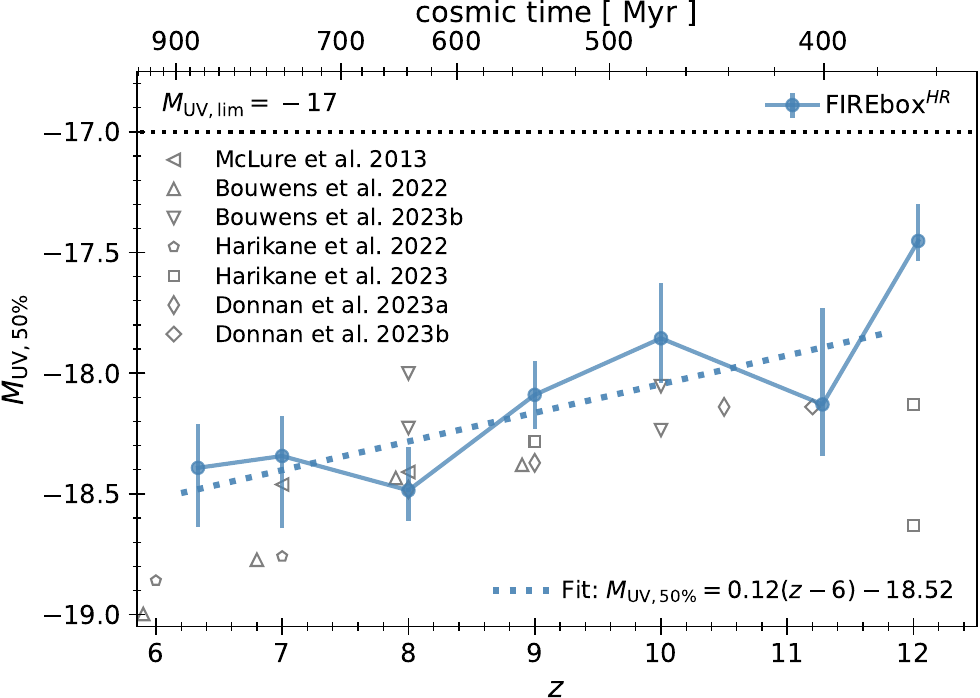} &
\includegraphics[width=85mm]{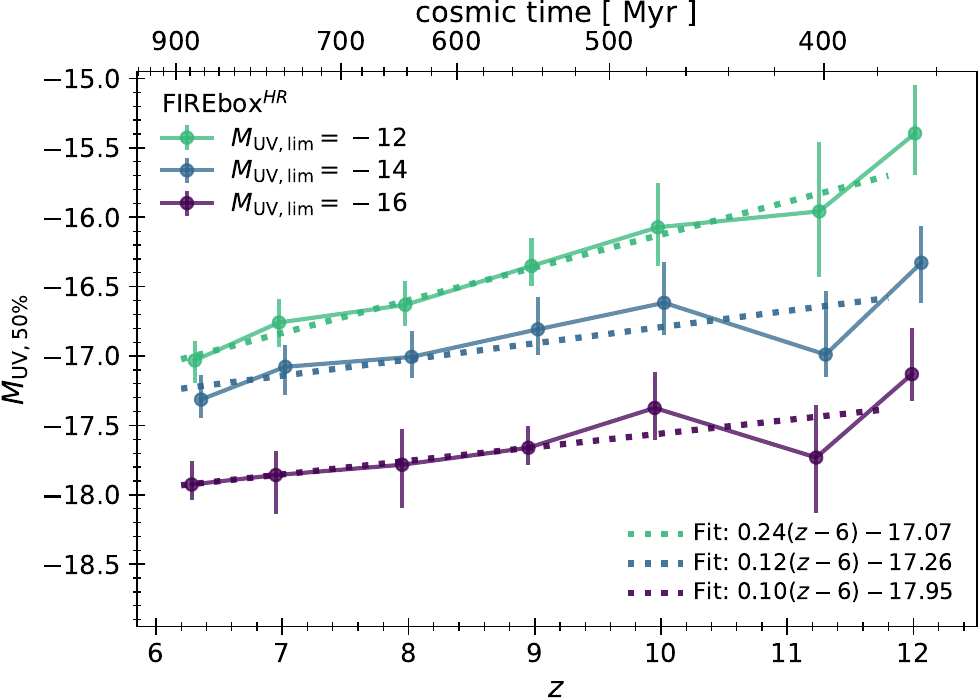}
\end{tabular}
\caption{Typical rest-frame ultraviolet (UV) magnitude, $M_{\rm UV,\,50\%}$, of galaxies dominating the cosmic UV luminosity density as function of redshift. By definition, galaxies brighter than $M_{\rm UV, \,50\%}$ produce the same luminosity density as galaxies with UV magnitudes between $M_{\rm UV, \,50\%}$ and the threshold value $M_{\rm UV, lim}$. 
(Left) Simulation results (blue circles and errorbars) are calculated from the  cumulative UV luminosity distribution of all galaxies in the simulation volume with $M_{\rm UV}\leq{}-17$ (including dust attenuation). Error bars are obtained via bootstrapping. The dotted line shows the result of a linear fit to the simulation data at $z<11.5$. Gray symbols show estimates based on integrating UV luminosity functions from the literature down to $M_{\rm UV}=-17$ including both Double-power-law and Schechter function fits (\citealt{McLure2013}, \citealt{Harikane2022, Harikane2023},  \citealt{Bouwens2023a}, \citealt{Donnan2023, Donnan2023a}). For  $M_{\rm UV, lim}=-17$, galaxies with UV magnitudes fainter than $-18.5$ (and with $M_{\rm halo}<3\times{}10^{10}\,M_\odot$, see Fig.~\ref{fig:MUV_Mhalo_Mstar}) dominate the cosmic UV luminosity density at $z\sim{}6-12$.
(Right) Analogous to left panel but showing predictions for $M_{\rm UV,\,50\%}$ for different choices of $M_{\rm UV, lim}$. The cosmic UV luminosity density in \hr{} at $z\sim{}6-11$ is dominated by faint galaxies ($-16$ to $-17$ for $M_{\rm UV, lim}=-12$).
}
\label{fig:Median_UV_luminosity}
\end{figure*}

According to Fig.~\ref{fig:UVLF}, the UV LFs predicted by \hr{} are in good agreement with observational estimates from JWST and HST at $z\sim{}6-14$ \citep{McLure2013, Harikane2022, Naidu2022, Bouwens2022, Donnan2023, Harikane2023, Perez-Gonzalez2023, Bouwens2023, Robertson2023b, Adams2024, Atek2024, Finkelstein2024, Donnan2024}. In particular, the LFs follow closely the observed double-power-law or Schechter function parametrization over a broad range in UV luminosity and with a normalization that decreases towards higher redshifts.

The simulations suggest a subtle change (`turn-over') in the LF at UV magnitudes just below that of the current observational threshold (around -14 to -15) at $z\sim{}6-8$, where they diverge from the Schechter function extrapolation to faint magnitudes based on observed data. Interestingly, a turn-over at such UV magnitudes appears to be required to avoid producing too many satellites around Milky-Way like galaxies in today's Universe \citep{Boylan-Kolchin2014, Boylan-Kolchin2015a}. The turn-over in the UV LFs of \hr{} appears at $z\lesssim{}10$, pointing to reionization as its physical origin (e.g., \citealt{Wu2024}). At earlier cosmic time, the faint end of the UV LFs resembles a power-law. We caution, however, that the faint end is dominated by low mass galaxies some of which are only marginally resolved in \hr{}. To illustrate the role of numerical resolution, we also show UV LFs once we restrict our sample to galaxies with $M_{\rm star}>10^5\,M_\odot$ ($12 - 15$ star particles) and $M_{\rm star}>10^{5.5}\,M_\odot$ ($35 - 50$ star particles).

Furthermore,  at $z=15$ the figure shows a flattening of the UV LF at the luminous end. We caution the reader, however, that low number statistics, star formation variability, and cosmic variance significantly affect our results at the luminous end and at the earliest cosmic times, while effects related to numerical resolution could potentially have an impact on the predicted faint-end of the UV LF in \hr{}.

The role of dust attenuation is limited to the bright end of our UV LFs. For instance, the number density of galaxies at $z=6.3-10$ with a UV magnitude of $-18.5$ increases by about $0.1-0.3$ dex if we consider intrinsic, i.e., not dust-attenuated UV luminosities, while the LF at UV magnitudes fainter than -17.5 is unchanged. The presence or absence of dust also has no impact on our UV LFs at $z>10$ given the dearth of massive, bright galaxies in the simulation at those redshifts. The limited dust attenuation in \hr{} galaxies reflects their low stellar masses and, consequently, low gas-phase metallicities. For a detailed analysis of the mass -- metallicity relation in high-redshift galaxies with FIRE-2 physics, see \cite{Marszewski2024}.

In this work we calculate the cosmic UV luminosity density by adding the UV luminosities of all galaxies brighter than a given UV magnitude limit $M_{\rm UV, lim}$ and then dividing the total luminosity by the simulation volume. This direct approach is feasible since \hr{} reproduces the number and luminosities of galaxies dominating the cosmic UV luminosity density. 
Specifically, as evidenced by Fig.~\ref{fig:Median_UV_luminosity}, half the dust-attenuated UV luminosity density in galaxies brighter than $M_{\rm UV, lim}=-17$, denoted as $M_{\rm UV, 50\%}$, arises from the subset of galaxies fainter than -18.4 at $z=6.3$ and fainter than about -18 at $z=10$. These UV luminosities correspond to galaxies with typical stellar masses of $M_{\rm star}\sim{}10^{7.5}-10^{8}\,M_\odot$ that reside in halos with $M_{\rm halo}\sim{}10^{10}-10^{10.5}\,M_\odot$, see Fig.~\ref{fig:MUV_Mhalo_Mstar} in the appendix. These stellar masses are in line with observations, e.g., \cite{Endsley2023} find that galaxies with $M_{\rm UV}\sim{}-18.4$ have typical stellar masses of $M_{\rm star}\sim{}10^{7.5}-10^{8}\,M_\odot$ at $z\sim{}6-7$.

A linear regression provides the following estimate of $M_{\rm UV, 50\%}$ for redshifts in the range $\sim{}6-12$
\begin{equation}
M_{\rm UV, 50\%} = 0.12(z-6) - 18.52.
\label{eq:MUV50}
\end{equation}
Our predictions are comparable to measured values based on integrating the observed UV LFs at $z\sim{}6-11$. For instance, we obtain $M_{\rm UV, 50\%}=-18.76$ based on the $z=7$ double-power-law UV LF by \cite{Harikane2022} and $M_{\rm UV, 50\%}\sim{}-18.1$ for the $z=10$ UV LF by \cite{Bouwens2023a}. According to these UV LFs, UV bright ($M_{\rm UV}<-20$) galaxies contain less than $18\%$ and $\sim{}10\%$ of the total UV luminosity density at $z=7$ and $z=10$ for $M_{\rm UV, lim}=-17$. We conclude that the star formation efficiency of massive, UV bright ($M_{\rm UV}<-20$) galaxies has little bearing on the evolution of the cosmic UV luminosity density for $M_{\rm UV, lim}=-17$.

Fig.~\ref{fig:Median_UV_luminosity} also explores how $M_{\rm UV, 50\%}$ changes if the UV magnitude limit is raised to include fainter galaxies. As expected, a fainter UV magnitude limit results in a fainter value for $M_{\rm UV, 50\%}$. In addition, $M_{\rm UV, 50\%}$ shows a stronger redshift evolution for $M_{\rm UV, lim}\geq{}-12$. 
For this faint UV magnitude limit, the cosmic UV luminosity density in \hr{} is dominated by galaxies with $M_{\rm UV,\,50\%}$ near -16 to -16.5 ($z\sim{}8-11$) and -17 ($z\sim{}6-7$), in qualitative agreement with observational estimates \citep{Atek2024, Mascia2024}. Typical stellar and halo masses of such galaxies are $M_{\rm star}\sim{}10^{6.8}-10^{7.5}\,M_\odot$ and $M_{\rm halo}\sim{}10^{9.4}-10^{10}\,M_\odot$, see Fig.~\ref{fig:MUV_Mhalo_Mstar}. Such galaxies also tend to have low dust attenuation, see Fig.~\ref{fig:MUV_SKIRT_vs_nodust}.

\begin{figure}
\begin{tabular}{c}
\includegraphics[width=83mm]{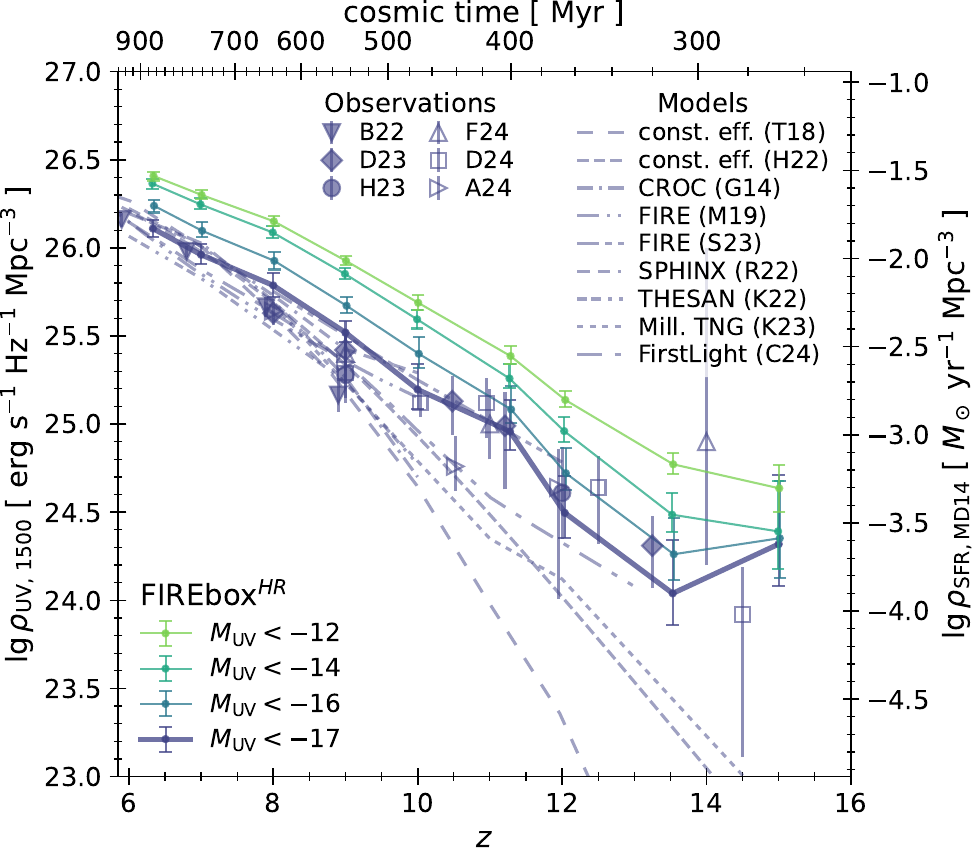}
\end{tabular}
\caption{Evolution of the dust-attenuated ultraviolet (UV) luminosity density in \hr{} at $z\sim{}6-15$ for different values of the UV magnitude limit. Uncertainties (16-84\%) in the UV luminosity density are estimated via bootstrapping. The simulation data is compared with observational data by \protect\cite{Bouwens2022}, \protect\cite{Donnan2023, Donnan2023a}, \protect\cite{Harikane2023}, \protect\cite{Adams2024}, \protect\cite{Finkelstein2024} and \protect\cite{Donnan2024}, as well as with theoretical predictions based on the empirical models by \protect\cite{Tacchella2018} and \protect\cite{Harikane2022}, the Cosmic re-ionization on computers (CROC) simulation suite \protect\citep{Gnedin2014a}, FIRE-2 zoom-in simulation suites by \protect\cite{Ma2019} and \protect\cite{Sun2023a}, the SPHINX$^{20}$ simulation \protect\citep{Rosdahl2022}, the THESAN-1 simulation \protect\citep{Kannan2022}, the Millennium-TNG simulation \protect\citep{Kannan2023}, and the FirstLight zoom-in suite \protect\citep{Ceverino2024}. A UV magnitude limit of -17 is adopted for all literature data. Small horizontal shifts ($\Delta{}z<0.05$) are applied to some data points to increase readability. The treatment of dust attenuation in the various models is discussed in the text. The UV luminosity density predicted by \hr{} agrees well with observational data, in particular those based on recent JWST observations, and it evolves more gradually at high redshift than predicted by many previous models. 
}
\label{fig:UVLD}
\end{figure}

Fig.~\ref{fig:UVLD} shows the cosmic UV luminosity density in \hr{} calculated as described above for different choices of $M_{\rm UV, lim}$. We compare the simulation predictions to observational data from JWST and HST for the common choice of $M_{\rm UV, lim}=-17$ \citep{Bouwens2015, Donnan2023, Donnan2023a, Harikane2023, Adams2024, Finkelstein2024, Donnan2024} finding overall good agreement. At $z\sim{}10-11$, the predicted UV luminosity density is in line with \cite{Donnan2023, Donnan2023a, Donnan2024} and \cite{Finkelstein2024} but higher ($\sim{}0.3$ dex) than the results reported by \cite{Adams2024}. The apparent upturn at $z=15$ in \hr{} is a result of low number statistics combined with a burst of star formation a few Myr earlier boosting UV luminosity. 

The figure also shows predictions for UV magnitude limits fainter than -17. At $z\lesssim{}10$, the cosmic UV luminosity density is not very sensitive to the choice of the limiting magnitude provided the latter is chosen to be fainter than about -14. For instance, the UV luminosity density at $z=8$ increases by 0.06 dex (0.30 dex) if  $M_{\rm UV, lim}$ is increased from -14 to -12 (from -17 to -14). We attribute the insensitivity of the cosmic UV luminosity to the UV magnitude limit at $z\lesssim{}10$ to the shallow slope (turn-over) of the faint end of the UV LFs shown in Fig.~\ref{fig:UVLF}. In contrast, at $z\gtrsim{}11$, i.e., before the onset of reionization, the cosmic UV luminosity depends much more noticeably on the chosen UV magnitude limit. For example, at $z=13.5$, the cosmic UV luminosity density changes by 0.29 dex if $M_{\rm UV, lim}$ is raised from -14 to -12.

Dust attenuation has a generally limited impact on UV luminosity density that further decreases as redshift increases. For $M_{\rm UV, lim}=-17$, the intrinsic UV luminosity density is higher than the dust-attenuated one by 0.22 dex at $z=6.3$, by 0.15 dex at $z=10$, and by less than 0.08 dex at $z=13.5$. The importance of dust attenuation is also reduced for fainter UV magnitude limits, e.g., for $M_{\rm UV, lim}=-14$ the shift is 0.13 (0.07, 0.03) dex at $z=6.3$ ($z=10$, $z=13.5$).

Fig.~\ref{fig:UVLD} also includes the UV luminosity density predicted by constant (i.e., non-evolving) efficiency models \citep{Tacchella2018, Harikane2022}, the Cosmic reionization on computers (CROC) simulation suite \citep{Gnedin2014a}, suites of zoom-in FIRE simulations by \cite{Ma2019} and \cite{Sun2023a}, the SPHINX$^{20}$ simulation \protect\citep{Rosdahl2022}, the THESAN-1 simulation \protect\citep{Kannan2022}, the Millenium-TNG simulation \citep{Kannan2023}, and the FirstLight zoom-in suite \citep{Ceverino2024}, all for $M_{\rm UV, lim}=-17$. The FLARES simulation \citep{Lovell2020, Wilkins2022} reports results only for galaxies brighter than about -18 and is not included in this comparison.

We convert the estimates of the cosmic SFR density by \cite{Tacchella2018} and \cite{Harikane2022} to a luminosity density with the help of the \cite{Madau2014} conversion factor. To account for dust attenuation, we first calculate the typical UV magnitude of galaxies contribution to the cosmic luminosity density via Eq. (\ref{eq:MUV50}). We then follow \cite{Tacchella2018} to derive the expected $A_{\rm V}$ of such galaxies via the method proposed by \cite{Smit2012} and reduce the UV luminosity density accordingly. This dust correction primarily affects results for $z\leq{}7$. For the CROC suite, we numerically integrate the UV LF functions shown in their Fig. 1 to a magnitude limit of -17. These LFs account for dust attenuation in an approximate fashion based on the overall stellar metallicity each galaxy and a redshift dependent dust column density calibrated against observations \citep{Bouwens2009}. The luminosity densities for SPHINX and THESAN are obtained similarly by sampling the provided UV LFs and integrating them to a magnitude limit of -17. The LFs in THESAN account for dust attenuation in a similar way as \cite{Gnedin2014a} but with the redshift-dependent dust-to-metal ratio from \cite{Vogelsberger2020a}, while those from SPHINX are calculated with dust-radiative transfer. The UV luminosity density provided by \cite{Ma2019} accounts for dust attenuation by post-processing with SKIRT. The results of \cite{Sun2023a} are based on integrating their fitted UV LFs at $z=8-12$ over the UV magnitude range -26 and -17 with dust attenuation calculated as described in their article, i.e., by combining the $A_{\rm UV}-\beta_{\rm UV}$ relation by \cite{Fudamoto2020} with the $\beta_{\rm UV}-M_{\rm UV}$ relation by \cite{Cullen2023}. Similarly, we obtain an estimate of the UV luminosity density in Millennium-TNG by integrating the provided Schechter functions fits of the dust-attenuated UV LFs over the UV magnitude range -26 and -17. \cite{Kannan2023} employ an empirical dust-attenuation model fitted to ALMA observations of galaxies at $z\sim{}4-7$ by \cite{Bouwens2016b}. Finally, \cite{Ceverino2024} adopt an empirical dust correction based on the observed relation between UV continuum slope and UV magnitude at $z\sim{}6-9$ \citep{Bouwens2014, Atek2023}.

While these models all predict a UV luminosity density in approximate agreement with both \hr{} and observational data at $z\sim{}6-9$, most of them increasingly underestimate the UV luminosity density towards higher redshift. Exceptions are the SPHINX$^{20}$ simulation \citep{Rosdahl2022}, which matches the predictions of \hr{} at $z=10$ but does not provide estimates for higher redshifts, the FIRE-2 zoom-in suites by \cite{Ma2019} and \cite{Sun2023a}, which match \hr{} out to $z=10$ and $z=12$, respectively, and the FirstLight suite \citep{Ceverino2024}, which underpredicts the observed UV luminosity density at $z\sim{}10-11$ with respect to \cite{Donnan2023, Donnan2023a, Donnan2024} and \cite{Finkelstein2024}, but is in approximate agreement with observations at $z\sim{}12-13$.

\subsection{The star formation efficiency of galaxies}
\label{sect:SFE}

\begin{figure*}
\begin{tabular}{c}
\includegraphics[width=140mm]{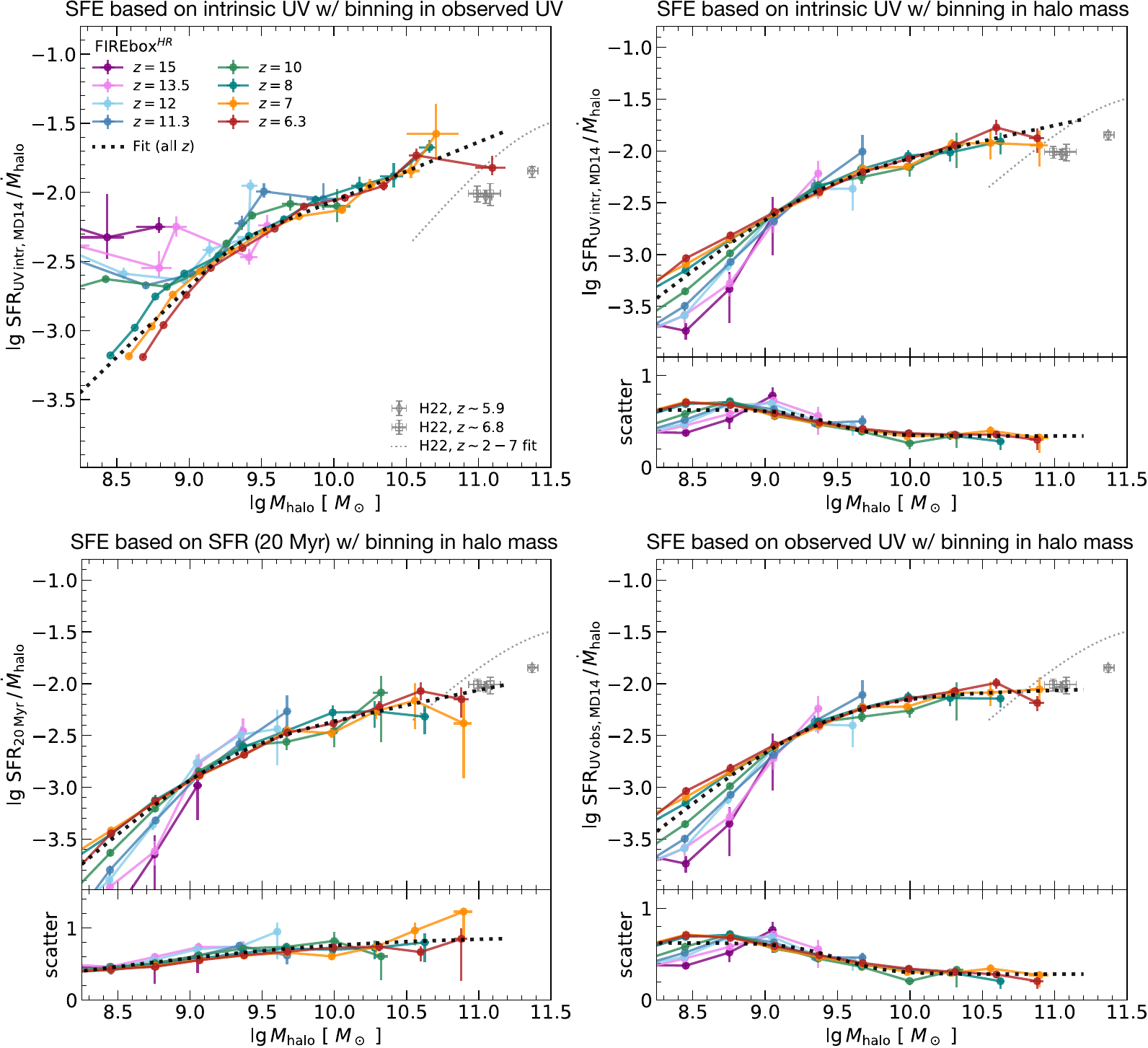}
\end{tabular}
\caption{Star formation efficiency (SFE), the ratio of star formation rate (SFR) and average halo growth rate, in \hr{} as function of redshift and for different halo masses calculated in four different ways. In all panels the halo growth rate is estimated from the halo mass via the fitting function by \protect\cite{Behroozi2015}. UV luminosities and SFRs are measured within 3 proper kpc from the halo center, see Section \ref{sect:postprocessing}. Simulation results are shown by colored symbols and lines. Dotted lines show fits to the \hr{} results for the combined redshift range with fit parameters provided in Tables~\ref{tab:SFEFitParams} and \ref{tab:SFEScatterParams}. Gray symbols and lines reproduce the empirical SFE estimates by \protect\cite{Harikane2022}. 
(Top left) Estimate of the SFE mimicking the observational approach by \protect\cite{Harikane2022}. The average halo mass is calculated in bins of the observed UV luminosity. SFRs are obtained from the intrinsic UV luminosity via a constant conversion factor \citep{Madau2014}, with the intrinsic UV luminosity being derived from the observed UV luminosity of \hr galaxies{} based on the $A_{\rm UV}(\beta)$ relation by \protect\cite{Meurer1999} and the $\beta(M_{\rm UV}, z)$ relation by \protect\cite{Bouwens2014}.
(Top right) Logarithm of the average SFE and the standard deviation (`scatter') of $\lg{}{\rm SFE}$ calculated from the intrinsic UV luminosity of galaxies in \hr{} in bins of halo mass and for different $z$. The scatter of $\lg{}{\rm SFE}$ is $\sim{}0.3$ in $M_{\rm halo}\sim{}10^{11}\,M_\odot$ halos and $\sim{}0.6-0.7$ in $M_{\rm halo}\sim{}10^{9}\,M_\odot$ halos. This scatter measures the variability of the intrinsic UV luminosity at fixed halo mass and does not include contributions from a variability in the halo growth rates.
(Bottom left) Same as top right except the the SFE is calculated from the time-averaged SFR of \hr{} galaxies over the past 20 Myr. Galaxies with zero SFRs are excluded from the calculation of the scatter.
(Bottom right) Same as top right except that the observed, not the intrinsic, UV luminosity is converted into SFR via the constant conversion factor by \protect\cite{Madau2014}.
All approaches to estimating the SFE are consistent with a largely non-evolving SFE -- $M_{\rm halo}$ relation, albeit with significant scatter, in $M_{\rm halo}>10^{9}\,M_\odot$ halos at $z\sim{}6-15$.
}
\label{fig:efficiency}
\end{figure*}

The star formation efficiency (SFE) is a dimensionless quantity that links the star formation activity of a galaxy to the mass growth rate\footnote{The SFE is sometimes defined relative to the \emph{baryonic} accretion rate of halos. In this case, the SFE is larger by a factor $1/f_{\rm bar, uni}$ ($\sim{}0.8$ dex) compared to the definition used in this work, where $f_{\rm bar, uni}\sim{}0.157$ is the universal baryon fraction.} of its halo. In this work we consider both time-averaged SFRs as well as effective SFRs, derived from both the intrinsic and the dust-attenuated UV luminosity, to quantify star formation activity. Average halo growth rates are calculated based on the current halo mass, using the approximation by \cite{Behroozi2015}, both for simplicity and to allow for a better comparison with the literature. The goal of this section is to characterize the SFE as function of halo mass and redshift as predicted by \hr{}.

It is well known that the SFE depends on halo mass, with low mass halos generally forming fewer stars per unit accreted halo mass (e.g., \citealt{Tacchella2018, Harikane2022}). Consequently, low mass halos tend to harbor galaxies with a low stellar mass to halo mass ratio (e.g., \citealt{Behroozi2019}), see Section \ref{sect:SHMR}. `Constant efficiency' models make the ansatz that the SFE -- halo mass relation does not evolve with redshift. These models match the observed evolution of the UV luminosity density at $z\sim{}4-8$ but seemingly predict too steep an evolution at the highest redshifts, see Fig.~\ref{fig:UVLD}. An increase of star formation efficiency with redshift at such early epochs has been suggested as a potential explanation for this discrepancy (e.g., \citealt{Harikane2023, Sipple2023, Ceverino2024}). Given that \hr{} reproduces observational constraints at $z\sim{}6-15$, we can use the simulation to test this hypothesis. 

Fig.~\ref{fig:efficiency} shows the SFE -- halo mass relation as function of redshift for galaxies in \hr{} using four different approaches. 
The first approach mimics the observational procedure by \cite{Harikane2022}. Galaxies are grouped in bins of their dust-attenuated, rest-frame UV luminosities and for each bin we calculate both the average halo mass and the average intrinsic UV luminosity. The latter is obtained from the dust-attenuated luminosities following the procedure by \cite{Smit2012}. The intrinsic UV luminosity is then converted into a SFR via a conversion factor $\kappa=1.15\times{}10^{-28}\,M_\odot\,{\rm yr}^{-1}\,{\rm erg}^{-1}\,{\rm s}\,{\rm Hz}$  \citep{Madau2014}. This conversion factor is a compromise value derived for a \cite{Salpeter1955} IMF and for a variety of assumed star formation and enrichment histories. In contrast, our luminosities are calculated for a \cite{Chabrier2003} IMF and for galaxies based on their actual star formation histories and stellar metallicities as predicted by the simulation. While we consequently find that a lower conversion factor of $6\times{}10^{-29}\,M_\odot\,{\rm yr}^{-1}\,{\rm erg}^{-1}\,{\rm s}\,{\rm Hz}$ better describes the mapping between spectral UV luminosity and SFR for \hr{} galaxies, we stick with the \cite{Madau2014} convention for simplicity. The so-derived SFRs may thus be better understood as renormalized UV luminosities.

Before comparing predictions by \hr{} with the observational data from \cite{Harikane2022}, we note the narrow overlap in halo mass and redshift. Specifically, while \hr{} analyzes the SFE for $M_{\rm halo}<10^{11}\,M_\odot$ halos at $z>6$, \cite{Harikane2022} studies $\gtrsim{}10^{11}\,M_\odot$ halos at $z<7$. At face value, \hr{} suggests an SFE in $\sim{}10^{11}\,M_\odot$ halos at $z\sim{}6$ that is slightly higher (0.2 dex) than the estimates by \cite{Harikane2022}. More importantly, however, is the different scaling of the SFE with halo mass. In \hr{}, the SFE declines only weakly with decreasing halo mass (for $M_{\rm halo}\sim{}10^{9.5}-10^{11}\,M_\odot$). As a consequence, such halos have a much higher SFE than would be expected from a low-mass extrapolation of the \cite{Harikane2022} result. We will discuss the implications of this result for the evolution of the UV luminosity density in Section \ref{sect:ModelImplicationsUV}.

The figure also shows that the SFE -- halo mass relation evolves little, if at all, with redshift for $M_{\rm halo}>10^{9.5}\,M_\odot$ halos. At $M_{\rm halo}<10^{9.5}\,M_\odot$ and $z\gtrsim{}10$ some evolution is apparent, which we attribute to the scatter in halo mass at fixed UV luminosity combined with the evolution of the halo mass function. Because of this scatter, halos with a wide range of masses can produce galaxies of a given luminosity. At high $z$, more massive halos become increasingly rare in the simulation volume, resulting in a smaller average halo mass at fixed UV luminosity.

In the second method galaxies are binned in halo mass and not in dust-attenuated UV luminosity. SFEs estimated this way provide a direct link between the evolution of the cosmic luminosity or SFR and the evolution of the halo mass function \citep{Trenti2010, Tacchella2013, Mason2015, Ferrara2023}. In this method, we also derive SFRs via the \cite{Madau2014} conversion factor but this time from the \emph{intrinsic} UV luminosity of galaxies in \hr{} to remove another source of uncertainty. SFEs are then calculated for each viewing angle of each galaxy and averaged in the given halo mass bin. The figure shows the common logarithm of the bin-averaged SFE.
Overall, we find that the SFE -- halo mass relation estimated this way is similar to the one obtained from the first method. The SFE increases with increasing halo mass for $M_{\rm halo}\sim{}10^{8.5}-10^{11}\,M_\odot$ halos. Furthermore, the SFE -- halo mass relation does not significantly evolve with redshift in halos with $M_{\rm halo}>10^9\,M_\odot$. In contrast to the results from the first method, the efficiency decreases with increasing redshift at the lowest halo masses ($M_{\rm halo}<10^9\,M_\odot$). Clearly, the binning strategy matters for low mass halos as they show large variations in the average SFEs with $z$ as well as a high scatter – the standard deviation of $\lg{\rm SFE}$ is in the range $\sim{}0.4-0.8$. The SFE scatter is considerably lower ($\sim{}0.3-0.4$ dex) in halos with $M_{\rm halo}\sim{}10^{10}-10^{11}\,M_\odot$.

The third method is similar to the second one but uses the actual time-averaged SFR of galaxies in \hr{}. Here, ${\rm SFR}_{20\,\rm{Myr}}$ is defined as the initial mass in stars formed in the last 20 Myr divided by 20 Myr. The resulting SFEs -- halo mass relation has the same shape as in the second method but is shifted downwards by about $0.3$ dex because of the use of the \cite{Madau2014} conversion factor in the previous methods. The scatter in the relation is somewhat larger ($\sim{}0.6-0.7$ dex) and almost independent of halo mass over the $M_{\rm halo}\sim{}10^{9}-10^{11}\,M_\odot$ range. Results for different choices of the averaging time scale are shown in Fig.~\ref{fig:efficiencySupp} in the appendix. They are qualitatively identical to the one for a 20 Myr averaging time aside from the reduction in scatter, and the small downward shift in normalization, with increasing SFR averaging time.

The fourth and final method is similar to the second method but SFRs are derived from dust-attenuated, not intrinsic, UV luminosities with the help of the \cite{Madau2014} conversion factor. The SFE -- halo mass relation obtained this way can be combined with the halo mass function to predict the dust-attenuated UV luminosity function and density. This fourth method results in a SFE -- halo mass relation that is slightly lower in more massive halos, a consequence of the dust-attenuation in such halos, but is virtually identical to the one obtained by the second method for low mass halos.

\begin{table}
\begin{tabular}{llccccc}
Binning & SFR tracer & $A$ & $x_{\rm b}$ & $\Delta$ & $\alpha_1$ & $\alpha_2$ \\ \hline
$M_{\rm halo}$ & UV obs & -2.370 & 9.406 & 0.341 & 1.127 & 0.032 \\
$M_{\rm halo}$ & UV intr ($\beta_{\rm UV}$) & -2.418 & 9.321 & 0.207 & 1.039 & 0.229 \\
$M_{\rm halo}$ & UV intr & -2.420 & 9.311 & 0.295 & 1.092 & 0.299 \\
$M_{\rm halo}$ & SFR (100 Myr) & -2.767 & 9.326 & 0.108 & 1.015 & 0.372 \\  
$M_{\rm halo}$ & SFR (20 Myr) & -2.752 & 9.216 & 0.286 & 1.204 & 0.285 \\
$M_{\rm halo}$ & SFR (5 Myr) & -2.629 & 9.384 & 0.318 & 1.057 & 0.287 \\
$M_{\rm UV}$ & UV intr ($\beta_{\rm UV}$) & -2.377 & 9.331 & 0.078 & 1.020 & 0.425 \\
\end{tabular}
\caption{Parametrization of the average star formation efficiency (SFE) -- halo mass relation predicted by \hr{} for $z\sim{}6-15$. The first column denotes whether galaxies are binned in halo mass or in observed (i.e., dust-attenuated) UV luminosity. The SFE -- halo mass relation is obtained by fitting Eq. (\ref{eq:SFEfit}) to the bin-averaged SFEs and halo masses. The entry in the second column refers to the tracer of star formation rate (SFR) which can either be the actual intrinsic UV luminosity (`UV intr'), the intrinsic UV luminosity inferred from the observed UV luminosity via the method by \protect\cite{Smit2012} (`UV intr ($\beta_{\rm UV}$)'), the observed UV luminosity (`UV obs'), or the time-averaged SFR. Columns 3-7 list the parameters of the broken power law fitting function given by Eq. (\ref{eq:SFEfit}).
Use of the fitting functions should be limited to $M_{\rm halo}\lesssim{}10^{11.2}\,M_\odot$ as more massive halos are not probed by \hr{}. %
The fitting functions shown in the top left, top right, bottom left, and bottom right panels of Fig.~\ref{fig:efficiency} correspond to rows 7, 3, 5, and 1.
}
\label{tab:SFEFitParams}
\end{table}

\begin{table}
\begin{tabular}{llccccc}
Binning & SFR tracer & $x_{\rm b}$ & $\Delta$ & $c_1$ & $c_2$ \\ \hline
$M_{\rm halo}$ & UV obs & 9.507 & 0.176 & 0.624 & 0.282 \\ 
$M_{\rm halo}$ & UV intr ($\beta_{\rm UV}$) & 9.428 & 0.138 & 0.623 & 0.329 \\ 
$M_{\rm halo}$ & UV intr & 9.452 & 0.153 & 0.624 & 0.342 \\
$M_{\rm halo}$ & SFR (100 Myr) & 9.851 & 0.101 & 0.484 & 0.384 \\
$M_{\rm halo}$ & SFR (20 Myr) & 8.427 & 0.909 & 0.000 & 0.894 \\
$M_{\rm halo}$ & SFR (5 Myr) & 8.496 & 0.686 & 0.000 & 0.836
\end{tabular}
\caption{Mass dependence of the scatter of the star formation efficiency (SFE) -- halo mass relation predicted by \hr{} for $z\sim{}6-15$. The scatter of the SFE as defined in this work measures the standard deviation of the logarithm of the UV luminosity or SFR, depending on tracer, at fixed halo mass. The first two columns are the same as in Table~\ref{tab:SFEFitParams}. The final four columns provide the parameters of the sigmoid fitting function given by Eq. (\ref{eq:SFEScatter}).
}
\label{tab:SFEScatterParams}
\end{table}

We fit the average SFE -- halo mass relation over the mass range $10^{8.2}\,M_\odot < M_{\rm halo} < 10^{11.2}\,M_\odot$ with a broken power-law
\begin{equation}
\lg{\rm SFE}(x) = A + \alpha_1(x-x_{\rm b}) + (\alpha_2 - \alpha_1)\,\Delta{}\,\left[\ln\left(1 + e^{\frac{x - x_{\rm b}}{\Delta{}}}\right) - \ln{}2\right],
\label{eq:SFEfit}
\end{equation}
with $x=\lg{}M_{\rm halo}$, and 5 free parameters that specify the pivot log halo mass $x_{\rm b}$, the amplitude $A$ of the relation at $x_{\rm b}$, the low-mass slope $\alpha_1$, the high-mass slope $\alpha_2$, and the smoothness of the slope change $\Delta{}$. The fit is based on the binned, average SFE -- halo mass relation shown in Fig.~\ref{fig:efficiency}, combining $z\sim{}6-15$, and accounts for the uncertainties of the SFE in each bin. Bins with fewer than 3 galaxies are excluded from the analysis. The fit is carried out with the \texttt{scipy.optimize.curve\_fit} function and the resulting fit parameters are provided in Table~\ref{tab:SFEFitParams}.

All methods infer an average SFE -- halo mass relation with the same qualitative properties, but with some differences in detail. First, the normalization is lower by 0.2-0.3 dex when the SFE is calculated based on the actual SFR instead of the UV luminosity. This difference is largely a result of using the \cite{Madau2014} conversion factor which is about 0.28 dex too high given the specific star formation histories and metallicities of galaxies in \hr{}. Secondly, using the observed UV luminosity instead of the intrinsic one as a SFR tracer results in a flatter high-mass slope ($\sim{}0.03$ vs $\sim{}0.3$) as a result of dust extinction affecting preferentially galaxies in more massive halos. Finally, binning in UV luminosity results in a moderately steeper high mass slope.

In a similar fashion we fit the mass dependence of the scatter of the SFE, i.e., the standard deviation of $\lg{}{\rm SFE}$, with a sigmoid function
\begin{equation}
{\rm std}(\lg{}{\rm SFE})(x)  = c_1 + (c_2 - c_1)\left[1 + e^{-\frac{x-x_{\rm b}}{\Delta{}}}\right]^{-1},
\label{eq:SFEScatter}
\end{equation}
with $x=\lg{}M_{\rm halo}$. The four fit parameters are the scatter $c_1$ and $c_2$ in the limit of low mass and high mass halos, the smoothness of the transition $\Delta{}$, and the log halo mass at the transition $x_{\rm b}$. Fit parameters can be found in Table~\ref{tab:SFEScatterParams}.

The scatter of the relation is of moderate size ($\sim{}0.3-0.7$ dex). For UV based estimates, we find a comparably small scatter ($\sim{}0.3$ dex) for massive halos ($M_{\rm halo}\sim{}10^{11}\,M_\odot$) and a higher scatter at lower masses (e.g., $\sim{}0.6-0.7$ dex for $M_{\rm halo}\sim{}10^{8.5}\,M_\odot$). For a SFR averaging-time of 100 Myr the scatter is approximately constant ($\sim{}0.4$ dex) over the same mass range, see also Fig.~\ref{fig:efficiencySupp} in the appendix. Shorter SFR averaging-times result generally both in higher scatter and a reversed mass dependence, e.g., for a 20 Myr averaging time scale the scatter is $\sim{}0.8$ dex at $M_{\rm halo}\sim{}10^{11}\,M_\odot$ and $0.4$ dex at $M_{\rm halo}\sim{}10^{8.5}\,M_\odot$. While averaging times of $\sim{}10-100$ Myr are often used to mimic SFR tracers based on far UV \citep{Leroy2012b, Feldmann2012c, FloresVelazquez2021}, the different halo mass dependence of the scatter shows that the variability of the UV luminosities of galaxies is not fully captured by averaging SFRs over fixed time-intervals.

In summary, all four approaches of quantifying the SFE of galaxies reveal a well defined SFE -- halo mass relation with a moderate amount of scatter (Fig.~\ref{fig:efficiency}). The SFE -- halo mass relation is approximately redshift independent for halos of intermediate mass ($M_{\rm halo}\sim{}10^{9}-10^{11}\,M_\odot$) at $z\sim{}6-10$ and our results are consistent with an extension of this redshift independence to higher $z$. We cannot rule out, however, potential changes in $M_{\rm halo}>10^{10}\,M_\odot$ halos at $z>11$, given the limitations in the number of massive halos at high redshift in \hr{}.
Furthermore, we find that the SFE -- halo mass relation is monotonically increasing with mass for low to intermediate mass halos ($M_{\rm halo}\lesssim{}10^{11}\,M_\odot$). The SFE of galaxies residing in these halos thus increases as the halos themselves grow over cosmic time, i.e., the SFEs of individual galaxies tend to \emph{decrease} with increasing redshift.

How can we understand that the average SFE -- halo mass relation is approximately independent of redshift and only weakly evolving with halo mass for intermediate mass halos? A speculative but intuitive explanation is based on the ansatz that the average SFR scales as
\begin{equation}
{\rm SFR} = \epsilon{}\frac{M_{\rm halo}}{t_{\rm ff}},
\label{eq:SFR_model1}
\end{equation}
where $\epsilon$ is a (mass dependent) efficiency factor and $t_{\rm ff}\approx{}0.06\,H^{-1}(z)$ is the halo free fall time \citep{Ferrara2023, Ferrara2024}. We can motivate this equation on dimensional grounds and via the equilibrium ansatz \citep{Bouche2010, Dave2012a} balancing gas inflow into galaxies with star formation and outflows from galaxies. In equilibrium, $\dot{M}_{\rm gas, in} = (1-R+\eta{})\,{\rm SFR}$, where $\dot{M}_{\rm gas, in}$ is the gas inflow rate into galaxies, $R$ is the mass return fraction from evolving stellar populations in the instantaneous recycling approximation, and $\eta$ is the mass loading factor of galactic outflows. Provided the gas inflow rate into galaxies scales proportional to $M_{\rm halo}/t_{\rm ff}$, we recover Eq.~(\ref{eq:SFR_model1}) with $\epsilon \propto{} 1/(1-R+\eta{})\sim{}1/\eta$ for low mass galaxies. The alternative choice $\dot{M}_{\rm gas, in}\propto{}\dot{M}_{\rm halo}$ is discussed further below.

In FIRE-2, gas ejection from high redshift galaxies of moderate mass is approximately `energy-driven'  ($\eta=V_{\rm halo}^{-2}$) \citep{Pandya2021}. Equivalently, we can equate the efficiency $\epsilon$ with the gas ejection efficiency \citep{Dayal2014} which scales proportional to the square of the escape velocity in the mass range of interest. Physically, this scaling arises from (supernova) feedback expelling gas from galaxies thus reducing their star formation activity.
Since $V^2_{\rm halo}\propto{}(1+z)M_{\rm halo}^{2/3}$ and $H(z)\propto(1+z)^{3/2}$ at $z\sim{}6-15$, Eq.~(\ref{eq:SFR_model1}) predicts
\begin{equation}
{\rm SFR} \propto M_{\rm halo}^{5/3} (1+z)^{5/2}.
\label{eq:SFR_model}
\end{equation}

The halo growth rate can be approximated as a power law of the halo mass and redshift \citep{Krumholz2012},
\begin{equation}
\dot{M}_{\rm halo}\propto{}M_{\rm halo}^{1.14}(1+z)^{2.4}.
\label{eq:dotMhalo_model}
\end{equation}
This expression is less accurate than the halo accretion rate formula by \cite{Behroozi2015} used in this work but useful here for illustrating purposes. 

Combining (\ref{eq:SFR_model}) with (\ref{eq:dotMhalo_model}), and using our definition of the SFE as ${\rm SFR}/\dot{M}_{\rm halo}$,
we obtain
\begin{equation}
{\rm SFE} \propto{} M_{\rm halo}^{0.53} (1+z)^{0.1}.
\label{eq:SFE_model}
\end{equation}
This simple model thus reproduces, at least qualitatively, the effective redshift independence of the SFE in \hr{} and the weak scaling of SFE with halo mass. The weak halo mass dependence is a consequence of the near halo mass independence of the specific halo growth rate $\dot{M}_{\rm halo}/M_{\rm halo}$ and the weak halo mass scaling of $\epsilon\propto{}V^2_{\rm halo}$. The redshift independence of the SFE arises from the canceling of the various redshift scalings of $t_{\rm ff}$, $\dot{M}_{\rm halo}$, and $\epsilon$. An alternative model assuming $\dot{M}_{\rm gas, in}\propto{}\dot{M}_{\rm halo}$ \citep{Furlanetto2017} predicts ${\rm SFE}\propto{}\epsilon\propto{}V_{\rm halo}^{2}\propto{}M_{\rm halo}^{2/3}(1+z)$, i.e., a similar scaling of the SFE with halo mass but a slightly stronger evolution with redshift ($\sim{}0.3$ dex over $z\sim{}6-14$).

\subsection{The stellar mass -- halo mass relation}
\label{sect:SHMR}

\begin{figure}
\begin{tabular}{c}
\includegraphics[width=80mm]{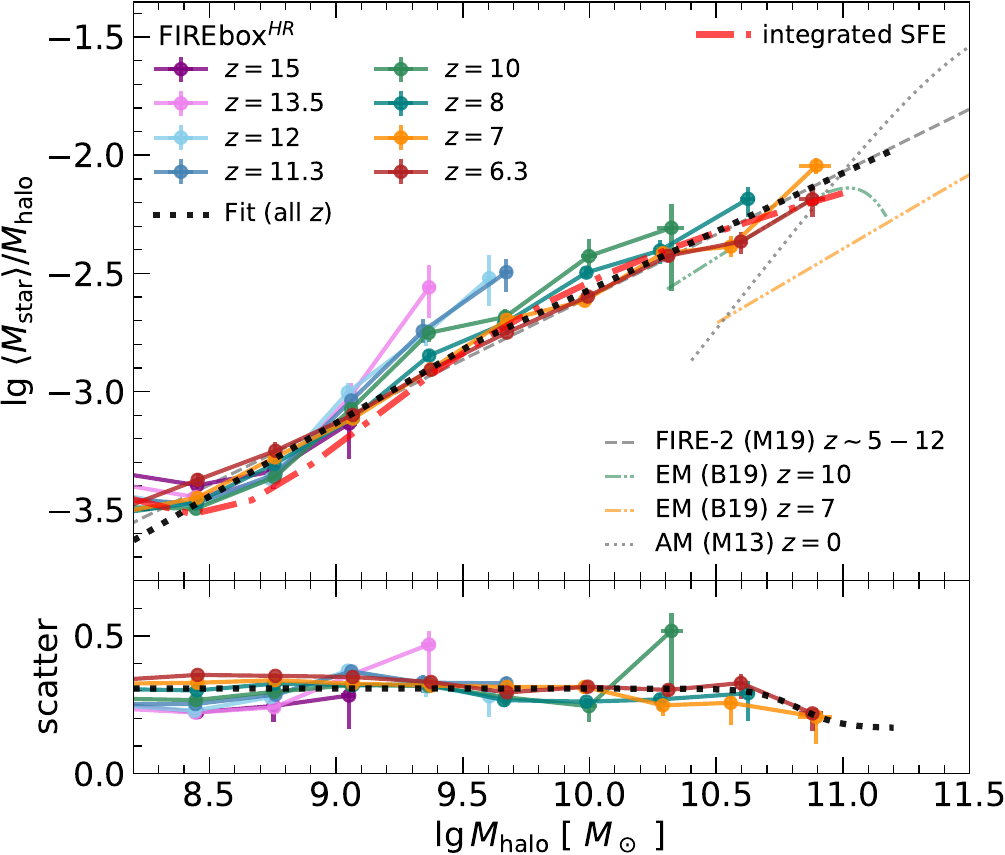}
\end{tabular}
\caption{Stellar mass -- halo mass relation (SHMR) in \hr{} at $z\sim{}6-15$ (symbols and solid lines). The upper panel shows the average stellar mass per halo mass in bins of the latter, while the lower panel reports the standard deviation of $\lg{}M_{\rm star}$ for each halo mass bin (the `scatter'). 
Halos that do not contain star particles ($<1\%$ of selected halos with $M_{\rm halo}\sim{}10^{8.5}\,M_\odot$ at $z=6.3$, and lower fractions at higher $z$) are excluded from the figure. Stellar masses are measured within radii of 3 proper kpc from the respective halo center.
The dashed line reproduces the SHMR for a suite of FIRE-2 zoom-in simulations reported by \protect\cite{Ma2019} (M19). Double-dot-dashed lines show predictions of the \textsc{UniverseMachine} empirical model by \protect\cite{Behroozi2019} (B19) for $z=7$ and $z=10$. For comparison, we also include the abundance matching prediction for the SHMR in today's Universe by \protect\cite{Moster2013} (M13), which agrees well with FIRE-2 predictions at $z=0$ \citep{Hopkins2018}.
The SHMR for intermediate mass halos does not appear to evolve strongly with redshift at $z>6$ but differs from the SHMR in the local Universe. Dotted lines show fits to the \hr{} results for the combined redshift range (see text).
The SHMR obtained from integrating the star formation efficiency -- halo mass relation (red dot-dashed line) agrees well with the actual SHMR in \hr{}.
}
\label{fig:SHMR}
\end{figure}

The stellar mass -- halo mass relation (SHMR) is a fundamental diagnostic of galaxy evolution that can in principle be constrained, e.g., via  abundance matching \citep{Kravtsov2004, Vale2004, Behroozi2010a, Moster2013} and empirical modeling (e.g., \citealt{Moster2018, Behroozi2019}). However, the scaling and redshift evolution of the SHMR at high $z$ is still debated with some observational and theoretical constraints pointing towards strong redshift evolution (e.g., \citealt{Harikane2016, Sun2016, Behroozi2019, Ceverino2024}) and others suggesting weak or no evolution (e.g., \citealt{Feldmann2017a, Tacchella2018, Ma2018b, Ma2019, Stefanon2021}).

Fig.~\ref{fig:SHMR} plots the SHMR in \hr{} for $M_{\rm halo}\sim{}10^{8.2}-10^{11.2}\,M_\odot$ halos at $z\sim{}6-15$. In agreement with previous results from FIRE-2 zoom-in simulations covering the $z\sim{}5-12$ redshift range \citep{Ma2018b, Ma2019}, the SHMR in \hr{} does not strongly evolve at such early cosmic epochs. Weak trends include an increase in the stellar mass with redshift (especially at $z>11$) in $M_{\rm halo}\sim{}10^9-10^{10}\,M_\odot$ halos and a slight decrease of the scatter with redshift at the lowest masses.
The SHMR in \hr{} at $z>6$ differs significantly, however, from the SHMR in FIRE-2 simulations at low redshift \citep{Hopkins2018}. Because of the much flatter slope of the relation at $z>6$ in \hr{}, galaxies residing in halos with $M_{\rm halo}<10^{11}\,M_\odot$ at early cosmic time have higher stellar masses compared with $z=0$ galaxies in halos of the same mass \citep{Moster2013}. The empirical model by \cite{Behroozi2019} also suggests a comparably flat slope, roughly in line with \hr{}, but predicts an evolution in the normalization at $z>6$. We caution that the SHMR at high $z$ is yet to be fully constrained observationally (see, e.g., \citealt{Sun2016, Rodriguez-Puebla2017, Moster2018}).

The redshift-combined relation between average stellar mass and halo mass in \hr{} is well fit by a (broken) power law function, see Eq. (\ref{eq:SFEfit}) with SFE replaced by $M_{\rm star}$, with the following parameters: $A=-2.787$, $x_{\rm b}=9.563$, $\Delta{}=0.003$, $\alpha_1=0.617$, $\alpha_2=0.496$, i.e., it is close to a single power law with a slope of $\sim{}0.5-0.6$.
The scatter of the relation is $\sim{}0.2-0.4$ dex and it increases weakly with decreasing halo mass over the considered mass range. The halo mass scaling of the scatter is approximated by a sigmoid function analogous to Eq. (\ref{eq:SFEScatter}) with parameters $x_{\rm b}=10.827$, $\Delta{}=0.083$, $c_1=0.308$, and $c_2=0.165$.

The SFE -- halo mass relation is closely linked to the stellar mass -- halo mass relation. If we assume that the change in stellar mass is $\dot{M}_{\rm star} = (1-R)\,{\rm SFR}$, where $R$ is the mass return fraction in the instantaneous recycling approximation (e.g., \citealt{Bouche2010, Lilly2013c, Feldmann2015a}), we obtain $\dot{M}_{\rm star} = (1-R)\,{\rm SFE}\,\dot{M}_{\rm halo}$ for a SFE defined in terms of SFR and halo growth rate $\dot{M}_{\rm halo}$. This equation can be easily integrated for a non-evolving SFE, resulting in
\begin{equation}
{M}_{\rm star}(M_{\rm halo}) = M_{\rm star, min} + (1-R) \int_{M_{\rm halo, min}}^{M_{\rm halo}} {\rm SFE}(M_{\rm halo}^\prime) dM_{\rm halo}^\prime,
\label{eq:MstarIntSFE}
\end{equation}
where $M_{\rm halo, min}$ is a chosen halo mass and $M_{\rm star, min}=M_{\rm star}(M_{\rm halo, min})$ is the average stellar mass of galaxies in halos of mass $M_{\rm halo, min}$. Since the fit of the SFE -- halo mass relation if obtained from halos with masses $10^{8.2}-10^{11.2}\,M_\odot$, we adopt $M_{\rm halo, min}=10^{8.2}\,M_\odot$ and set $M_{\rm star, min}= 3.5\times{}10^{-4}\,M_{\rm halo, min}\sim{}10^{4.7}\,M_\odot$. The alternative choice of setting $M_{\rm halo, min}=M_{\rm star, min}=0$ lowers the predicted stellar mass in $\sim{}10^{8.5}\,M_\odot$ halos by about $0.26$ dex and has a negligible effect ($\lesssim{}0.05$ dex) on the stellar mass in $M_{\rm halo}\gtrsim{}10^{9}\,M_\odot$ halos.

In Fig.~\ref{fig:SHMR} we show the stellar fraction obtained by integrating the SFE in \hr{} via Eq.~(\ref{eq:MstarIntSFE}) for a 5 Myr SFR averaging time (row 6 in Table~\ref{tab:SFEFitParams}). A short past averaging time is selected to reduce the bias in estimating the SFRs while the latter are quickly rising (e.g., \citealt{Tacchella2018}). Given the low masses and young stellar ages of many of the galaxies in our sample, we set $R=0.1$ which corresponds to the return fraction of a 20 Myr old single stellar population with $Z\lesssim{}0.007$ in the FIRE-2 model. The predicted SHMR from Eq.~(\ref{eq:MstarIntSFE}) closely matches the fit to the actual SHMR in \hr{}, differing by less than $\sim{}0.1$ dex.

One interesting aspect is the scaling of this predicted stellar fraction with halo mass at the massive end. 
Provided $M_{\rm star, min}\ll{}M_{\rm star}$, the halo mass scaling of the SFE is closely related to the scaling of the SHMR. Specifically, for a power-law ${\rm SFE}\propto{}M_{\rm halo}^\alpha$ (with $\alpha>-1$), $M_{\rm star}\propto{}M_{\rm halo}^{1+\alpha}$ and the stellar fraction $f_{\rm star}=M_{\rm star}/M_{\rm halo}\propto{}M_{\rm halo}^\alpha$, i.e., the `integrated SFE' $f_{\rm star}$ has the same scaling with halo mass as the instantaneous SFE.
A broken power-law fit to the $f_{\rm star}$ -- halo mass relation obtained via Eq.~(\ref{eq:MstarIntSFE}) returns a high mass slope of $0.33$, in line with the expectation from the halo mass scaling of the SFE in \hr{}, see Table~\ref{tab:SFEFitParams}. This slope is, however, lower than the high mass slope ($\sim{}0.50$) of the actual $f_{\rm star}$ -- halo mass relation in \hr{}.

Several factors could account for this difference in the high-mass slope. Firstly, Eq.~(\ref{eq:MstarIntSFE}) does not consider the contribution of mergers to stellar mass growth, an effect that tends to increase with mass (e.g., \citealt{Behroozi2015, Behroozi2019}) and potentially steepens the SHMR beyond what star formation alone would suggest. Secondly, the equation does not account for the scatter of the SFE at given halo mass, with the impact of the scatter likely depending both on its magnitude and the autocorrelation time of the star formation rate. Thirdly, the SFE is calculated by dividing the SFR of a halo of mass $M_{\rm halo}$ by an average halo growth rate of halos of that mass, thereby ignoring variations in accretion rates at fixed halo mass. Lastly, the average halo growth rates in \hr{} may differ slightly from those in \cite{Behroozi2015}, as the latter are based on simulations without baryonic effects (e.g., \citealt{Sawala2013, Beltz-Mohrmann2021}). For instance, removing the weak halo mass dependence of the specific halo growth rate at $z>6$ would increase the high-mass slope of the SFE -- halo mass relation by about 0.1, aligning it more closely with the SHMR slope in \hr{}. Discrepancies in the average halo growth rates change the nominal SFE fitting parameters and the predictions of Eq.~(\ref{eq:MstarIntSFE}) but have otherwise no impact on this work's predictions as the latter involve the product of the SFE and the average halo growth rate, thus depending only on the underlying SFR or corresponding UV luminosity.

\subsection{A theoretical model for the UV luminosity function and density at $z>6$}
\label{sect:TheoreticalModel}

As discussed in Section \ref{sect:UVLFdens}, \hr{} predicts UV luminosity densities at $z>10$ in good agreement with observational data by JWST. In this section, we introduce a simple model that provides a theoretical explanation for this result. Our approach builds on the ideas of previous work \citep{Trenti2010, Tacchella2013, Mason2015, Furlanetto2017, Tacchella2018, Sabti2022a, Harikane2022, Lovell2023a, Ferrara2023} but differs in its use of a SFE -- halo mass relation derived from a cosmological, hydrodynamical simulation.
Given that the FIRE-2 physics model is not explicitly tuned to specific observables, this model is thus theoretical, and not empirical, in nature. This theoretical framework demonstrates that the gradual evolution of the UV luminosity density can be directly attributed to the baryonic processes, including stellar feedback and gas accretion, that take place in galaxies at high redshifts.

The core component of the model is the SFE -- halo mass relation predicted by \hr{}. Since we are interested in estimating the \emph{observed} UV luminosity density $\mathcal{L}$, we define the SFE $\mathcal{S}$ as the ratio between $\mathcal{L}$, converted to units of SFR by multiplication with a constant conversion factor $\kappa$, and the averaged halo accretion rate\footnote{In this section we denote halo mass as $M$ instead of $M_{\rm halo}$ for notational simplicity.} $\dot{M}$, i.e.,
\begin{equation}
\mathcal{S}=\kappa{}\mathcal{L}/\dot{M}.
\label{eq:lum_to_SFE}
\end{equation}
Analogous definitions may be used to define $\mathcal{S}$ in terms of, e.g.,  the intrinsic UV luminosity density or the actual SFR density. The function $\dot{M}$ measures the averaged and smoothed accretion rate for halos of mass $M$ at redshift $z$ \citep{Behroozi2015}. We adopt the constant $\kappa=1.15\times{}10^{-28}\,M_\odot\,{\rm yr}^{-1}\,{\rm erg}^{-1}\,{\rm s}\,{\rm Hz}$ as the conversion factor between luminosity and SFR \citep{Madau2014}. However, this choice has no impact on the model predictions because the same factor is used to constrain the average SFE -- halo mass relation from the UV luminosities of \hr{} galaxies.

We model the SFE of galaxies residing in halos of mass $M$ at redshift $z$ as a log-normally distributed random variable, i.e.,
\begin{equation}
\begin{split}
\mathcal{S} & \sim{}{\rm Lognormal}(\mu, \sigma^2),\,{\rm with} \\
\mu &=
\ln\langle{}\mathcal{S}\rangle{} - \sigma^2 / 2, \\
\sigma &= {\rm std}(\lg{}\mathcal{S})\,\ln{10},
\end{split}
\label{eq:LognormalModel}
\end{equation}
where $\langle{}\mathcal{S}\rangle{}$ denotes the average SFE and ${\rm std}(\lg{}\mathcal{S})$ is the scatter of the SFE in dex. Both are inferred directly from \hr{} and they may, in principle, vary with both $M$ and $z$. An advantage of the above parametrization, in contrast to alternatives based on, e.g., the median SFE \citep{Shen2023a}, is that it cleanly distinguishes between changes to the average SFE and changes to the scatter of the SFE. This is useful because absent a UV magnitude limit, the cosmic UV luminosity density depends on the \emph{average} SFE -- halo mass relation (see below) and not on its scatter.

It follows from Eq.~(\ref{eq:lum_to_SFE}) that $\mathcal{L}$ is also a log-normally distributed random variable with $\langle{}\mathcal{L}\rangle{}=\langle{}\mathcal{S}\rangle{}\dot{M}/\kappa$. In contrast, the rest-frame UV magnitude defined as
\begin{equation}
\mathcal{M} = \mathrm{Mag}(\mathcal{L}) \equiv{} -2.5 \lg{}\left(\frac{\mathcal{L}}{4\pi\,(10\,{\rm pc})^2}\right) - 48.6
\label{eq:lum_to_mag}
\end{equation}
is normally distributed at a given $M$ and $z$ with mean and standard deviation given by
\begin{equation}
\langle{}\mathcal{M}\rangle{} =  \mathrm{Mag}(e^\mu \dot{M} / \kappa)\,\,{\rm and}\,\,{\rm std}(\mathcal{M}) = \frac{2.5}{\ln{10}}\sigma.
\label{eq:mag_pdf_params}
\end{equation} 

According to the model, the cosmic UV luminosity density in galaxies brighter than some limit luminosity $l$ is
\begin{equation}
\begin{split}
\rho_{\rm UV}(z) &=\int_0^{\infty} dM \frac{d n}{d M}(M, z) \int_{l}^{\infty} p(\mathcal{L}=L\vert{}M, z) L\, dL \\
&=\int_0^{\infty} dM \frac{d n}{d M} \langle{}\mathcal{L}\rangle{}\,\Phi{}\left(\frac{\ln(\langle{}\mathcal{L}\rangle{}/l) + \sigma^2/2}{\sigma}\right) \\
&=\frac{1}{\kappa}\int_{-\infty}^{\infty} d\lg{}M \frac{d n}{d\lg{}M} \dot{M} \langle{}\mathcal{S}\rangle{}_{\rm eff}.
\end{split}
\label{eq:Model_rho_UV}
\end{equation}
In the equation above, $\Phi$ is the cumulative distribution function of the standard normal distribution, $p$ is the probability density function (pdf) of $\mathcal{L}$, $dn/dM$ is the halo mass function, and $\langle{}\mathcal{S}\rangle{}_{\rm eff}$ is an effective SFE defined by
\begin{equation}
\langle{}\mathcal{S}\rangle{}_{\rm eff} = \langle{}\mathcal{S}\rangle{}\,\Phi{}\left(\frac{\ln(\langle{}\mathcal{S}\rangle{}\dot{M}/(\kappa{}l)) + \sigma^2/2}{\sigma}\right).
\label{eq:SFE_eff}
\end{equation}
The effective SFE is $\langle{}\mathcal{S}\rangle{}/2$ if $\langle{}\mathcal{S}\rangle{}=\kappa{}l\exp(-\sigma^2/2)/\dot{M}$.
In case of vanishing scatter ($\sigma\rightarrow{}0$), $\langle{}\mathcal{S}\rangle{}_{\rm eff}=\langle{}\mathcal{S}\rangle{}$ if $\langle{}\mathcal{S}\rangle{}>\kappa{}l/\dot{M}$ and $\langle{}\mathcal{S}\rangle{}_{\rm eff}=0$ otherwise. In case of a vanishing limit luminosity ($l\rightarrow{}0$), $\langle{}\mathcal{S}\rangle{}_{\rm eff}=\langle{}\mathcal{S}\rangle{}$ and $\rho_{\rm UV}(z)$ is independent of $\sigma$.

We similarly obtain the number density of galaxies brighter than some luminosity $l$ or magnitude $m$ as
\begin{equation}
\begin{split}
n(>l, z) &=\int_0^{\infty} dM \frac{d n}{d M} \,\Phi{}\left(\frac{\ln(\langle{}\mathcal{L}\rangle{}/l) - \sigma^2/2}{\sigma}\right),\,{\rm and} \\
n(<m, z) &=\int_0^{\infty} dM \frac{d n}{d M} \,\Phi{}\left(\frac{m - \langle{}\mathcal{M}\rangle{}}{{\rm std}(\mathcal{M})}\right).
\end{split}
\label{eq:Model_num_UV}
\end{equation}

The UV luminosity function predicted by the model is
\begin{equation}
\begin{split}
 \phi(m, z) &= \frac{dn(<m, z)}{dm} = n p(\mathcal{M}=m \vert{} z) \\
                &= \int_0^{\infty} dM \frac{d n}{d M} p(\mathcal{M}=m\vert{}M, z),
 \label{eq:Model_UV_LF}
 \end{split}
 \end{equation}
 where $p$ is the pdf of $\mathcal{M}$ with parameters given by (\ref{eq:mag_pdf_params}).
We evaluate the integrals in Eqs.~(\ref{eq:Model_rho_UV}, \ref{eq:Model_num_UV}, \ref{eq:Model_UV_LF}) numerically with halo mass functions (HMFs) provided by the \textsc{hmf} module for Python \citep{Murray2013a, Murray2014}. Appendix \ref{sect:FaintEndUVLF} provides an analytical solution of Eq. (\ref{eq:Model_UV_LF}) in the case of a power-law HMF and SFE -- halo mass relation.

The discussion so far has been quite general. The non-evolving SFE model makes the additional assumption, based on the results of the previous sections, that the average SFE and its scatter in halos of a given mass do not vary with redshift at $z>6$.

\begin{figure}
\begin{tabular}{c}
\includegraphics[width=80mm]{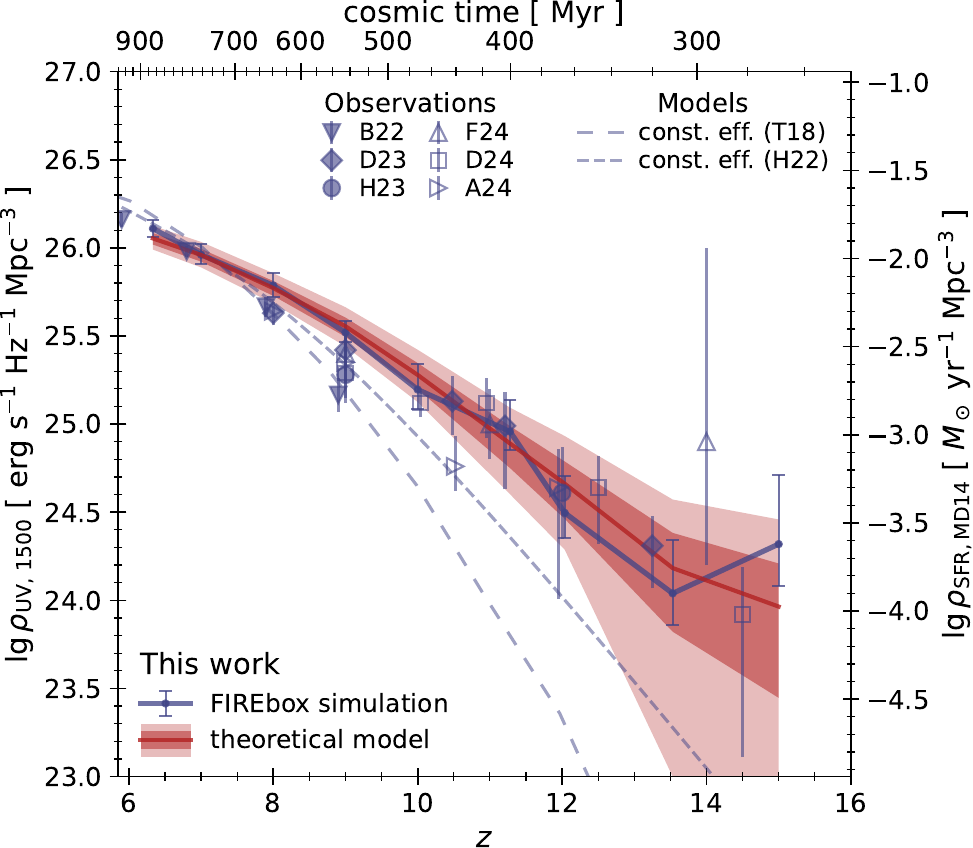}
\end{tabular}
\caption{Evolution of the ultraviolet (UV) luminosity density down to a limiting observed rest-frame magnitude of -17 as predicted by \hr{} (blue solid line and error bars) and by a theoretical model with a non-evolving star formation efficiency (SFE) -- halo mass relation inferred from the simulation (red line and shaded areas). The red line shows the UV luminosity density in galaxies brighter than -17 as predicted by the theoretical model for the same halo mass distribution as in \hr{}, while dark and light shaded areas indicate how the scatter in the SFE affects the UV luminosity in the given volume at the 68 and 95 percent confidence level. Uncertainties (68\% confidence level) of the \hr{} predictions are calculated via bootstrapping and account for both the scatter in SFE and variations in halo numbers. Long and short dashed lines reproduce the predictions of empirical models by \protect\cite{Tacchella2018} and \protect\cite{Harikane2022}, and large symbols with error bars show observational estimates by \protect\cite{Bouwens2022}, \protect\cite{Donnan2023, Donnan2023a}, \protect\cite{Harikane2023}, \protect\cite{Adams2024}, \protect\cite{Finkelstein2024} and \protect\cite{Donnan2024}. The theoretical model reproduces the evolution of the UV luminosity density found by \hr{} within the uncertainties.}
\label{fig:UVLD_Model}
\end{figure}

Fig.~\ref{fig:UVLD_Model} compares the evolution of the (dust-attenuated) UV luminosity density in \hr{}, for $M_{\rm UV}<-17$, with the predictions of the theoretical model under the assumption of a non-evolving SFE -- halo mass relation. To this end, we use the SFE -- halo mass relation inferred directly from \hr{} for dust-attenuated UV luminosities and halo mass binning. Specifically, the average SFE -- halo mass relation is approximated by a broken-power law form, see Eq. (\ref{eq:SFEfit}), and the halo mass dependence of the scatter follows Eq. (\ref{eq:SFEScatter}), with fit parameters as provided in the first row of Tables \ref{tab:SFEFitParams} and \ref{tab:SFEScatterParams}.

For a fairer comparison that accounts for the finite box size of the simulation, we calculate the UV luminosity density not via integrating Eq. (\ref{eq:Model_rho_UV}) but by sampling. Specifically, for each simulated halo of mass $M$ and $z$ we first draw a UV luminosity from the appropriate log-normal distribution, see eqs. (\ref{eq:lum_to_SFE}, \ref{eq:LognormalModel}). Next, we add the luminosities of all halos with a magnitude brighter than -17 and divide the sum by the simulation volume. By repeatedly resampling with multiple redraws, we estimate the uncertainty of the UV luminosity density in the simulation volume, attributable to the scatter in the SFE -- halo mass relation. This uncertainty estimate does not include any variance due to fluctuations in the halo numbers and is thus generally smaller than the uncertainty in \hr{} obtained via bootstrapping.

As the figure demonstrates, the theoretical model accurately captures the evolution of the cosmic UV luminosity density in \hr{}. The simulation results are within the 95\% confidence interval of the model’s predictions. Given that the uncertainties in the theoretical model stem from the scatter in the SFE -- halo mass relation, we conclude that the fluctuations of the UV luminosity with redshift found in \hr{} can be attributed to this variability. This includes the data point at $z\sim{}15$ where a strong burst in star formation just prior to $z=15$ elevates the UV luminosity density in the simulation snapshot. Moreover, the simulation results are entirely consistent (often within their 68\% confidence interval obtained via bootstrapping) with the expectation from the theoretical model.
We conclude that the theoretical model introduced in this section reproduces the UV luminosity density predicted by \hr{} and matches corresponding observational constraints at $z>6$.
 
\subsection{Implications of a non-evolving SFE -- halo mass relation: UV luminosities of galaxies at the EoR and Cosmic Dawn}
\label{sect:ModelImplicationsUV}

\begin{figure*}
\begin{tabular}{cc}
\includegraphics[width=80mm]{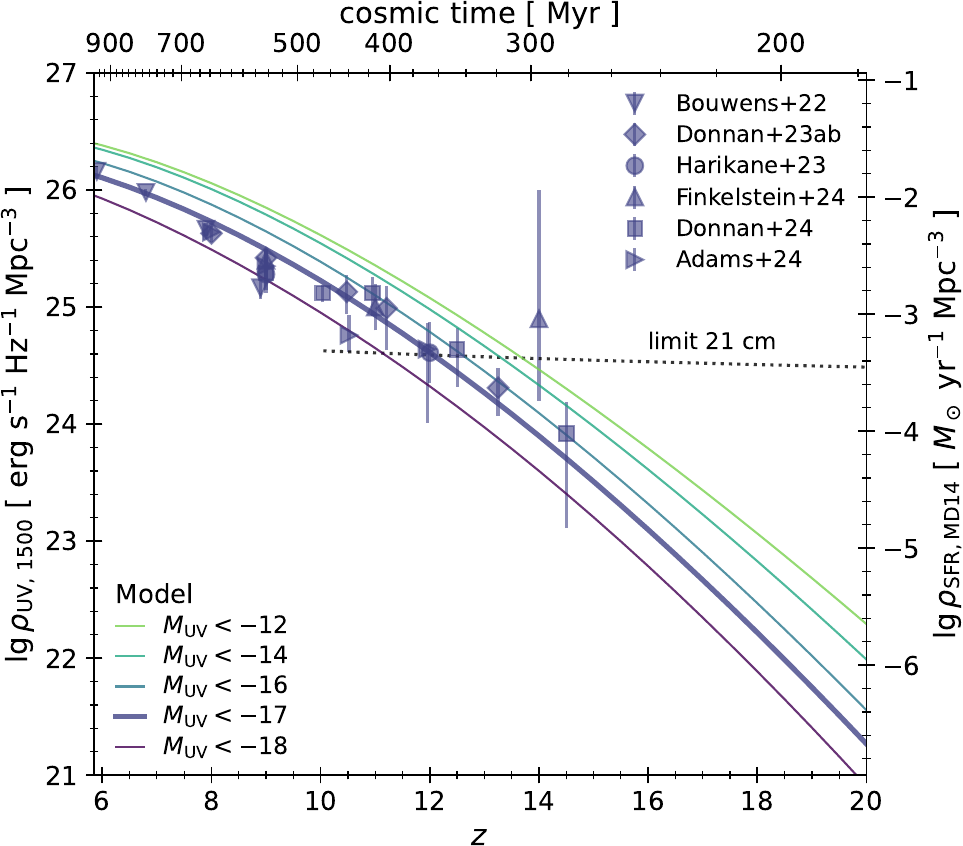} &
\includegraphics[width=73mm]{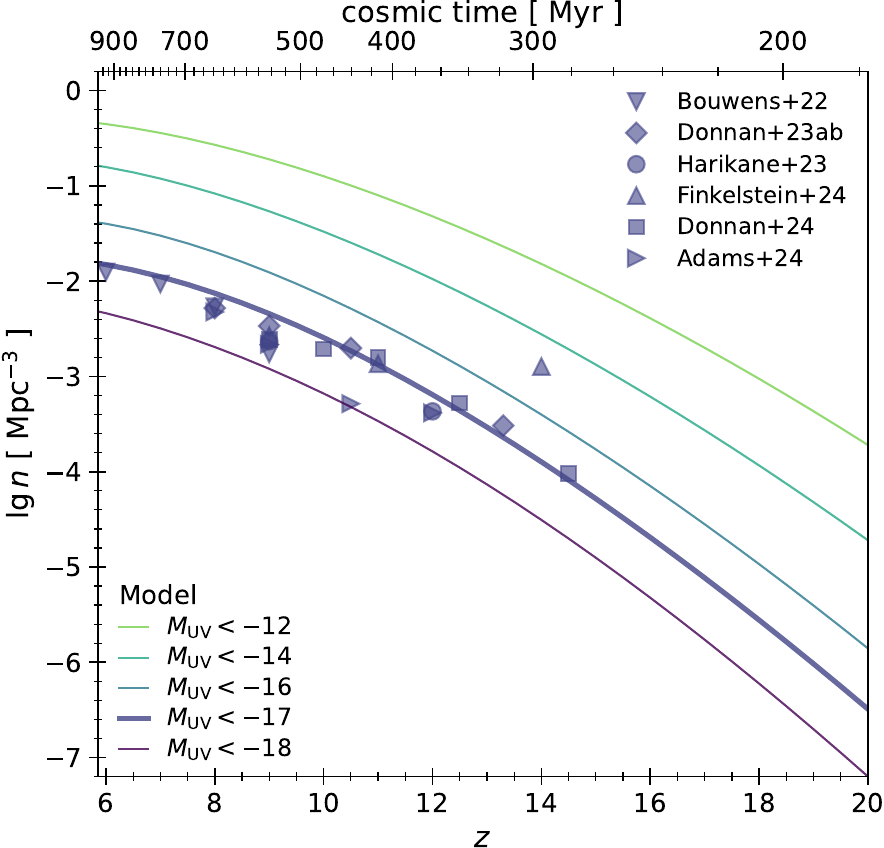}
\end{tabular}
\caption{Predictions of the theoretical model for the cosmic ultraviolet (UV) luminosity density (left panel) and the corresponding galaxy number density (right panel) at $z\sim{}6-20$. The predictions (shown by solid lines) assume a non-evolving star formation efficiency -- halo mass relation with mass-dependent scatter with parameters provided in Tables \ref{tab:SFEFitParams} and \ref{tab:SFEScatterParams}, a \protect\cite{Tinker2008} halo mass function with the \protect\cite{Behroozi2013c} extension to high redshift, and Planck 2018 cosmology (\protect\citealt{Aghanim2020}). Line colors indicate the adopted limiting magnitude (see legend). Symbols show observational estimates of the UV luminosity density and their uncertainties for $M_{\rm UV, lim}=-17$ (left panel) and galaxy number densities obtained by integrating the reported UV luminosity functions to $M_{\rm UV, lim}=-17$ \protect\citep{Bouwens2022, Donnan2023, Donnan2023a, Harikane2023, Adams2024, Finkelstein2024, Donnan2024}. The dotted line in the left panel is the UV luminosity density needed for a 21 cm absorption signal in the ‘minimal coupling’ regime \citep{Madau2018}. The UV luminosity density and number density predicted by the theoretical model for $M_{\rm UV, lim}=-17$  matches current observations and shows a steeper evolution at the highest redshift.}
\label{fig:UVLD_Model_highz}
\end{figure*}

The theoretical model introduced in the previous section allows us to study the evolution of the UV luminosity density and UV LF during reionization and late Cosmic Dawn without being limited by simulation volume. To integrate eqs. (\ref{eq:Model_rho_UV}, \ref{eq:Model_UV_LF}), we use the halo number densities provided by \textsc{hmf} version 3.4.4 for a Planck 2018 cosmology \citep{Aghanim2020} with $h=0.6766$, $\Omega_0=0.3111$, $\Omega_{\rm b}=0.04897$, and $\sigma_8=0.8102$, a CAMB transfer function with\footnote{Setting $k_{\rm max}$ to a sufficiently large value is critical as \textsc{hmf}'s default setting, at least in the code version we used, overestimates (e.g., by about 0.5 dex for $10^8\,M_\odot$ halos at $z=10$) the number density of low mass halos at high redshift.} $k_{\rm max}=150$, and the \cite{Tinker2008} halo mass function with the \cite{Behroozi2013c} extension to high redshift. 
To approximately account for the impact of baryonic effects on the halo mass (e.g., \citealt{Sawala2013, Velliscig2014, Cui2014, Schaller2015, Beltz-Mohrmann2021}), we reduce the halo masses provided by \textsc{hmf} by 0.07 dex, i.e., $M_{\rm halo}\rightarrow{}M_{\rm halo}^\prime = 10^{-0.07}M_{\rm halo}$ and $n(>M_{\rm halo}) \rightarrow{} n^\prime(>M_{\rm halo}^\prime) = n(>M_{\rm halo})$. For the fiducial set up, we adopt the non-evolving SFE -- halo mass relation with a mass-dependent scatter and with parameters provided in Tables \ref{tab:SFEFitParams} and \ref{tab:SFEScatterParams} for dust-attenuated UV luminosities and halo mass binning. Variations of this set-up will be discussed in the text when needed. We stress that some model predictions involve extrapolations of the non-evolving SFE -- halo mass ansatz to lower halo masses and higher redshifts than reliably probed by \hr{}. 
Additionally, the predictions assume no fundamental changes in baryonic physics and star formation processes at high $z$, such as contributions from Population III stars (see Section  \ref{sect:Caveats}).

The left panel of Fig.~\ref{fig:UVLD_Model_highz} presents the UV luminosity density and UV LFs predicted by the theoretical model for $z\sim{}6-20$. The model accurately captures the observed evolution of the UV luminosity density over $z\sim{}6-13.5$. Notably, $\log{}\rho_{\rm UV}$  does not evolve linearly with redshift at $z\geq{}6$ (cf. \citealt{Ceverino2024}) showing instead a steeper decline in UV luminosity density towards higher redshifts. For instance, the UV luminosity density at $z=16$ (for a limiting magnitude of -17) is predicted to be about two orders of magnitude lower than at $z=11$, highlighting a challenge for observational studies targeting galaxies at those redshifts (e.g., \citealt{Robertson2023b}). Extending this model speculatively to $z=20$ suggests a further reduction in luminosity density by an additional two orders of magnitude. Recent tentative observational evidence may suggest a sudden steepening of the UV luminosity density evolution between $z\sim{}12.5$ and $14.5$ \citep{Robertson2023b, Donnan2024}. While a steeper evolution at higher $z$ is in line with our findings, our model predicts a gradual steepening over $z\sim{}6-15$, not a sudden transition between two separate evolutionary regimes.

\begin{figure*}
\begin{tabular}{cc}
\includegraphics[width=80mm]{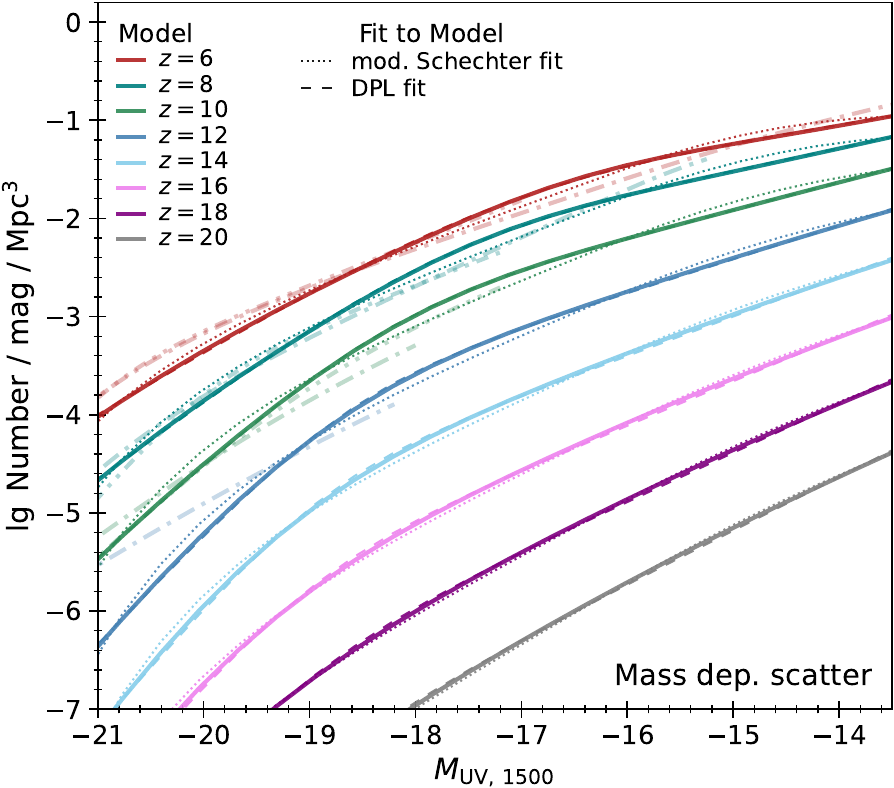} &
\includegraphics[width=80mm]{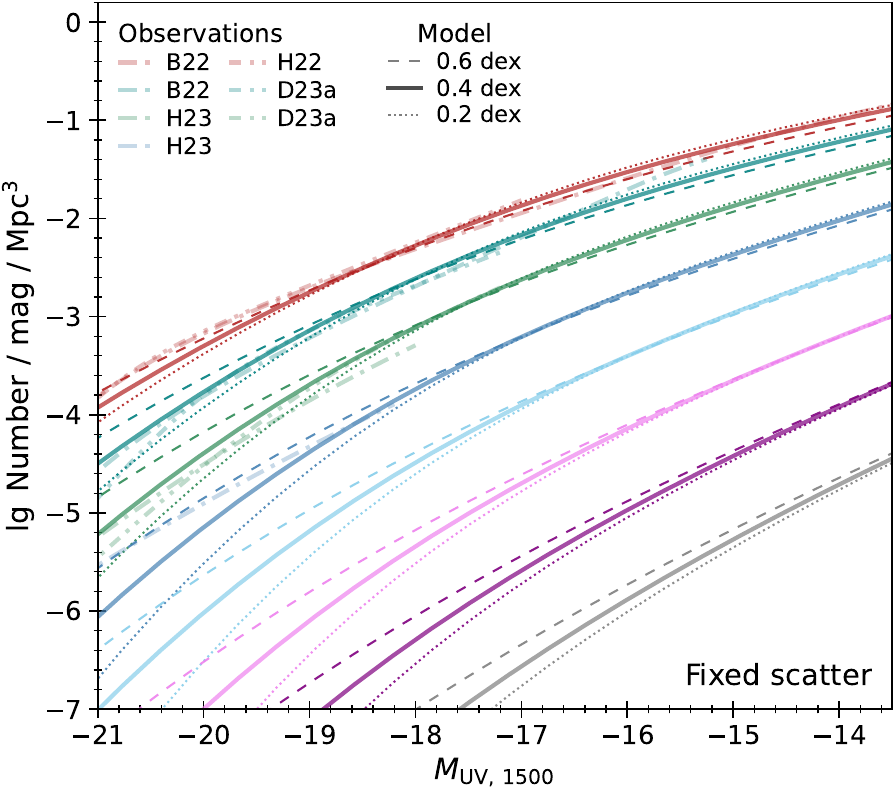}
\end{tabular}
\caption{Predictions of the theoretical model for the ultraviolet luminosity functions (UV LFs) at $z\sim{}6-20$. All predictions assume a non-evolving SFE -- halo mass relation, a \protect\cite{Tinker2008} halo mass function with the Behroozi extension to high redshift \protect\citep{Behroozi2013c}, and Planck 2018 cosmology (\protect\citealt{Aghanim2020}). Solid lines of various colors indicate the model predictions for different redshifts. Dot-dashed and double dot-dashed lines show observational estimates of the UV LF at $z\sim{}6-12$ \protect\citep{Bouwens2022, Harikane2022, Donnan2023, Harikane2023}.
(Left) Predictions for the fiducial model with a mass dependent scatter, see Table \ref{tab:SFEScatterParams}. Dotted lines (dashed lines) show fits of the model UV LFs to a modified Schechter (a double power law) function \protect\citep{Bouwens2017, Harikane2022}. While both fitting functions describe the UV LFs of the theoretical model reasonably well, a double power law provides a significantly better fit, especially at $z\leq{}14$.
(Right) Similar to left panel but for a mass-independent scatter of the SFE -- halo mass relation. Solid (dotted, dashed) lines correspond to a scatter of 0.4 (0.2, 0.6) dex. A larger scatter increases the number density of UV luminous galaxies.
In addition to a rapid decline in overall normalization, the theoretical model predicts a steepening in the faint-end of the UV LFs towards higher redshift. Fitting parameters for a modified Schechter function and double power law are provided in appendix \ref{sect:FitParamsUVLF}.
}
\label{fig:UVLF_Model_highz}
\end{figure*}

Constraining the evolution of the cosmic UV luminosity density at high $z$ is of critical importance for the large number of experiments aiming to detect and characterize the distribution of neutral hydrogen (${\rm H_I}$) during cosmic dawn and reionization (e.g., HERA \citealt{Deboer2017, Abdurashidova2022},  LOFAR \cite{Patil2017, Mertens2020}, NenuFAR \citealt{Munshi2024}, MWA \citealt{Barry2019}), including the future Square Kilometer Array Observatory (SKAO; \citealt{Mellema2013, Koopmans2015}).  Ly-$\alpha$ radiation emitted from galaxies during these early times ($z\lesssim{}20$) can decouple the spin temperature of ${\rm H_I}$ from the CMB background, rendering neutral hydrogen accessible to observations via its 21 cm line \citep{Wouthuysen1952, Field1958}. Provided that faint, metal-poor galaxies are the dominant Ly-$\alpha$ contributors, detectability of the 21 cm line requires a cosmic UV luminosity density exceeding $\rho_{\rm UV, 21cm}(z)\sim{}10^{24.5} (18/(1+z))^{1/2}\,{\rm erg}\,{\rm s}^{-1}\,{\rm Mpc}^{-3}\,{\rm Hz}^{-1}$ \citep{Madau2018}.

Our theoretical model predicts that, for $M_{\rm UV, lim}=-12$, $\rho_{\rm UV}(z)$ exceeds $\rho_{\rm UV, 21cm}(z)$ only at $z\lesssim{}14$.
The tentative detection of a 21 cm signal corresponding to $z\sim{}16-19$ reported by the EDGES collaboration \citep{Bowman2018} would thus be at odds with the predictions of our theoretical model. We note that a subsequent measurement of the radio sky in the same band (55-85 MHz) ruled out an astrophysical origin of the EDGES signal with 95\% confidence \citep{Singh2022}. Instead, various measurement and instrument systematics have been suggested as culprits \citep{Hills2018, Singh2019, Bradley2019, Sims2020, Bevins2021}. Our predictions for the detectability of an EDGES-like signal in 21 cm are thus much more pessimistic than expectations based on linearly extrapolating $\log\rho_{\rm UV}(z)$ as function of $z$ from $z\sim{}4-9$ to cosmic dawn (\citealt{Madau2018, Hassan2023}, see also \citealt{Bera2023}). However, the UV luminosity density at higher redshift could potentially be boosted by a larger contribution of Population III stars that are not included in our model (e.g., \citealt{Munoz2022}). In addition, low mass ($M_{\rm halo}<10^8\,M_\odot$) halos could provide a significant contribution to the cosmic star formation activity before reionization \citep{Barkana2000, Harikane2023}.

The right panel of Fig.~\ref{fig:UVLD_Model_highz} show the number density of galaxies with rest-frame UV magnitudes brighter than a limiting value, $M_{\rm UV, lim}$, as predicted by our theoretical model. The predicted number densities match well observational estimates derived by integrating UV luminosity functions down to $M_{\rm UV, lim}=-17$. Deeper observations would dramatically increase the number density of detectable galaxies, e.g., by one order of magnitude when going from $M_{\rm UV, lim}=-17$ to $M_{\rm UV, lim}=-14$ at $z=6$. In contrast, the UV luminosity density is much less sensitive to the UV magnitude limit, increasing by only 0.3 dex at $z=6$ for the same change in $M_{\rm UV, lim}$.

The UV LFs predicted by the theoretical model are also in approximate agreement with observations, see Fig.~\ref{fig:UVLF_Model_highz}. Deviations at the luminous end partly arise due to the lack of massive halos in \hr{} (given its modest box size) needed to constrain the high mass behavior of the SFE -- halo mass relation. In addition, it is possible that \hr{} underestimates the scatter in the SFE in more massive halos. A larger scatter would indeed increase the number density of UV luminous galaxies as shown by the figure (see also \citealt{Sun2023a}). For instance, a SFE scatter of 0.6 dex (compared with the lower $\sim{}0.3$ dex found in \hr{} for $M_{\rm halo}\sim{}10^{10}-10^{11}\,M_\odot$ halos) would be sufficient to match the reported UV LF at $z=12$ by \cite{Harikane2023}. However, applying the same scatter to $z=10$ would then lead to an overprediction of the observed UV LFs at this earlier redshift \citep{Donnan2023}. 
The dependence of the UV LFs on the scatter is further discussed in Appendix~\ref{sect:FaintEndUVLF}.
A peculiar feature of the model with mass-dependent scatter is that it introduces a much more pronounced change in slope of the UV LF at moderately faint luminosities (e.g., near $-16.5$ at $z=6$) compared to the model with mass-independent scatter.

We fit the predicted UV LFs both with a modified Schechter function, see Section 4.3 in \cite{Bouwens2017}, and a double power law (e.g., \citealt{Harikane2022}). The modified Schechter function includes a multiplicative term $10^{-0.4 \delta{} (M_{\rm UV}+16)^2}$ to model a roll-over of the LF at the faint end. The fitting parameters are provided in appendix \ref{sect:FitParamsUVLF}. While not expected a priori, our model predictions are well fit by either functional form. In either case, and independently of the chosen scatter of the SFE -- halo mass relation, fits to the UV LFs confirm the visually apparent decrease in the overall normalization and the steepening in the faint end slope with increasing redshift.

We now compare the parameters from the modified Schechter fit to the UV LFs predicted by the theoretical model with mass-dependent scatter to those derived from the observational analysis by \cite{Bouwens2022}. At $z=6$, we find decent agreement with a characteristic magnitude $M_{\rm UV}^*$ that is about 0.3 mag lower, a similar normalization, and a faint end slope ($-2.0$ vs $-1.9$). At $z=8$, the agreement is less favorable with the theoretical model predicting an $M_{\rm UV}^*$ that is about 0.8 mag too faint and a slope that is slightly too shallow ($-2.0$ in the theoretical model vs $-2.2$ found by \citealt{Bouwens2022}). The theoretical model is in somewhat better agreement with the observational results by \cite{Bouwens2015}.

The roll-over parameter obtained by the fit is positive ($\delta{}\sim{}0.09-0.17$), suggesting that the UV LF deviates from a traditional Schechter function at lower luminosities. This is further indicated by the turn-over magnitude $M_{\rm UV, T}\equiv{} -16 - 0.5(1+\alpha)/\delta{}$, which increases with redshift -- from $-13.2$ at $z=6$ to $-11.2$ at $z=12$. These findings suggest that observations at $z\lesssim{}6$ are likely better suited to identifying deviations from a traditional Schechter function.

While a modified Schechter function is able to approximate the UV LF reasonably well, we find that the double power law generally provides a superior fit, see the left panel of Fig.~\ref{fig:UVLF_Model_highz}. The positive value of $\delta$ may thus reflect a limitation on the fitting function rather than a true turn-over in the UV LFs at low luminosities. Moreoever, the double power law fit reveals that the faint end appears to have a slope that is shallower than the one suggested by the modified Schechter fit. Specifically, at $z=6$, the faint end slope is $\alpha=-1.4$ (compared to $\alpha=-2.0$ for the modified Schechter fit), and $\alpha=-1.9$ at $z=12$ (vs $\alpha=-2.1$).

\begin{figure*}
\begin{tabular}{cc}
\includegraphics[width=80mm]{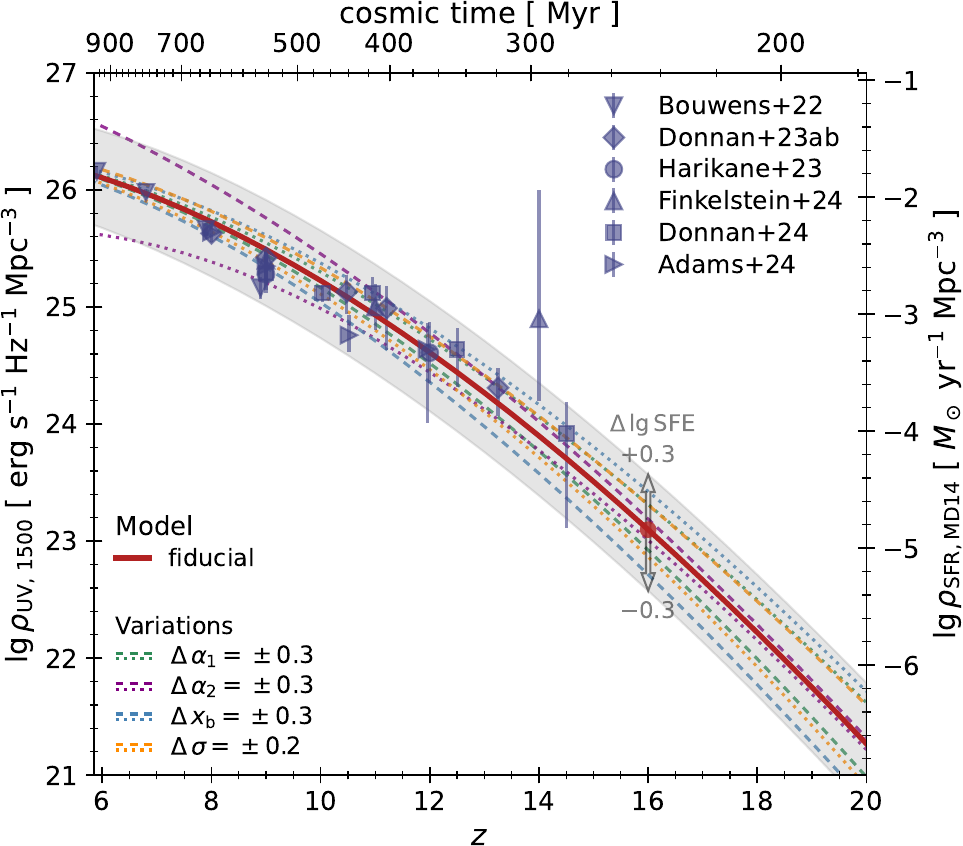} &
\includegraphics[width=80mm]{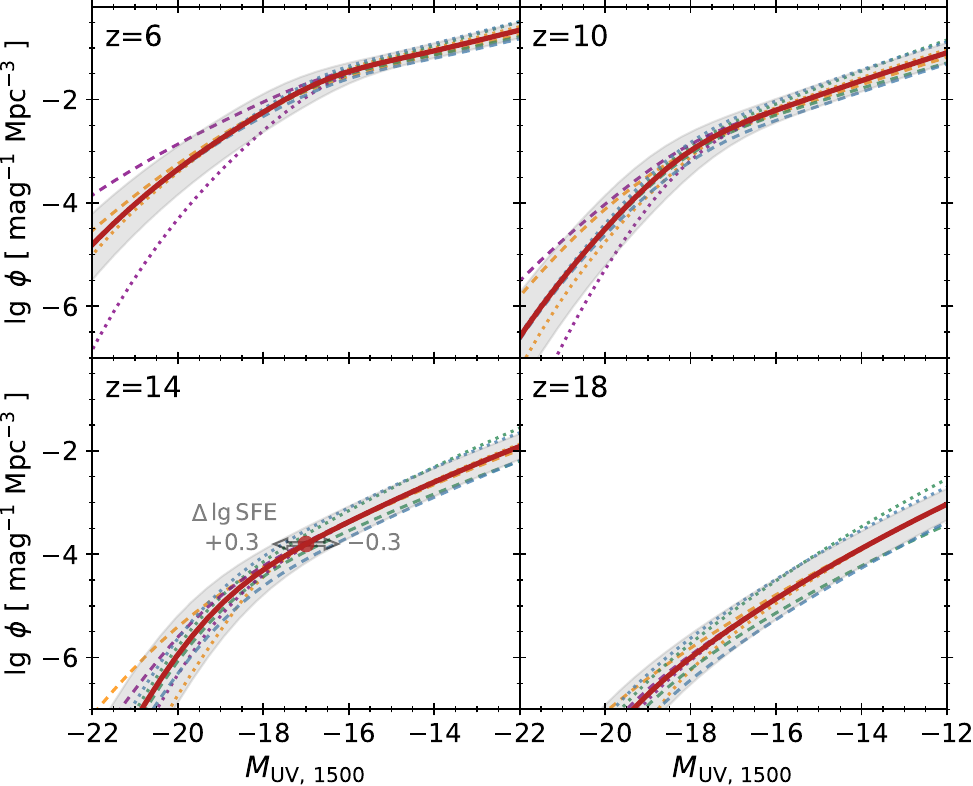}
\end{tabular}
\caption{The effect of varying the fit parameters of the non-evolving star formation efficiency (SFE) model on the predicted ultraviolet (UV) density (left panel) and the UV luminosity function (right four panels). Only galaxies with a rest-frame UV magnitude brighter than -17 are included in the UV luminosity density. 
In each panel, model predictions for the fiducial set of parameters are shown by solid red lines. Predictions for different low mass ($\alpha_1$) and high-mass ($\alpha_2$) slopes of the SFE -- halo mass relation are indicated by green and purple lines while a change in the scatter ($\sigma$) of the SFE --- halo mass relation is shown by orange lines. In each case, dashed lines (dotted lines) show the result for an increase (decrease) of the corresponding parameter. The gray shaded region illustrates how the model predictions would change if the overall normalization of the SFE -- halo mass relation is increased or decreased by up to 0.3 dex. 
Varying the scatter of the SFE -- halo mass relation between 0.2 dex to 0.6 dex, a range consistent with the scatter predicted by \hr{} (see the bottom right panel of Fig.~\ref{fig:efficiency}), has only a weak effect on the UV luminosity function or UV luminosity density. Instead, the shape of the UV luminosity function at low and high luminosities is largely controlled by the low mass and high mass slope of the SFE -- halo mass relation. A change in the high mass slope also has a noticeable impact on the UV luminosity density at $z<10$. In contrast, at $z>10$, the UV luminosity density of galaxies brighter than -17 is primarily determined by the SFE in $M_{\rm halo}\sim{}10^9-10^{10}\,M_\odot$ halos, i.e., by the SFE of halos near the pivot mass of the SFE -- halo mass parametrization, see Table~\ref{tab:SFEFitParams}.}
\label{fig:UVLD_ModelVariations}
\end{figure*}

The faint end slope of the UV LF, $\alpha$, is directly related to a combination of the low mass slope $\alpha_1$ of the SFE -- halo mass relation, the slope $\alpha_M$ of the halo mass function over the relevant halo mass range, and the slope $\gamma$ of the halo growth rate. Specifically,
\begin{equation}
\alpha = \frac{1 + \alpha_M}{\alpha_1 + \gamma} - 1,
\label{eq:alpha}
\end{equation}
assuming that the SFE, halo mass function, and halo growth rate follow power law relations with halo mass, and the scatter of the SFE is mass-independent, see Appendix~\ref{sect:FaintEndUVLF}. 
For a given $\Lambda{}$CDM cosmology, the faint end slope of the UV LF  thus directly probes the low mass slope of the SFE -- halo mass relation. Alternatively, if the SFE -- halo mass relation can be determined independently, then the faint end slope can serve as a constraint on the low-mass end of the halo mass function. In particular, if the SFE -- halo mass relation is redshift independent, as suggested by \hr{}, the observed redshift evolution of $\alpha$ (e.g., \citealt{Bouwens2012e, Finkelstein2015, Bouwens2022}) can be explained by the steepening (i.e., the decrease) of the slope $\alpha_M$ of the halo mass function for halos corresponding to moderately faint galaxies ($M_{\rm UV}\sim{}-18$ to $-16$) with increasing redshift.

So far, we discussed model predictions mainly for a specific SFE -- halo mass relation obtained from \hr{}. In Fig.~\ref{fig:UVLD_ModelVariations}, we address the question how sensitively the UV luminosity density and LFs depend on the parameters of the SFE -- halo mass relation. This analysis will also allow us to better understand the comparably gradual evolution of the UV luminosity density with cosmic time in our model. Specifically, we explore the effect of varying (i) the overall normalization by $\pm{}0.3$ dex, (ii) the low mass slope $\alpha_1$ by $\pm{}0.3$, (iii) the high mass slope $\alpha_2$ by $\pm{}0.3$, (iv) the log pivot mass $x_{\rm b}$ by $\pm{}0.3$, and (v) the scatter of the SFE -- halo mass relation by $\pm{}0.2$ dex. Each of these variations is much larger than the statistical uncertainty ($<0.1$) in the corresponding fit parameter reported in Tables \ref{tab:SFEFitParams} and \ref{tab:SFEScatterParams}.

A change in the normalization of the SFE -- halo mass relation by 0.3 dex has the expected effect of shifting the UV LFs horizontally by a corresponding amount (0.75 magnitudes). The UV luminosity density in galaxies brighter than -17 is affected at the level of about 0.4 dex at $z\sim{}6-8$, and 0.5 dex at $z\sim{}13-15$, i.e., the change is not a strong function of redshift. The models by \cite{Tacchella2018} and \cite{Harikane2022} predict UV luminosity densities that match observations at $z\sim{}6-7$ but are $1.2-2.4$ dex lower than observations at $z\sim{}12$. Bringing these models in agreement with observations would thus require boosting the SFE of the galaxies responsible for the cosmic UV luminosity by a redshift dependent factor. Hence, a change in the normalization of a non-evolving SFE -- halo mass relation alone cannot explain the observed gradual evolution of the UV luminosity density. 

The value of the low mass slope $\alpha_1$ has little impact on the UV luminosity density of galaxies brighter than -17 at $z<12$ although it changes the faint end slope of the UV LFs. The luminosity density is not strongly affected because, at $z<12$, $M_{\rm UV}=-17$ galaxies reside typically in halos more massive than the pivot mass. The faint end of the UV LF becomes shallower with a steeper (i.e., larger) $\alpha_1$ because increasing $\alpha_1$, while keeping the normalization fixed at the pivot mass, reduces the SFE and UV luminosity of low mass halos. Generally, the faint end slope gains in importance with increasing redshift as the average halo mass per given UV magnitude decreases, see appendix \ref{sect:MUV_vs_MhaloMstar}. However, while varying $\alpha_1$ affects the UV luminosity density at $z>12$, these changes are comparably moderate. Hence, we do not attribute the gradual evolution of the UV luminosity density in \hr{} to this parameter.

\begin{figure*}
\begin{tabular}{cc}
\includegraphics[width=80mm]{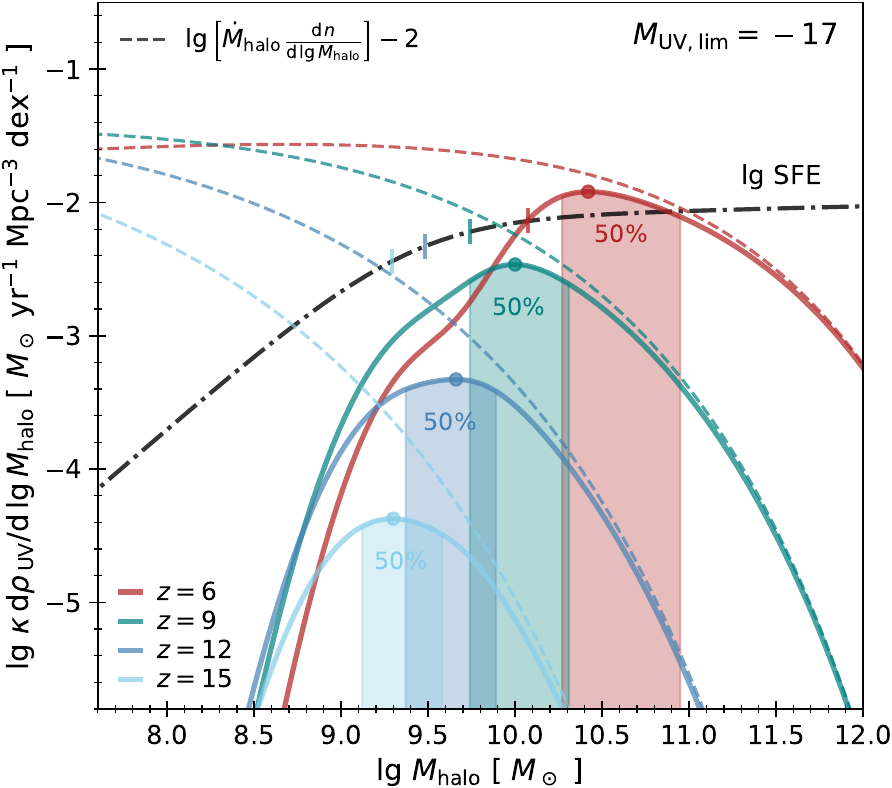} &
\includegraphics[width=80mm]{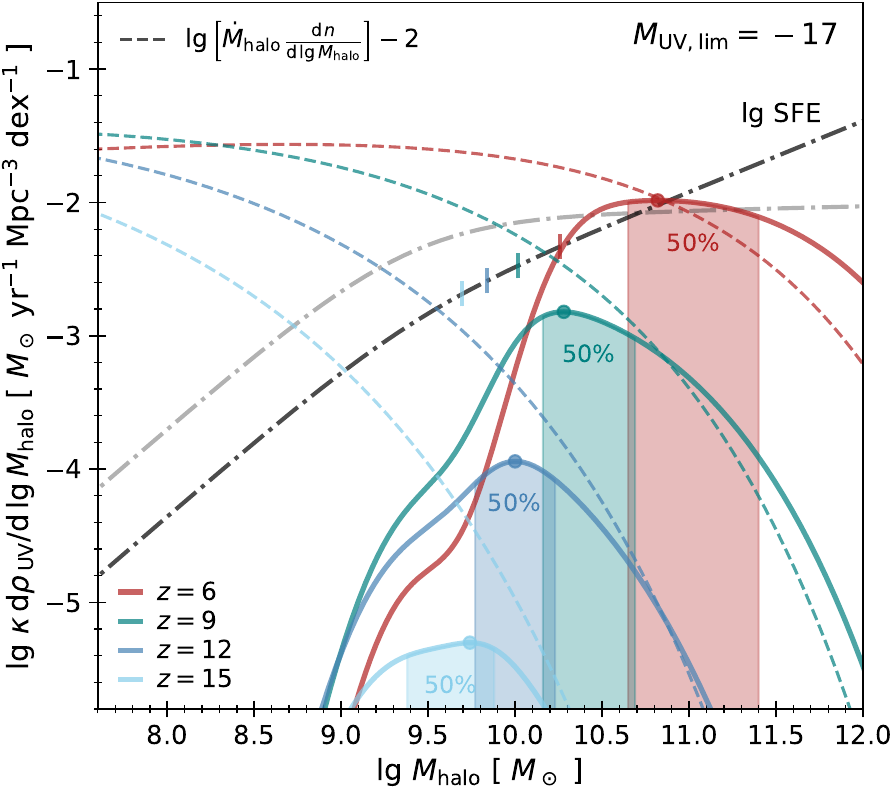}
\end{tabular}
\caption{
Contribution of halos of different mass to the UV luminosity density at redshifts $z=6-15$ according to Eq. (\ref{eq:CSFH_components}) for a UV magnitude limit of $M_{\rm UV, lim}=-17$.  Solid curves show $\kappa{}\,{\rm d}\rho_{\rm UV}/{\rm d}\lg{}M_{\rm halo}$, i.e., the UV luminosity density times the constant $\kappa$ per dex in halo mass, in units of $M_\odot\,{\rm yr}^{-1}\,{\rm cMpc}^{-3}$. 
(Left) Predictions of the theoretical model for the non-evolving star formation efficiency (SFE) -- halo mass relation from the bottom right panel of Fig.~\ref{fig:efficiency} with high mass slope $\alpha_2=0.032$ and normalization of $A=-2.370$ (dot-dashed line). (Right) Same as the left panel but for a SFE -- halo mass relation with a larger high-mass slope $\alpha_2=0.532$ and a lower normalization $A=-3.027$ (black dot-dashed line). The SFE -- halo mass relation from the left panel is shown by a gray dot-dashed line for comparison. By construction, both relations result in the same UV luminosity density for galaxies brighter than $M_{\rm UV, lim}=-17$ at $z=6$.
In either panel, halo masses corresponding to $\langle{}\mathcal{S}\rangle{}_{\rm eff}=0.5\langle{}\mathcal{S}\rangle{}$ are indicated by short vertical lines on the SFE -- halo mass relation. Dashed lines depict the halo mass growth rate density per dex in halo mass, $\dot{M}_{\rm halo}{\rm d}n/{\rm d}\lg{}M_{\rm halo}$ in units $M_\odot\,{\rm yr}^{-1}\,{\rm cMpc}^{-3}$, rescaled by a factor of 1/100 for visual clarity. Circles indicate the halo mass corresponding to the maximum value of ${\rm d}\rho_{\rm UV}/{\rm d}\lg{}M_{\rm halo}$, while shaded areas labeled `50\%' depict the halo mass range contributing to half the total UV luminosity density down to the magnitude limit. As detailed in the text, a shallower slope of the SFE -- halo mass relation implies a larger downward shift of this halo mass range resulting in a more gradual evolution of the UV luminosity density.
}
\label{fig:CSFH_components}
\end{figure*}

Instead, the figure shows that the high mass slope $\alpha_2$ of the SFE -- halo mass relation significantly affects the evolution of the UV luminosity density of galaxies brighter than -17 at $z\sim{}6-10$. In addition, the slope also strongly influences the shape of the bright end of the UV LFs. Provided we keep the normalization of the relation fixed at the pivot mass ($M_{\rm halo}\sim{}10^{9.5}\,M_\odot$), a steeper $\alpha_2$ results in a larger SFE in halos more massive than the pivot mass. A steeper slope thus increases the cosmic UV luminosity density as well as the number density of UV bright galaxies. Hence, provided we have an appropriate normalization of the SFE -- halo mass relation such that $\rho_{\rm UV}$ matches observations at some lower redshift (e.g., for $z=6$), a shallower (steeper) high mass slope leads to a shallower (steeper) evolution of the cosmic UV luminosity density. Indeed, the slope of the SFE -- halo mass relation predicted by \hr{} for galaxies in $M_{\rm halo}=10^{9.5}-10^{10.5}\,M_\odot$ halos is much lower than the slope by \cite{Harikane2022} extrapolated to the same mass regime while approximately matching at the high mass end ($M_{\rm halo}\sim{}10^{11}\,M_\odot$). \hr{} thus predicts a much higher SFE for galaxies that dominate the cosmic UV luminosity density compared with \cite{Harikane2022}. We note that the UV luminosity above which the LF is significantly affected by a change in high-mass slope increases with increasing redshift (from about $-17$ at $z=6$ to $-19$ at $z=14$). Consequently, the impact of $\alpha_2$ on the UV luminosity density decreases with increasing $z$ and the theoretical model always predicts a rather steep evolution of $\rho_{\rm UV}$ at the highest redshifts.

Varying the pivot mass, i.e., the mass separating the low mass and high mass regime of the broken-power law fit to the SFE -- halo mass relation in \hr{}, has a similar effect as changing the low mass slope. Increasing $x_{\rm b}$, while keeping the normalization fixed, reduces the number density of halos above the pivot mass and thus lowers the UV luminosity density. This effect is more pronounced at higher $z$ when a larger proportion of the UV luminosity density arises from halos near the pivot mass.

Finally, the figure analyzes how scatter in the SFE -- halo mass relation, and the corresponding variability in the UV magnitudes, affects the cosmic UV luminosity density and the UV LFs. Specifically, we vary the mass-dependent scatter by $\pm{}0.2$ dex finding that the cosmic UV luminosity density is largely insensitive to this scatter. This finding is in qualitative agreement with a recent study by \cite{Gelli2024} who concluded that scatter alone does not explain the high UV luminosity density at $z>12$. Scatter in the SFE -- halo mass relation has a significant impact, however, on the bright end of the UV LFs. As discussed before, a higher UV variability boosts the number density of UV bright galaxies given their much lower number density compared to faint ones. However, unless the scatter is increased to a value much larger than found in \hr{}, the number density of galaxies which dominate the cosmic UV luminosity density is only weakly affected. For a further discussion of the importance of the scatter on the UV LF, see Appendix~\ref{sect:FaintEndUVLF}.

Our finding contrasts with alternative models that aim to explain the relatively shallow evolution of the UV luminosity density at high $z$ with a large and redshift-dependent scatter of the SFE (e.g., \citealt{Shen2023a}). We parametrize the SFE -- halo mass relation in terms of both the mean relation and its log-normal scatter.  As discussed in Section \ref{sect:TheoreticalModel}, an alternative apporach is to parametrize the SFE -- halo mass relation in terms of the \emph{median} relation and its scatter. However, for a log-normally distributed SFE, increasing the scatter while holding the median SFE fixed results in an increase in the mean SFE. A scatter that grows with redshift, while maintaining a fixed median SFE -- halo mass relation, consequently raises the normalization of the average SFE -- halo mass relation and causes a weaker decline in UV luminosity density with redshift. In contrast, our theoretical model successfully matches the observed evolution of the UV luminosity density without requiring an explicit redshift dependence of the average SFE -- halo mass relation.

To better understand the comparably shallow evolution of the UV luminosity function at $z\lesssim{}12$, we analyze how halos of different mass contribute to the cosmic UV luminosity density for different UV magnitude limits. According to Eq. (\ref{eq:Model_rho_UV}), the luminosity density per dex in halo mass
\begin{equation}
\frac{{\rm d} \rho_{\rm UV}}{{\rm d}\lg{}M_{\rm halo}} = \left[\dot{M}_{\rm halo}\frac{{\rm d}n}{{\rm d}\lg{}M_{\rm halo}}\right] \langle{}\mathcal{S}\rangle{}_{\rm eff}\frac{1}{\kappa}
\label{eq:CSFH_components}
\end{equation}
is proportional to the halo mass growth rate density per dex in halo mass (the term in brackets) and the effective SFE $\langle{}\mathcal{S}\rangle{}_{\rm eff}$. While the former is related to gravitational processes, the latter encapsulates the baryonic physics of galaxy formation. While both terms are functions of halo mass and redshift, the effective SFE also depends on the chosen UV magnitude limit.

In the left panel of Fig.~\ref{fig:CSFH_components}, we plot ${\rm d} \rho_{\rm UV}/{\rm d}\lg{}M_{\rm halo}$ as function of mass and redshift as predicted by our theoretical model for a mass-dependent scatter. According the figure, the majority of the cosmic UV luminosity at $z\sim{}6-15$ is produced by galaxies in halos of moderate mass ($M_{\rm halo}\sim{}10^9-10^{11}\,M_\odot$). For instance, at $z=6$ half the UV luminosity arises in halos with $M_{\rm halo}\sim{}10^{10.3}-10^{11}\,M_\odot$  if a UV magnitude limit of $M_{\rm UV, lim}=-17$ is chosen. The mass range of halos responsible for the majority of the UV luminosity shifts towards lower masses with increasing redshift, e.g., it is $M_{\rm halo}\sim{}10^{9.1}-10^{9.6}\,M_\odot$ at $z=15$. For $M_{\rm UV, lim}=-14$, the corresponding halo mass ranges are $M_{\rm halo}\sim{}10^{9.7}-10^{10.7}\,M_\odot$ and $M_{\rm halo}\sim{}10^{8.8}-10^{9.3}\,M_\odot$. 
The downward shift of the halo mass range is partly driven by the evolution of the halo mass function which reduces the number density of massive halos at higher $z$. Furthermore, halos of a given mass have a higher growth rate at higher redshift \citep{Behroozi2015, Rodriguez-Puebla2016a} and thus, for a non-evolving SFE -- halo mass relation, harbor brighter galaxies ($\mathcal{L}\propto{}\mathcal{S}\dot{M}_{\rm halo}$). Consequently, a larger number of galaxies in low mass halos is UV detectable at higher $z$.

In order to analyze the impact a change in slope of the SFE -- halo mass relation has on the UV luminosity density, we need to clarify how we adjust the normalization of the relation. In the following we choose a normalization such that the UV luminosity density at $z=6$ remains unchanged. For instance, if we increase the fiducial value of the (high-mass) slope $\alpha_2$ of the relation by 0.5, we need to decrease the normalization $A$ by 0.66. Incidentally, the SFE in halos of $\sim{}10^{10.7}\,M_\odot$ remains approximately constant which is near the lower end halo mass range probed by \cite{Harikane2022}. In this scenario, galaxies in halos of a given mass below $10^{10.7}\,M_\odot$ have a higher SFE, and thus a higher UV luminosity, when the slope $\alpha_2$ of the SFE -- halo mass relation is lower. Also, a smaller slope allows additional galaxies, namely those from comparably low mass halos, to contribute to the UV luminosity density, see left vs right panel of Fig.~\ref{fig:CSFH_components}. Hence, the UV luminosity density at $z>6$ generally increases as $\alpha$ is decreased. The effect is stronger at higher $z$ because then more of the UV luminosity arises from lower mass galaxies (see previous paragraph). 

For a more quantitative discussion of the role of the slope of the SFE -- halo mass relation it is useful to consider the contributions to the UV luminosity density from different halo masses via Eq. (\ref{eq:CSFH_components}). As discussed above, reducing the slope implies that additional UV luminosity is provided by galaxies in low mass halos. The `typical mass' of halos dominating the total UV luminosity, which we may define to be represented by the mode in the ${\rm d}\rho_{\rm UV}/{\rm d}\lg{}M_{\rm halo}$ distribution, thus decreases with decreasing $\alpha_2$ at fixed $z$. As this typical halo mass changes with redshift, both the halo mass growth rate density $\dot{M}_{\rm halo}{\rm d}n/{\rm d}\lg{}M_{\rm halo}$ and the effective SFE of such halos evolves as well.
Specifically, in our fiducial model, the effective SFE at this typical halo mass decreases by about $0.6$ dex between $z=6$ and $z=15$, the halo mass growth rate density decreases by about $1.8$ dex, and the size of the halo mass range contributing half the total UV luminosity decreases by about 0.2 dex. Hence, the UV luminosity density should evolve by about 2.6 dex between $z=6$ and $z=15$, which is consistent with Fig.~\ref{fig:UVLD_Model_highz}. By contrast, if we increase $\alpha_2$ by 0.5, the effective SFE of the typical halo mass decreases by $0.9$ dex, the halo mass growth rate density plummets by about $2.4$ dex, and the width of the halo mass is reduced by $0.2$ dex, resulting in a much steeper redshift evolution of the UV luminosity density ($3.5$ dex between $z=6$ and $z=15$).

In conclusion, we attribute the gradual evolution of the UV luminosity density in \hr{}, compared with other theoretical studies, to the more accurate accounting of baryonic processes in the FIRE-2 physics model. At $z\sim{}6-15$, these baryonic processes give rise to a redshift-independent SFE -- halo mass relation with a comparably shallow slope at intermediate halo masses ($M_{\rm halo}\sim{}10^{9}-10^{11}\,M_\odot$). As described above, a shallower slope results in an additional number of UV detectable galaxies in low mass halos at higher $z$, and, hence, a more gradual evolution of the cosmic UV luminosity density.

\subsection{Implications of a non-evolving SFE -- halo mass relation: The role of galaxies for the cosmic reionization history}
\label{sect:ModelImplicationsReionization}

During cosmic reionization ($z\sim{}6-10$), the neutral hydrogen component of the intergalactic medium (IGM) is converted into its ionized state (e.g., \citealt{Fan2006a, Robertson2010}). The primary physical driver of this phase transition is thought to be Lyman-continuum radiation escaping low-mass galaxies \citep{Madau1999, Gnedin2000a, Robertson2015, Mascia2023, Simmonds2024, Atek2024, Dayal2024}. In this section, we will explore the implications of the theoretical model implemented as described in Section \ref{sect:ModelImplicationsUV} by comparing its predictions with available observational constraints on the ionization of the IGM. Specifically, by linking the predicted UV luminosity to the observed ionized hydrogen fraction of the Universe, we will constrain the product of the escape fraction $f_{\rm esc}$ of ionizing radiation and the ionizing photon production efficiency $\xi_{\rm ion}$ of faint galaxies.

We model the reionization history of the IGM with the help of a commonly employed analytical technique that keeps track of the number density of ionizing photons \citep{Madau1999, Kuhlen2012a, Gnedin2022}. Specifically, we follow the approach by \cite{So2014} and write the time evolution of the volume fraction $Q$ of ionized hydrogen as
\begin{equation}
\frac{dQ}{dt} = \frac{\gamma}{\delta{}}\frac{\dot{n}_{\rm ion}}{\bar{n}_{\rm H}} - \frac{Q}{\bar{t}_{\rm rec}},
\label{eq:Q}
\end{equation}
where $\dot{n}_{\rm ion} = f_{\rm esc}\,\xi_{\rm ion}\,\rho_{\rm UV}$ is the injection rate of ionizing photons per volume, $f_{\rm esc}$ is the fraction of ionizing photons escaping from galaxies, $\xi_{\rm ion}$ is the rate of ionizing photons per spectral UV luminosity, $\rho_{\rm UV}$ is the intrinsic spectral UV luminosity density, $\bar{n}_{\rm H}$ is the average number density of hydrogen nuclei, and $\bar{t}_{\rm rec}$ is an averaged recombination timescale. Compared to \cite{Madau1999}, this equation contains the extra factors $\gamma$ and $\delta{}$. Multiplying $\dot{n}_{\rm ion}$ by $\gamma$ converts the injection rate density of ionizing photons to the ionization rate density, while multiplying $\bar{n}_{\rm H}$ by $\delta$ accounts for the slightly higher density of ionized bubbles compared to the cosmic average.
We adopt the following parameterizations for $\gamma$, $\delta$, and $\bar{t}_{\rm rec}$, obtained by fitting to the results of radiative transfer calculations \citep{So2014}
\begin{equation}
\gamma = 1 - 0.91 Q^{2.44},\,\delta = \frac{1+0.02}{Q^{0.12} + 0.02},\,\bar{t}_{\rm rec}=5.6\times{}10^{12}\,(1+z)^{-4.35}\,{\rm yr}.
\label{eq:GammaDelta}
\end{equation}
While the use of this parametrization is not without caveats, e.g., the simulations by \cite{So2014} only cover volumes of 20 cMpc$^3$ and employ a different (WMAP year 7) cosmology, the predictions of Eq.~(\ref{eq:Q}) are similar if $\gamma=\delta=1$ is chosen, except near $z\sim{}6-6.75$, where this choice results in earlier and more abrupt completion of reionization.

\begin{figure}
\begin{tabular}{c}
\includegraphics[width=80mm]{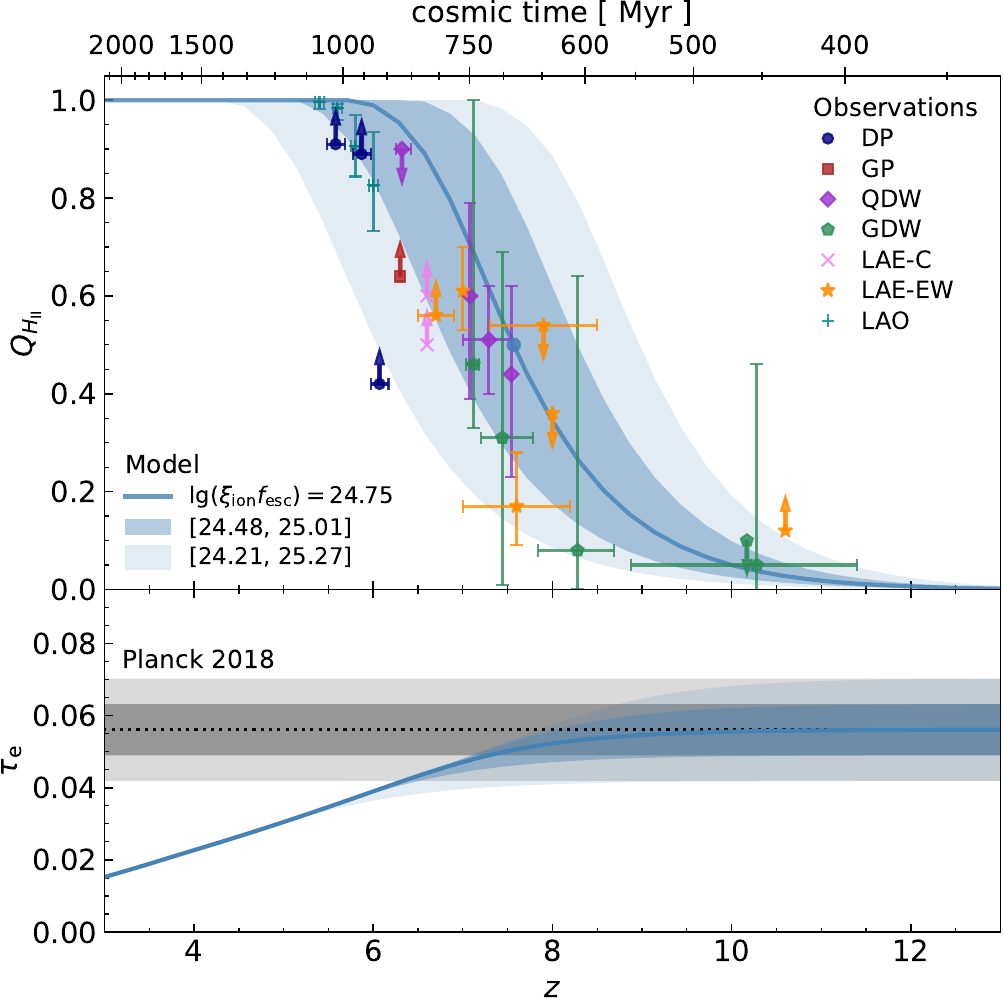}
\end{tabular}
\caption{Reionization history predicted by the theoretical model for a non-evolving star formation efficiency -- halo mass relation inferred from \hr{}. In each panel, results are shown for a range of values for $f_{\rm esc}\xi_{\rm ion}$, see legend, and for a limiting magnitude of $M_{\rm UV}=-12$. (Top) Volume-weighted ionized hydrogen fraction predicted by the model. Overplotted are observational constraints based on the Lyman-$\alpha$ dark pixel fraction (DP; \protect\citealt{McGreer2015}), gap-peak statistics in quasar absorption spectra (GP; \protect\citealt{Gallerani2008}), Lyman-$\alpha$ damping wing absorption of quasars (QDW; \protect\citealt{Schroeder2013, Greig2017, Banados2018, Greig2022}), Lyman-$\alpha$ damping wing absorption of galaxies (GDW; \protect\citealt{Hsiao2023, Umeda2023}), clustering of Lyman-$\alpha$ emitters (LAE-C; \protect\citealt{Ouchi2010, Sobacchi2015}), the fractions and equivalent width distributions of Lyman-$\alpha$ emitters (LAE-EW; \protect\citealt{Schenker2014, Mason2019a, Bolan2022, Bruton2023}), and Lyman-$\alpha$ opacity statistics (LAO; \citealt{Gaikwad2023}), see legend. Upper and lower limits are indicated by arrows and correspond to the 95 percent confidence level, if reported, otherwise to the 68 percent confidence level. Horizontal errorbars indicate the redshift range of the observational sample. Vertical errorbars show uncertainties in the estimated ionized volume fraction at the 68 percentile confidence level. The blue circle at the center of the upper panel indicates the mid-point of re-ionization at $z_{\rm re}=7.57$ as predicted by the theoretical model for $f_{\rm esc}\xi_{\rm ion}=10^{24.74}\,{\rm Hz\,erg}^{-1}$. (Bottom) Optical depth of cosmic microwave background (CMB) photons due to Thomson scattering off free electrons as predicted by the model. The values for $f_{\rm esc}\xi_{\rm ion}$ are chosen to match the measured optical depth by the Planck Collaboration (dotted line; \protect\citealt{Aghanim2020}), obtained from CMB power spectra combined with CMB lensing reconstruction and baryonic acoustic oscillations, and the boundaries of the 68 and 95 percentile confidence interval (gray shaded regions). The model is in good agreement with observational data provided $f_{\rm esc}\xi_{\rm ion}$ lies in the range $\sim{}10^{24.5}-10^{24.7}\,{\rm Hz\,erg}^{-1}$.}
\label{fig:reionization}
\end{figure}

For simplicity, we assume that $f_{\rm esc}\,\xi_{\rm ion}$ is a constant. Incidentally, this model then implies that comparably faint galaxies (fainter than $M_{\rm UV}\sim{}-16$ to $-17$, see Fig.~\ref{fig:Median_UV_luminosity}) dominate the reionization budget at $z\sim{}6-11$ (e.g., \citealt{Wise2014, Katz2019, Lewis2020}), not more luminous galaxies with $M_{\rm UV}<-18$ (cf. \citealt{Naidu2020}). 
We use in this section the SFE -- halo mass relation derived from the intrinsic UV luminosity of \hr{} galaxies (third row in Tables~\ref{tab:SFEFitParams} and \ref{tab:SFEScatterParams}). However, we find no qualitative, and only minor quantitative, changes if we adopt SFEs based on dust-attenuated UV luminosities.

Given $f_{\rm esc}\,\xi_{\rm ion}$ and a cosmic UV luminosity density $\rho_{\rm UV}(z)$ provided by our theoretical model, we solve eqs.~(\ref{eq:Q}, \ref{eq:GammaDelta}) numerically with the \textsc{scipy.integrate.solve\_ivp} function starting from a small, positive value for $Q$ at $t=0$ (we use $10^{-7}$). Values of $Q>1$ are clipped to $Q=1$. A limiting magnitude of $M_{\rm UV, lim}=-12$ is used to calculate $\rho_{\rm UV}(z)$. The reionization history predicted by this model is insensitive to the choice of $M_{\rm UV, lim}\geq{}-14$.

We set the value of $f_{\rm esc}\,\xi_{\rm ion}$ by comparing the predicted optical depth $\tau_{\rm e}$ of CMB photons due to free electron scattering to the observed optical depth $\tau_{\rm e, Planck}=0.0561$ \citep{Aghanim2020}. To this end, we first calculate the optical depth between redshifts 0 and $z$ \citep{Kuhlen2012a}
\begin{equation}
\tau_{\rm e}(z)= \sigma_{\rm T}\bar{n}_{\rm H}c \int_0^z \frac{1+z^2}{H(z)}Q(z)\left(1 + \eta{}\frac{Y}{4(1-Y)}\right)dz,
\end{equation}
where $\sigma_{\rm T}$ is the Thomson cross section, $c$ is the speed of light, $H(z)$ is the Hubble parameter, $Q(z)$ is the reionization history computed numerically via eqs.~(\ref{eq:Q}, \ref{eq:GammaDelta}),
$Y=0.243$ is the cosmic Helium mass fraction \citep{Aver2015, Peimbert2016, Aghanim2020}, and $\eta=1$ ($\eta=2$) assumes singly (doubly) ionized Helium at $z\geq{}4$ ($z<4$). We then find a value for $f_{\rm esc}\,\xi_{\rm ion}$ such that $\tau_{\rm e}(z=14) = \tau_{\rm e, Planck}$ with the help of the \textsc{scipy.optimize.root\_scalar} function. To represent the uncertainty in the Planck 2018 measurement, we also constrain $f_{\rm esc}\,\xi_{\rm ion}$ such that $\tau_{\rm e}(z=14)$ equals $\tau_{\rm e, Planck}\pm{}0.0071$ and $\tau_{\rm e, Planck}\pm{}2\times{}0.0071$.

Fig.~\ref{fig:reionization} compares the reionization history $Q(z)$ predicted by our theoretical model with a variety of observational data based on Lyman-$\alpha$ dark pixel fractions (\citealt{McGreer2015}), gap-peak statistics in quasar absorption spectra \citep{Gallerani2008}, Lyman-$\alpha$ damping wing absorption of quasars and galaxies \citep{Schroeder2013, Greig2017, Banados2018, Greig2022, Hsiao2023, Umeda2023}, clustering of Lyman-$\alpha$ emitters \citep{Ouchi2010, Sobacchi2015}, fractions and equivalent width distributions of Lyman-$\alpha$ emitters \citep{Schenker2014, Mason2019a, Bolan2022, Bruton2023}, and Lyman-$\alpha$ opacity statistics \citep{Gaikwad2023}. The reionization history predicted by the theoretical model is in good agreement with observational data for $f_{\rm esc}\,\xi_{\rm ion}$ values that result in optical depths close to $\tau_{\rm e, Planck}$. Once possible exception is the lower limit of 0.12 at $z=10.6$ reported by \cite{Bruton2023} which, however, conflicts with the upper limit of 0.1 at $z=10.17$ reported by \cite{Hsiao2023}. Replacing Eq. (\ref{eq:GammaDelta}) with the simpler ansatz $\gamma=\delta=1$ leads to reionization histories that are near identical to those shown in the figure when $Q<0.7$ but also to a much faster evolution of the ionized hydrogen fraction during the final stages of reionization.

As highlighted by the figure, a larger (smaller) optical depth goes along with an earlier (later) start, mid-point, and finish of reionization. This trend puts strong limits on the allowed optical depth in the context of our model. For instance, an optical depth similar to the Planck 2015 best fit value of 0.066, which is about $1.4\sigma$ above the 2018 estimate, would appear to be inconsistent with several observational constraints \citep{Schroeder2013, Schenker2014, Bolan2022}. For $\tau_{\rm e}=0.0561\pm{}0.0071$, the mid point of reionization is predicted to take place at $z_{\rm re}=7.57^{+0.71}_{-0.75}$, which is consistent with the estimate $z_{\rm re}=7.82\pm{}0.71$ by Planck 2018 \citep{Aghanim2020}. 
Our model predicts an extended duration of reionization $\Delta{}z_{\rm re}=z_{10}-z_{90}\sim{}2.8$ and $\tilde{\Delta}z_{\rm re}=z_{25}-z_{75}\sim{}1.4$ with only a weak dependence on the choice of $f_{\rm esc}\,\xi_{\rm ion}$. Here, $z_{f}$ is the redshift at which $Q=f/100$. 
While current estimates are still rather uncertain, an extended duration of reionization ($\Delta{}z_{\rm re}\sim{}2-4$, $\tilde{\Delta}z_{\rm re}\sim{}1-2$), potentially driven by the large contribution of UV faint galaxies \citep{Kuhlen2012a}, is at least consistent with cosmological radiative transfer simulations \citep{Rosdahl2018a, Villasenor2021, Kannan2022}, semi-numerical approaches \citep{Maity2022, Bera2023}, and observational data \citep{Adam2016, Reichardt2021, Gorce2022}. 
The figure also shows that the evolution of $Q(z)$ is not symmetric in redshift around the mid-point, i.e., it is not well described by a shifted hyperbolic tangent function that is sometimes adopted by models. 
Table~\ref{tab:reionization} lists the reionization mid-point and duration predicted by our theoretical model for different choices of the limiting UV magnitude $M_{\rm UV, lim}$ and values of $f_{\rm esc}\,\xi_{\rm ion}$.

\begin{table}
\begin{tabular}{cccccccccc}
$\zeta$ & $\tau_{\rm e}$ & $z_{\rm re}$ & $\Delta{}z_{\rm re}$ & $\tilde{\Delta{}}z_{\rm re}$ & $z_{10}$ & $z_{25}$ & $z_{75}$ & $z_{90}$ \\
\hline \multicolumn{9}{c}{$M_{\rm UV, lim}=-12$} \\ \hline
24.21 & 0.0419 & 6.04 & 2.84 &1.40 & 7.84 & 6.85 & 5.44 & 5.00\\
24.48 & 0.0490 & 6.82 & 2.80 &1.40 & 8.58 & 7.62 & 6.22 & 5.78\\
24.75 & 0.0561 & 7.57 & 2.78 &1.38 & 9.31 & 8.35 & 6.97 & 6.53\\
25.01 & 0.0632 & 8.28 & 2.75 &1.37 & 9.99 & 9.06 & 7.69 & 7.24\\
25.27 & 0.0703 & 8.97 & 2.72 &1.35 & 10.66 & 9.73 & 8.39 & 7.94\\
\hline \multicolumn{9}{c}{$M_{\rm UV, lim}=-14$} \\ \hline
24.24 & 0.0419 & 6.05 & 2.77 &1.38 & 7.81 & 6.84 & 5.46 & 5.03\\
24.52 & 0.0490 & 6.83 & 2.74 &1.37 & 8.55 & 7.61 & 6.25 & 5.81\\
24.80 & 0.0561 & 7.58 & 2.70 &1.35 & 9.27 & 8.34 & 6.99 & 6.57\\
25.07 & 0.0632 & 8.29 & 2.67 &1.33 & 9.95 & 9.04 & 7.72 & 7.28\\
25.33 & 0.0703 & 8.98 & 2.62 &1.30 & 10.61 & 9.71 & 8.41 & 7.98\\
\end{tabular}
\caption{Statistics of the reionization history as predicted by the theoretical model for a non-evolving star formation efficiency -- halo mass relation inferred from \hr{} for given UV magnitude limits $M_{\rm UV, lim}$ (row header) and values of $\zeta\equiv{}\lg(f_{\rm esc}\,\xi_{\rm ion}\,[{\rm Hz}\,{\rm erg}^{-1}])$ (first column). The values of $\zeta$ are chosen such that the electron scattering optical depth $\tau_{\rm e}$ (second column) predicted by the model matches the optical depth reported by Planck 2018 ($\tau_{\rm e, Planck}=0.0561$) or the lower and upper values of its 68\% and 95\% confidence intervals. Columns 3-9 list the redshift of the mid-point of reionization ($z_{\rm re}\equiv{}z_{50}$), the duration of reionization defined as  $\Delta{}z_{\rm re}=z_{10}-z_{90}$ or $\tilde{\Delta}z_{\rm re}=z_{25}-z_{75}$, and the redshifts $z_{f}$ corresponding to $Q(z_{f})=f/100$ for $f=10, 25, 75, 90$.}
\label{tab:reionization}
\end{table}

Quantitatively, we find that values of 
\begin{equation}
\zeta\equiv{}\lg{}(f_{\rm esc}\,\xi_{\rm ion}[{\rm Hz}\,{\rm erg}^{-1}])\sim 24.5-24.7
\end{equation}
result in good agreement with most observational data. To convert constraints on $\zeta$ into an estimate of the escape fraction, we need to divide by the ionizing photon production efficiency $\xi_{\rm ion}$. At $z\sim{}4-6$ the efficiency $\lg{}(\xi_{\rm ion}\,[{\rm Hz}\,{\rm erg}^{-1}])$ is $\sim{}25-25.5$ (e.g., \citealt{Robertson2013, Bouwens2016, Naidu2020, Castellano2023d, Saldana-Lopez2023, Simmonds2023}), with a similar range in ionizing efficiency also predicted by numerical models (e.g., \citealt{Wilkins2016c, Ceverino2019, Seeyave2023}). Recent estimates with JWST, however, suggest somewhat larger values up to $25.8$ in galaxies at $z\sim{}7-9$ likely as a result of their younger ages and lower metallicities \citep{Tang2023, Fujimoto2023, Endsley2023a, Atek2024, Simmonds2024}. Adopting the upper value of $\lg{}(\xi_{\rm ion}\,[{\rm Hz}\,{\rm erg}^{-1}])=25.8$, we infer escape fractions of $f_{\rm esc}=0.05-0.08$, while lower ionization efficiencies of $25.4$ and $25.0$ would imply correspondingly larger escape fractions of $f_{\rm esc}=0.12-0.20$ and $f_{\rm esc}=0.30-0.50$.

The escape fractions of galaxies that dominate cosmic reionization have yet to be fully constrained. Numerical predictions vary significantly depending on included physics, resolution, redshift, and halo mass (e.g., \citealt{Kimm2014a, Wise2014, Sharma2016, Rosdahl2022}). For instance, cosmological simulations with FIRE-2 physics suggest average escape fractions $f_{\rm esc}\sim{}20\%$ \citep{Ma2020} that peak in $M_{\rm halo}\sim{}10^{9.5}-10^{11}\,M_\odot$ halos, while other recent works suggest somewhat smaller values ($f_{\rm esc}\sim{}3-10\%$, e.g., \citealt{Katz2018, Lewis2020, Rosdahl2022, Yeh2023}) with a different mass dependence.
Observational constraints can only be indirect given that Lyman continuum photons emitted from galaxies during reionization are fully absorbed by intervening neutral hydrogen \citep{Inoue2014}. Scaling relations based on local analogs suggest escape fractions during reionization in the range of 10\%-15\% \citep{Mascia2023, Lin2023}, while measurements of escape fractions at $z\sim{}2-5$ are often found to be in the range 5-10\% \citep{Matthee2017a, Pahl2021, Saldana-Lopez2023} but with a large scatter for individual objects sometimes showing much larger escape fractions \citep{Leethochawalit2016, Shapley2016a, Naidu2017}. 

In summary, a non-evolving SFE -- halo halo mass relation produces an ionization history in agreement with observations provided $\lg{}(f_{\rm esc}\,\xi_{\rm ion}[{\rm Hz}\,{\rm erg}^{-1}])$ is about $24.5-24.7$. Large ionizing efficiencies $\lg{}(\xi_{\rm ion}\,[{\rm Hz}\,{\rm erg}^{-1}])\sim{}25.8$ thus imply small escape fractions of $\sim{}5-8\%$, see also \cite{Munoz2024}. Vice versa, larger escape fractions of $\sim{}20\%$ as predicted by FIRE-2 zoom-in simulations at $z\sim{}6-10$ \citep{Ma2020} require lower average ionization efficiencies of $25.2-25.4$ (cf. \citealt{Endsley2023}) to avoid overproducing ionizing photons resulting in too early an reionization of the Universe. Future measurements of the ionization efficiency during $z\sim{}6-10$ for large representative samples with JWST should provide more clarity.

\subsection{Caveats}
\label{sect:Caveats}

A caveat of the present study is that it is limited to a single physical model (FIRE-2). Fortunately, the chosen model is able to reproduce successfully the galaxy properties of interest, i.e., the UV luminosity functions and the evolution of the UV luminosity density, lending credence to the inferences drawn from it. Furthermore, the model has been extensively validated in previous studies from the FIRE collaboration \citep{Hopkins2018, Faucher-Giguere2018a}.

A further concern is the mass range of galaxies captured by \hr{}. While the high resolution of the simulation allows us to resolve faint galaxies, the high mass end ($M_{\rm star}>10^9\,M_\odot$, $M_{\rm halo}>10^{11}\,M_\odot$) is not well sampled given the modest box size ($L\sim{}22.1$ cMpc). However, as we demonstrated explicitly, the lack of such massive galaxies does not affect the UV luminosity density at a significant level (Fig.~\ref{fig:Median_UV_luminosity}).
Furthermore, given the focus on low to intermediate mass galaxies, the lack of AGN feedback in \hr{} does not appear to pose a significant limitation. We plan to revisit the role of AGN feedback with future simulations covering larger volumes. 

A recent study by \cite{Borrow2023} suggests that the use of a spatially uniform cosmic ionizing background may overestimate the UV luminosity in faint, low mass galaxies ($M_{\rm UV}\gtrsim{}-13$, $M_{\rm star}\lesssim{}10^6\,M_\odot$) compared with a fully self-consistent radiation-hydrodynamical simulation. However, such galaxies do not strongly influence the cosmic UV luminosity density (Figs.~\ref{fig:Median_UV_luminosity}, \ref{fig:UVLD}). Moreover, while a uniform background is used to capture the collective radiation from distant galaxies, \hr{} also accounts explicitly, albeit approximately, for local inhomogeneities in the radiation field, see Section \ref{sect:Sim}.

Our estimates also do not account for a potential Population III contribution with a top-heavy IMF which could boost the UV luminosities significantly \citep{Zackrisson2011, Harikane2023, Yung2023}. More generally, the physics of star formation, gas cooling, and feedback, might noticeably change at the earliest cosmic times leading to different SFEs. Our work suggests, however, that such strong changes are not present at $z<14$. We leave a deeper exploration of this topic for future work.

\section{Summary and Conclusions}
\label{sect:Summary}

We have introduced \hr{}, a high-resolution cosmological volume simulation from the \emph{Feedback in Realistic Environments} (FIRE) project, to explore galaxy formation from late Cosmic Dawn through the end of reionization—a period that continues to attract intense observational and theoretical interest. The simulation covers the same volume ($L=22.1$ Mpc) as the original \fb{} run \citep{Feldmann2023} down to $z=6.3$, employing the established FIRE-2 physics model, but with significantly improved numerical resolution ($m_{\rm b}\sim{}7800\,M_\odot$). \hr{} resolves a substantial number of faint, low mass galaxies (e.g., about 2000 galaxies with $M_{\rm UV}<-14$ or with $M_{\rm star}>10^{6.4}\,M_\odot$ at $z=6.3$) making it well suited to examine the evolution of the faint end of the galaxy population that may have played a pivotal role in cosmic reionization (e.g., \citealt{Robertson2015, Mascia2023, Simmonds2024, Atek2024, Dayal2024}). In this initial application of \hr{}, we analyzed the predicted ultraviolet (UV) luminosities and star formation efficiencies (SFEs) of high redshift galaxies ($z\sim{}6-15$) and discussed implications for early galaxy formation. We plan to investigate other properties of \hr{} galaxies, such their metallicities, sizes, and shapes, in future work. Our main findings are as follows:

\begin{itemize}
\item The UV luminosity functions (LFs) predicted by \hr{} are in excellent agreement with observations at $z\sim{}6-12$ (Fig.~\ref{fig:UVLF}). The UV LFs are well fit by a Schechter function or double power law for $M_{\rm UV}$ brighter than about -15. At  magnitudes near or just below current observational limits at $z\sim{}6-8$, the predicted LFs deviate from the traditional Schechter form indicative of a turn-over.  

\item The cosmic UV luminosity density at $z\sim{}6-12$ is dominated by galaxies of moderate luminosity, i.e., by galaxies with $M_{\rm UV}\sim{}-18$ to $-18.5$ for a UV magnitude limit $M_{\rm UV, lim}=-17$ and by galaxies with $M_{\rm UV}\sim{}-16$ to $-17$ for $M_{\rm UV, lim}=-12$, depending on redshift (Fig.~\ref{fig:Median_UV_luminosity}). These galaxies reside in halos of intermediate mass $M_{\rm halo}\sim{}10^9-10^{11}\,M_\odot$.

\item In contrast to other theoretical and empirical studies (e.g., \citealt{Tacchella2018, Harikane2022, Kannan2023}), \hr{} reproduces the comparably gradual evolution of the UV luminosity density out to $z\sim{}14$ found in recent JWST observations (Fig.~\ref{fig:UVLD}). \hr{} demonstrates that such a comparably shallow evolution of the UV luminosity density is not in conflict to galaxy formation theory but a natural outcome of galaxy evolution in the context of the FIRE-2 physics model.

\item The SFE -- halo mass relation of \hr{} galaxies does not significantly evolve with redshift and is only weakly dependent on mass for intermediate mass halos ($M_{\rm halo}\sim{}10^9-10^{11}\,M_\odot$), see Fig.~\ref{fig:efficiency}. The slope for $M_{\rm halo}\sim{}10^{10}-10^{11}\,M_\odot$ halos is $\sim{}0.03$ if dust-attenuated UV luminosities are used as SFR tracers and $\sim{}0.3$ if based on intrinsic UV luminosities or time-averaged SFRs. The scatter of the relation, which as defined here measures the variability of the UV luminosity or SFR, shows no obvious dependence on redshift and it is too small ($\sim{}0.3-0.6$ dex over the same halo mass range) to substantially affect the UV luminosity density for $M_{\rm UV, lim}=-17$. The comparably gradual evolution of the UV luminosity density at $z\sim{}6-14$ is thus neither explained by galaxies at higher redshift becoming more efficient on average in forming stars nor by an increase in the scatter of the SFE. Indeed, given the monotonic increase of the SFE with halo mass for $M_{\rm halo}\sim{}10^9-10^{11}\,M_\odot$ halos, individual galaxies tend to become less efficient in forming stars at higher redshift. A parametrization of the SFE -- halo mass relation in \hr{} is provided in Table~\ref{tab:SFEFitParams}.

\item The stellar mass -- halo mass relation (SHMR) predicted from the non-evolving SFE -- halo mass relation is in line with the actual SHMR in \hr{} (Fig.~\ref{fig:SHMR}). The slope ($\sim{}0.5$) of the SHMR for $M_{\rm halo}\sim{}10^{10}-10^{11}\,M_\odot$ halos is steeper than the slope ($\sim{}0.03$) of the (dust-attenuated UV luminosity based) SFE -- halo mass relation  largely, but not completely, because of the dust-attenuation. Possible explanations for the remaining difference in slope are discussed in Section \ref{sect:SHMR}.

\item A simple theoretical model based on the inferred SFE -- halo mass relation well reproduces the observed, gradual evolution of the UV luminosity density (Fig.~\ref{fig:UVLD_Model}). Applying this model up to $z=20$, we observe a pronounced decrease in cosmic UV luminosity density at higher redshifts (Fig.~\ref{fig:UVLD_Model_highz}). As a result, detecting neutral hydrogen in absorption via Ly-$\alpha$ activation of its 21 cm line appears not to be possible at redshifts as high as suggested by the 21 cm interpretation of the EDGES signal. Formally, $\rho_{\rm UV}(z)$ exceeds the minimum required UV luminosity density \citep{Madau2018} only at $z\lesssim{}14$ (for $M_{\rm UV, lim}=-12$), i.e., at a time around which cosmic gas was presumably (pre-)heated by stellar and X-ray sources (e.g., \citealt{Eide2020, Abdurashidova2023}).

\item The evolution of the cosmic UV luminosity density depends on the slope and normalization of the SFE -- halo mass relation for $M_{\rm halo}\sim{}10^9-10^{11}\,M_\odot$ halos but not on its scatter provided the latter is in a range ($\sim{}0.2-0.6$ dex) consistent with \hr{} (Fig.~\ref{fig:UVLD_ModelVariations}).

\item Further inspection of the theoretical model reveals that the comparably gradual evolution of the cosmic UV luminosity density is a consequence of the weak mass dependence (and the redshift independence) of the SFE -- halo mass relation in the intermediate halo mass regime (Fig.~\ref{fig:CSFH_components}). When reducing the slope of the relation while keeping the UV luminosity density at $z=6$ fixed, galaxies in $M_{\rm halo}\lesssim{}10^{10.7}\,M_\odot$ halos tend to become more UV luminous. Also, additional galaxies in low mass halos may exceed the adopted UV magnitude limit and thus contribute to the UV luminosity density. The resulting increase in the cosmic UV luminosity density is stronger at higher $z$ as the typical halo mass harboring galaxies contributing to the cosmic UV luminosity density becomes smaller. Hence, a lower (i.e., shallower) slope of the SFE -- halo mass relation in the intermediate halo mass range leads to a more gradual evolution of the UV luminosity density. In contrast, the large slopes ($\sim{}1-1.2$) adopted by previous non-evolving SFE models \citep{Tacchella2018, Harikane2022} lead to a steep evolution of $\rho_{\rm UV}(z)$ at $z\sim{}6-14$.

\item The faint end slope of the UV LF steepens with increasing redshift (Fig.~\ref{fig:UVLF_Model_highz}). Deviations from a pure Schechter function may not necessarily point to a turn-over at low luminosities but could reflect a double power law form of the UV LFs. Changes to the low mass and high mass slope of the SFE -- halo mass relation affect the UV LF in expected ways.

\item The theoretical model based on the non-evolving SFE -- halo mass relation predicts a reionization history in line with both observational data at $z\sim{}5.5-10$ and the optical depth reported by Planck 2018 provided $\zeta\equiv{}\lg{}(f_{\rm esc}\,\xi_{\rm ion}[{\rm Hz}\,{\rm erg}^{-1}])$ is in the range $\sim 24.5-24.7$ (Fig.~\ref{fig:reionization}). Here, $f_{\rm esc}$ is the escape fraction of ionizing radiation and $\xi_{\rm ion}$ is the ionizing photon production efficiency. Table~\ref{tab:reionization} lists the reionization mid-point and duration for different choices of $\zeta$.

\item The constraint $\zeta\sim{}24.5-24.7$ translates into escape fractions of $f_{\rm esc}=0.05-0.08$ ($f_{\rm esc}=0.12-0.20$, $f_{\rm esc}=0.30-0.50$) for ionizing efficiencies $\lg{}(\xi_{\rm ion}\,[{\rm Hz}\,{\rm erg}^{-1}])=25.8$ ($25.4$, $25.0$). Adopting $f_{\rm esc}\sim{}0.2$ as predicted by FIRE-2 zoom-in simulations at $z\sim{}6-10$ \citep{Ma2020} implies an average ionizing efficiency of $\lg{}(\xi_{\rm ion}\,[{\rm Hz}\,{\rm erg}^{-1}])=25.2-25.4$.

\end{itemize}

The present work leaves open several key follow-up questions for future studies. First, it remains to be understood how the physical processes included in the simulation lead to a SFE -- halo mass relation for $M_{\rm halo}\sim{}10^9-10^{11}\,M_\odot$ halos at $z\sim{}6-15$ that is only weakly dependent on halo mass and redshift. Here, a thorough exploration of the dependence of the SFE -- halo mass relation on either the adopted physical model or on secondary galaxy or halo properties should prove fruitful. Secondly, we may ask how the SFE -- halo mass relation evolves at the massive end ($M_{\rm halo}>10^{11}\,M_\odot$), which is not probed by \hr{} because of its modest box size. 
Empirical models predict a potential reduction of the SFE at masses above $M_{\rm halo}\sim{}10^{11.2}-10^{11.7}\,M_\odot$ and perhaps some redshift evolution \citep{Tacchella2018, Harikane2022}. Simulations of moderately large cosmological volumes with AGN feedback will likely be needed to properly explore this mass regime. Such simulations should not only help to constrain the physical parameters of AGN feedback by comparing with available JWST observations of UV bright galaxies, but also provide insights into the growth of massive, luminous galaxies with star formation efficiencies (e.g., \citealt{Lovell2023a, Casey2023, Bassini2023, Glazebrook2023, Carnall2024}). Thirdly, understanding how galaxies reionize intergalactic gas and influence its topology is a key focus of many observational programs, including those planned with the upcoming SKAO. Given that the SFE -- halo mass relation encapsulates well the star formation activity of galaxies during the epoch of reionization, it appears promising to combine the theoretical model outlined in this paper with approximate numerical methods based on one-dimensional radiative transfer calculations (e.g., \citealt{Ghara2018, Schaeffer2023}), as a computationally cheaper alternative to radiative-hydrodynamical simulations (e.g., \citealt{Rosdahl2018a, Ocvirk2020, Lewis2022, Kannan2022}). A better understanding of the SFE -- halo mass relation also holds great value for semi-numerical models that predict line intensities of 21 cm \citep{Park2019} and other emission lines (e.g., \citealt{Sun2023c}).

Many predictions from this study will be testable through forthcoming observations. The stronger evolution of $\rho_{\rm UV}(z)$ at $z\gtrsim{}13$, for example, could be explored with deeper JWST observations, potentially aided by strong gravitational lensing. Such observational studies may also be able to test the continued steepening of the faint end slope of the UV LFs with increasing redshift beyond the epochs currently probed. Additionally, the shape of the UV LF at $M_{\rm UV}\sim{}-13$ (e.g., turn-over vs power-law) may be more robustly quantified with deeper observations at $z\sim{}5-6$. Finally, the SFE in low mass halos may be measured directly by analyzing the clustering of UV faint galaxies. These future observations will not only test our current understanding of galaxy formation but also impose critical constraints on the involved physics during the first billion years of cosmic time.

\section*{Acknowledgements}

The authors are grateful to the referee for their thoughtful suggestions, which improved the manuscript. RF thanks Andrey Kravtsov for insightful discussions which inspired the addition of Appendix D.
RF acknowledges financial support from the Swiss National Science Foundation (grants PP00P2\_194814 and 200021\_188552). 
MBK acknowledges support from NSF CAREER award AST-1752913, NSF grants AST-1910346 and AST-2108962, and HST-GO-16686, HST-AR-17028, and HST-AR-17043 from the Space Telescope Science Institute, which is operated by AURA, Inc., under NASA contract NAS5-26555.
MBK and JSB acknowledge supported from NASA grant 80NSSC22K0827.
CAFG was supported by NSF through grants AST-1715216, AST-2108230, and CAREER award AST-1652522; by NASA through grant 17-ATP17-0067; by STScI through grant HST-AR-16124.001-A; and by the Research Corporation for Science Advancement through a Cottrell Scholar Award.
DK was supported by NSF grant AST-210834.
AL was supported by NASA grant 80NSSSC20K1469.
PO acknowledges funding from the Swiss State Secretariat for Education, Research and Innovation (SERI) under contract number MB22.00072, as well as from the Swiss National Science Foundation (SNSF) through project grant 200020\_207349. The Cosmic Dawn Center (DAWN) is funded by the Danish National Research Foundation under grant DNRF140.
This work was supported in part by a Simons Investigator award from the Simons Foundation (EQ) and by NSF grant AST-2107872. 
GS was supported by a CIERA Postdoctoral Fellowship.
This research was supported in part by the International Space Science Institute (ISSI) in Bern, through ISSI International Team project \#562 (First Light at Cosmic Dawn: Exploiting the James Webb Space Telescope Revolution) and by grant NSF PHY-1748958 to the Kavli Institute for Theoretical Physics (KITP). This work also benefited from discussions during the Nordita Program "Cosmic Dawn at High Latitudes".
We acknowledge PRACE for awarding us access to MareNostrum at the Barcelona Supercomputing Center (BSC), Spain. This research was partly carried out via the Frontera computing project at the Texas Advanced Computing Center. Frontera is made possible by National Science Foundation award OAC-1818253. This work was supported in part by an allocation from the Swiss National Supercomputing Centre (CSCS) under project IDs s697 and s698. We acknowledge access to Piz Daint and Alps/Eiger at the Swiss National Supercomputing Centre, Switzerland under the University of Zurich's share with the project ID uzh18. This work made use of infrastructure services provided by S3IT (www.s3it.uzh.ch), the Service and Support for Science IT team at the University of Zurich. All plots were created with the Matplotlib library for visualization with Python \citep{Hunter2007}. This research has made use of NASA's Astrophysics Data System.

\section*{Data availability}
The data supporting the plots within this article are available on reasonable request to the corresponding author. A public version of the GIZMO code is available at \url{http://www.tapir.caltech.edu/~phopkins/Site/GIZMO.html}. FIRE data releases are publicly available at \url{http://flathub.flatironinstitute.org/fire}.

%%%%%%%%%%%%%%%%%%%%%%%%%%%%%%%%%%%%%%%%%%%%%%%%%%


\bibliographystyle{mnras}

\begin{thebibliography}{}
\makeatletter
\relax
\def\mn@urlcharsother{\let\do\@makeother \do\$\do\&\do\#\do\^\do\_\do\%\do\~}
\def\mn@doi{\begingroup\mn@urlcharsother \@ifnextchar [ {\mn@doi@}
  {\mn@doi@[]}}
\def\mn@doi@[#1]#2{\def\@tempa{#1}\ifx\@tempa\@empty \href
  {http://dx.doi.org/#2} {doi:#2}\else \href {http://dx.doi.org/#2} {#1}\fi
  \endgroup}
\def\mn@eprint#1#2{\mn@eprint@#1:#2::\@nil}
\def\mn@eprint@arXiv#1{\href {http://arxiv.org/abs/#1} {{\tt arXiv:#1}}}
\def\mn@eprint@dblp#1{\href {http://dblp.uni-trier.de/rec/bibtex/#1.xml}
  {dblp:#1}}
\def\mn@eprint@#1:#2:#3:#4\@nil{\def\@tempa {#1}\def\@tempb {#2}\def\@tempc
  {#3}\ifx \@tempc \@empty \let \@tempc \@tempb \let \@tempb \@tempa \fi \ifx
  \@tempb \@empty \def\@tempb {arXiv}\fi \@ifundefined
  {mn@eprint@\@tempb}{\@tempb:\@tempc}{\expandafter \expandafter \csname
  mn@eprint@\@tempb\endcsname \expandafter{\@tempc}}}

\bibitem[\protect\citeauthoryear{Abdurashidova et~al.,}{Abdurashidova
  et~al.}{2022}]{Abdurashidova2022}
Abdurashidova Z.,  et~al., 2022, \mn@doi [Astrophys. J.]
  {10.3847/1538-4357/ac1c78}, 925, 221

\bibitem[\protect\citeauthoryear{Abdurashidova et~al.,}{Abdurashidova
  et~al.}{2023}]{Abdurashidova2023}
Abdurashidova T. H. C.~Z.,  et~al., 2023, \mn@doi [Astrophys. J.]
  {10.3847/1538-4357/acaf50}, 945, 124

\bibitem[\protect\citeauthoryear{Adam et~al.,}{Adam et~al.}{2016}]{Adam2016}
Adam R.,  et~al., 2016, \mn@doi [Astron. Astrophys.]
  {10.1051/0004-6361/201628897}, 596, A108

\bibitem[\protect\citeauthoryear{Adams et~al.,}{Adams et~al.}{2024}]{Adams2024}
Adams N.~J.,  et~al., 2024, \mn@doi [Astrophys. J.] {10.3847/1538-4357/ad2a7b},
  965, 169

\bibitem[\protect\citeauthoryear{Ade et~al.,}{Ade
  et~al.}{2016}]{PlanckCollaboration2015a}
Ade P. A.~R.,  et~al., 2016, \mn@doi [Astron. Astrophys.]
  {10.1051/0004-6361/201525830}, 594, A13

\bibitem[\protect\citeauthoryear{Aghanim et~al.,}{Aghanim
  et~al.}{2020}]{Aghanim2020}
Aghanim N.,  et~al., 2020, \mn@doi [Astron. Astrophys.]
  {10.1051/0004-6361/201833910}, 641, A6

\bibitem[\protect\citeauthoryear{Atek, Richard, Kneib  \& Schaerer}{Atek
  et~al.}{2018}]{Atek2018}
Atek H.,  Richard J.,  Kneib J.~P.,   Schaerer D.,  2018, \mn@doi [Mon. Not. R.
  Astron. Soc.] {10.1093/mnras/sty1820}, 479, 5184

\bibitem[\protect\citeauthoryear{Atek et~al.,}{Atek et~al.}{2023}]{Atek2023}
Atek H.,  et~al., 2023, \mn@doi [Mon. Not. R. Astron. Soc.]
  {10.1093/mnras/stad1998}, 524, 5486

\bibitem[\protect\citeauthoryear{Atek et~al.,}{Atek et~al.}{2024}]{Atek2024}
Atek H.,  et~al., 2024, \mn@doi [Nature] {10.1038/s41586-024-07043-6}, 626, 975

\bibitem[\protect\citeauthoryear{Aver, Olive  \& Skillman}{Aver
  et~al.}{2015}]{Aver2015}
Aver E.,  Olive K.~A.,   Skillman E.~D.,  2015, \mn@doi [J. Cosmol. Astropart.
  Phys.] {10.1088/1475-7516/2015/07/011}, 2015

\bibitem[\protect\citeauthoryear{Ba{\~{n}}ados et~al.,}{Ba{\~{n}}ados
  et~al.}{2018}]{Banados2018}
Ba{\~{n}}ados E.,  et~al., 2018, \mn@doi [Nature] {10.1038/nature25180}, 553,
  473

\bibitem[\protect\citeauthoryear{Barkana \& Loeb}{Barkana \&
  Loeb}{2000}]{Barkana2000}
Barkana R.,  Loeb A.,  2000, \mn@doi [Astrophys. J.] {10.1086/309229}, 539, 20

\bibitem[\protect\citeauthoryear{Barry et~al.,}{Barry et~al.}{2019}]{Barry2019}
Barry N.,  et~al., 2019, \mn@doi [Astrophys. J.] {10.3847/1538-4357/ab40a8},
  884, 1

\bibitem[\protect\citeauthoryear{Bassini, Feldmann, Gensior, Hayward,
  Faucher-Gigu{\`{e}}re, Cenci, Liang  \& Bernardini}{Bassini
  et~al.}{2023}]{Bassini2023}
Bassini L.,  Feldmann R.,  Gensior J.,  Hayward C.~C.,  Faucher-Gigu{\`{e}}re
  C.-A.,  Cenci E.,  Liang L.,   Bernardini M.,  2023, \mn@doi [Mon. Not. R.
  Astron. Soc.] {10.1093/mnras/stad2617}, 525, 5388

\bibitem[\protect\citeauthoryear{Bassini, Feldmann, Gensior,
  Faucher-Gigu{\`{e}}re, Cenci, Moreno, Bernardini  \& Liang}{Bassini
  et~al.}{2024}]{Bassini2024}
Bassini L.,  Feldmann R.,  Gensior J.,  Faucher-Gigu{\`{e}}re C.-A.,  Cenci E.,
   Moreno J.,  Bernardini M.,   Liang L.,  2024, \mn@doi [Mon. Not. R. Astron.
  Soc. Lett.] {10.1093/mnrasl/slae036}, 532, L14

\bibitem[\protect\citeauthoryear{Behroozi \& Silk}{Behroozi \&
  Silk}{2015}]{Behroozi2015}
Behroozi P.~S.,  Silk J.,  2015, \mn@doi [Astrophys. J.]
  {10.1088/0004-637X/799/1/32}, 799

\bibitem[\protect\citeauthoryear{Behroozi, Conroy  \& Wechsler}{Behroozi
  et~al.}{2010}]{Behroozi2010a}
Behroozi P.~S.,  Conroy C.,   Wechsler R.~H.,  2010, \mn@doi [Astrophys. J.]
  {10.1088/0004-637X/717/1/379}, 717, 379

\bibitem[\protect\citeauthoryear{Behroozi, Wechsler  \& Conroy}{Behroozi
  et~al.}{2013}]{Behroozi2013c}
Behroozi P.~S.,  Wechsler R.~H.,   Conroy C.,  2013, \mn@doi [Astrophys. J.]
  {10.1088/0004-637X/770/1/57}, 770, 57

\bibitem[\protect\citeauthoryear{Behroozi, Wechsler, Hearin  \&
  Conroy}{Behroozi et~al.}{2019}]{Behroozi2019}
Behroozi P.,  Wechsler R.~H.,  Hearin A.~P.,   Conroy C.,  2019, \mn@doi [Mon.
  Not. R. Astron. Soc.] {10.1093/mnras/stz1182}, 488, 3143

\bibitem[\protect\citeauthoryear{Beltz-Mohrmann \& Berlind}{Beltz-Mohrmann \&
  Berlind}{2021}]{Beltz-Mohrmann2021}
Beltz-Mohrmann G.~D.,  Berlind A.~A.,  2021, \mn@doi [Astrophys. J.]
  {10.3847/1538-4357/ac1e27}, 921, 112

\bibitem[\protect\citeauthoryear{Bera, Hassan, Smith, Cen, Garaldi, Kannan  \&
  Vogelsberger}{Bera et~al.}{2023}]{Bera2023}
Bera A.,  Hassan S.,  Smith A.,  Cen R.,  Garaldi E.,  Kannan R.,
  Vogelsberger M.,  2023, \mn@doi [Astrophys. J.] {10.3847/1538-4357/ad05c0},
  959, 2

\bibitem[\protect\citeauthoryear{Bevins, Handley, Fialkov, {De Lera Acedo},
  Greenhill  \& Price}{Bevins et~al.}{2021}]{Bevins2021}
Bevins H.~T.,  Handley W.~J.,  Fialkov A.,  {De Lera Acedo} E.,  Greenhill
  L.~J.,   Price D.~C.,  2021, \mn@doi [Mon. Not. R. Astron. Soc.]
  {10.1093/mnras/stab152}, 502, 4405

\bibitem[\protect\citeauthoryear{Bolan et~al.,}{Bolan et~al.}{2021}]{Bolan2022}
Bolan P.,  et~al., 2021, \mn@doi [Mon. Not. R. Astron. Soc.]
  {10.1093/mnras/stac1963}, 517, 3263

\bibitem[\protect\citeauthoryear{Borrow, Kannan, Garaldi, Smith, Vogelsberger,
  Pakmor, Springel  \& Hernquist}{Borrow et~al.}{2023}]{Borrow2023}
Borrow J.,  Kannan R.,  Garaldi E.,  Smith A.,  Vogelsberger M.,  Pakmor R.,
  Springel V.,   Hernquist L.,  2023, \mn@doi [Mon. Not. R. Astron. Soc.]
  {10.1093/mnras/stad2523}, 525, 5932

\bibitem[\protect\citeauthoryear{Bouch{\'{e}} et~al.,}{Bouch{\'{e}}
  et~al.}{2010}]{Bouche2010}
Bouch{\'{e}} N.,  et~al., 2010, \mn@doi [Astrophys. J.]
  {10.1088/0004-637X/718/2/1001}, 718, 1001

\bibitem[\protect\citeauthoryear{Bouwens et~al.,}{Bouwens
  et~al.}{2009}]{Bouwens2009}
Bouwens R.~J.,  et~al., 2009, \mn@doi [Astrophys. J.]
  {10.1088/0004-637X/705/1/936}, 705, 936

\bibitem[\protect\citeauthoryear{Bouwens et~al.,}{Bouwens
  et~al.}{2012}]{Bouwens2012e}
Bouwens R.~J.,  et~al., 2012, \mn@doi [Astrophys. J.]
  {10.1088/2041-8205/752/1/L5}, 752, L5

\bibitem[\protect\citeauthoryear{Bouwens et~al.,}{Bouwens
  et~al.}{2014}]{Bouwens2014}
Bouwens R.~J.,  et~al., 2014, \mn@doi [Astrophys. J.]
  {10.1088/0004-637X/793/2/115}, 793, 115

\bibitem[\protect\citeauthoryear{Bouwens et~al.,}{Bouwens
  et~al.}{2015}]{Bouwens2015}
Bouwens R.~J.,  et~al., 2015, \mn@doi [Astrophys. J.]
  {10.1088/0004-637X/803/1/34}, 803, 34

\bibitem[\protect\citeauthoryear{Bouwens, Smit, Labb{\'{e}}, Franx, Caruana,
  Oesch, Stefanon  \& Rasappu}{Bouwens et~al.}{2016a}]{Bouwens2016}
Bouwens R.~J.,  Smit R.,  Labb{\'{e}} I.,  Franx M.,  Caruana J.,  Oesch P.,
  Stefanon M.,   Rasappu N.,  2016a, \mn@doi [Astrophys. J.]
  {10.3847/0004-637X/831/2/176}, 831, 176

\bibitem[\protect\citeauthoryear{Bouwens et~al.,}{Bouwens
  et~al.}{2016b}]{Bouwens2016b}
Bouwens R.,  et~al., 2016b, \mn@doi [Astrophys. J.]
  {10.3847/1538-4357/833/1/72}, 833, 72

\bibitem[\protect\citeauthoryear{Bouwens, Oesch, Illingworth, Ellis  \&
  Stefanon}{Bouwens et~al.}{2017}]{Bouwens2017}
Bouwens R.~J.,  Oesch P.~A.,  Illingworth G.~D.,  Ellis R.~S.,   Stefanon M.,
  2017, \mn@doi [Astrophys. J.] {10.3847/1538-4357/aa70a4}, 843, 129

\bibitem[\protect\citeauthoryear{Bouwens, Illingworth, Ellis, Oesch  \&
  Stefanon}{Bouwens et~al.}{2022}]{Bouwens2022}
Bouwens R.~J.,  Illingworth G.,  Ellis R.~S.,  Oesch P.,   Stefanon M.,  2022,
  \mn@doi [Astrophys. J.] {10.3847/1538-4357/ac86d1}, 940, 55

\bibitem[\protect\citeauthoryear{Bouwens, Illingworth, Oesch, Stefanon, Naidu,
  van Leeuwen  \& Magee}{Bouwens et~al.}{2023a}]{Bouwens2023a}
Bouwens R.,  Illingworth G.,  Oesch P.,  Stefanon M.,  Naidu R.,  van Leeuwen
  I.,   Magee D.,  2023a, \mn@doi [Mon. Not. R. Astron. Soc.]
  {10.1093/mnras/stad1014}, 523, 1009

\bibitem[\protect\citeauthoryear{Bouwens et~al.,}{Bouwens
  et~al.}{2023b}]{Bouwens2023}
Bouwens R.~J.,  et~al., 2023b, \mn@doi [Mon. Not. R. Astron. Soc.]
  {10.1093/mnras/stad1145}, 523, 1036

\bibitem[\protect\citeauthoryear{Bowman, Rogers, Monsalve, Mozdzen  \&
  Mahesh}{Bowman et~al.}{2018}]{Bowman2018}
Bowman J.~D.,  Rogers A. E.~E.,  Monsalve R.~A.,  Mozdzen T.~J.,   Mahesh N.,
  2018, \mn@doi [Nature] {10.1038/nature25792}, 555, 67

\bibitem[\protect\citeauthoryear{Boylan-Kolchin}{Boylan-Kolchin}{2023}]{Boylan-Kolchin2023}
Boylan-Kolchin M.,  2023, \mn@doi [Nat. Astron.] {10.1038/s41550-023-01937-7},
  7, 731

\bibitem[\protect\citeauthoryear{Boylan-Kolchin, Bullock  \&
  Garrison-Kimmel}{Boylan-Kolchin et~al.}{2014}]{Boylan-Kolchin2014}
Boylan-Kolchin M.,  Bullock J.~S.,   Garrison-Kimmel S.,  2014, \mn@doi [Mon.
  Not. R. Astron. Soc. Lett.] {10.1093/mnrasl/slu079}, 443, 44

\bibitem[\protect\citeauthoryear{Boylan-Kolchin, Weisz, Johnson, Bullock,
  Conroy  \& Fitts}{Boylan-Kolchin et~al.}{2015}]{Boylan-Kolchin2015a}
Boylan-Kolchin M.,  Weisz D.~R.,  Johnson B.~D.,  Bullock J.~S.,  Conroy C.,
  Fitts A.,  2015, \mn@doi [Mon. Not. R. Astron. Soc.] {10.1093/mnras/stv1736},
  453, 1503

\bibitem[\protect\citeauthoryear{Bradley, Tauscher, Rapetti  \& Burns}{Bradley
  et~al.}{2019}]{Bradley2019}
Bradley R.~F.,  Tauscher K.,  Rapetti D.,   Burns J.~O.,  2019, \mn@doi
  [Astrophys. J.] {10.3847/1538-4357/ab0d8b}, 874, 153

\bibitem[\protect\citeauthoryear{Bruton, Lin, Scarlata  \& Hayes}{Bruton
  et~al.}{2023}]{Bruton2023}
Bruton S.,  Lin Y.-H.,  Scarlata C.,   Hayes M.~J.,  2023, \mn@doi [Astrophys.
  J. Lett.] {10.3847/2041-8213/acd5d0}, 949, L40

\bibitem[\protect\citeauthoryear{Bryan \& Norman}{Bryan \&
  Norman}{1998}]{Bryan1998}
Bryan G.~L.,  Norman M.~L.,  1998, \mn@doi [Astrophys. J.] {10.1086/305262},
  495, 80

\bibitem[\protect\citeauthoryear{Camps \& Baes}{Camps \&
  Baes}{2020}]{Camps2020}
Camps P.,  Baes M.,  2020, \mn@doi [Astron. Comput.]
  {10.1016/j.ascom.2020.100381}, 31, 100381

\bibitem[\protect\citeauthoryear{Carnall et~al.,}{Carnall
  et~al.}{2024}]{Carnall2024}
Carnall A.~C.,  et~al., 2024, \mn@doi [Mon. Not. R. Astron. Soc.]
  {10.1093/mnras/stae2092}, 534, 325

\bibitem[\protect\citeauthoryear{Carniani et~al.,}{Carniani
  et~al.}{2024}]{Carniani2024}
Carniani S.,  et~al., 2024, \mn@doi [Nature] {10.1038/s41586-024-07860-9}, 633,
  318

\bibitem[\protect\citeauthoryear{Casey et~al.,}{Casey et~al.}{2024}]{Casey2023}
Casey C.~M.,  et~al., 2024, \mn@doi [Astrophys. J.] {10.3847/1538-4357/ad2075},
  965, 98

\bibitem[\protect\citeauthoryear{Castellano, Belfiori, Pentericci,
  Calabr{\`{o}}, Mascia, Napolitano  \& Caro}{Castellano
  et~al.}{2023}]{Castellano2023d}
Castellano M.,  Belfiori D.,  Pentericci L.,  Calabr{\`{o}} A.,  Mascia S.,
  Napolitano L.,   Caro F.,  2023, \mn@doi [Astron. Astrophys.]
  {10.1051/0004-6361/202346069}, 121, 1

\bibitem[\protect\citeauthoryear{Castellano et~al.,}{Castellano
  et~al.}{2024}]{Castellano2024}
Castellano M.,  et~al., 2024, \mn@doi [Astrophys. J.]
  {10.3847/1538-4357/ad5f88}, 972, 143

\bibitem[\protect\citeauthoryear{Ceverino, Klessen  \& Glover}{Ceverino
  et~al.}{2019}]{Ceverino2019}
Ceverino D.,  Klessen R.~S.,   Glover S. C.~O.,  2019, \mn@doi [Mon. Not. R.
  Astron. Soc.] {10.1093/mnras/stz079}, 484, 1366

\bibitem[\protect\citeauthoryear{Ceverino, Nakazato, Yoshida, Klessen  \&
  Glover}{Ceverino et~al.}{2024}]{Ceverino2024}
Ceverino D.,  Nakazato Y.,  Yoshida N.,  Klessen R.~S.,   Glover S. C.~O.,
  2024, \mn@doi [Astron. Astrophys.] {10.1051/0004-6361/202450224}, 689, A244

\bibitem[\protect\citeauthoryear{Chabrier}{Chabrier}{2003}]{Chabrier2003}
Chabrier G.,  2003, \mn@doi [Publ. Astron. Soc. Pacific] {10.1086/376392}, 115,
  763

\bibitem[\protect\citeauthoryear{Chakraborty \& Choudhury}{Chakraborty \&
  Choudhury}{2024}]{Chakraborty2024}
Chakraborty A.,  Choudhury T.~R.,  2024, \mn@doi [J. Cosmol. Astropart. Phys.]
  {10.1088/1475-7516/2024/07/078}, 2024, 078

\bibitem[\protect\citeauthoryear{Chan, Kere{\v{s}}, Wetzel, Hopkins,
  Faucher-Gigu{\`{e}}re, El-Badry, Garrison-Kimmel  \& Boylan-Kolchin}{Chan
  et~al.}{2018}]{Chan2018}
Chan T.~K.,  Kere{\v{s}} D.,  Wetzel A.,  Hopkins P.~F.,  Faucher-Gigu{\`{e}}re
  C.-A.,  El-Badry K.,  Garrison-Kimmel S.,   Boylan-Kolchin M.,  2018, \mn@doi
  [Mon. Not. R. Astron. Soc.] {10.1093/mnras/sty1153}, 478, 906

\bibitem[\protect\citeauthoryear{Chemerynska et~al.,}{Chemerynska
  et~al.}{2024}]{Chemerynska2024}
Chemerynska I.,  et~al., 2024, \mn@doi [Mon. Not. R. Astron. Soc.]
  {10.1093/mnras/stae1260}, 531, 2615

\bibitem[\protect\citeauthoryear{Conselice et~al.,}{Conselice
  et~al.}{2024}]{Conselice2024}
Conselice C.~J.,  et~al., 2024 (\mn@eprint {arXiv} {2407.14973})

\bibitem[\protect\citeauthoryear{Cui, Borgani  \& Murante}{Cui
  et~al.}{2014}]{Cui2014}
Cui W.,  Borgani S.,   Murante G.,  2014, \mn@doi [Mon. Not. R. Astron. Soc.]
  {10.1093/mnras/stu673}, 441, 1769

\bibitem[\protect\citeauthoryear{Cullen et~al.,}{Cullen
  et~al.}{2023}]{Cullen2023}
Cullen F.,  et~al., 2023, \mn@doi [Mon. Not. R. Astron. Soc.]
  {10.1093/mnras/stad073}, 520, 14

\bibitem[\protect\citeauthoryear{Curtis-Lake et~al.,}{Curtis-Lake
  et~al.}{2023}]{Curtis-Lake2023}
Curtis-Lake E.,  et~al., 2023, \mn@doi [Nat. Astron.]
  {10.1038/s41550-023-01918-w}, 7, 622

\bibitem[\protect\citeauthoryear{Dav{\'{e}}, Finlator  \&
  Oppenheimer}{Dav{\'{e}} et~al.}{2012}]{Dave2012a}
Dav{\'{e}} R.,  Finlator K.,   Oppenheimer B.~D.,  2012, \mn@doi [Mon. Not. R.
  Astron. Soc.] {10.1111/j.1365-2966.2011.20148.x}, 421, 98

\bibitem[\protect\citeauthoryear{Dayal}{Dayal}{2019}]{Dayal2019}
Dayal P.,  2019, \mn@doi [Proc. Int. Astron. Union]
  {10.1017/s1743921320001106}, 15, 43

\bibitem[\protect\citeauthoryear{Dayal, Ferrara, Dunlop  \& Pacucci}{Dayal
  et~al.}{2014}]{Dayal2014}
Dayal P.,  Ferrara A.,  Dunlop J.~S.,   Pacucci F.,  2014, \mn@doi [Mon. Not.
  R. Astron. Soc.] {10.1093/mnras/stu1848}, 445, 2545

\bibitem[\protect\citeauthoryear{Dayal et~al.,}{Dayal et~al.}{2024}]{Dayal2024}
Dayal P.,  et~al., 2024, pp~1--9 (\mn@eprint {arXiv} {2401.11242})

\bibitem[\protect\citeauthoryear{Deboer et~al.,}{Deboer
  et~al.}{2017}]{Deboer2017}
Deboer D.~R.,  et~al., 2017, \mn@doi [Publ. Astron. Soc. Pacific]
  {10.1088/1538-3873/129/974/045001}, 129, 1

\bibitem[\protect\citeauthoryear{Dekel, Sarkar, Birnboim, Mandelker  \&
  Li}{Dekel et~al.}{2023}]{Dekel2023}
Dekel A.,  Sarkar K.~C.,  Birnboim Y.,  Mandelker N.,   Li Z.,  2023, \mn@doi
  [Mon. Not. R. Astron. Soc.] {10.1093/mnras/stad1557}, 523, 3201

\bibitem[\protect\citeauthoryear{Donnan et~al.,}{Donnan
  et~al.}{2023a}]{Donnan2023}
Donnan C.~T.,  et~al., 2023a, \mn@doi [Mon. Not. R. Astron. Soc.]
  {10.1093/mnras/stac3472}, 518, 6011

\bibitem[\protect\citeauthoryear{Donnan, McLeod, McLure, Dunlop, Carnall,
  Cullen  \& Magee}{Donnan et~al.}{2023b}]{Donnan2023a}
Donnan C.~T.,  McLeod D.~J.,  McLure R.~J.,  Dunlop J.~S.,  Carnall A.~C.,
  Cullen F.,   Magee D.,  2023b, \mn@doi [Mon. Not. R. Astron. Soc.]
  {10.1093/mnras/stad471}, 520, 4554

\bibitem[\protect\citeauthoryear{Donnan et~al.,}{Donnan
  et~al.}{2024}]{Donnan2024}
Donnan C.~T.,  et~al., 2024, \mn@doi [Mon. Not. R. Astron. Soc.]
  {10.1093/mnras/stae2037}, 533, 3222

\bibitem[\protect\citeauthoryear{Draine et~al.,}{Draine
  et~al.}{2007}]{Draine2007j}
Draine B.~T.,  et~al., 2007, \mn@doi [Astrophys. J.] {10.1086/518306}, 663, 866

\bibitem[\protect\citeauthoryear{Dwek}{Dwek}{1998}]{Dwek1998b}
Dwek E.,  1998, \mn@doi [Astrophys. J.] {10.1086/305829}, 501, 643

\bibitem[\protect\citeauthoryear{Eide, Ciardi, Graziani, Busch, Feng  \& {Di
  Matteo}}{Eide et~al.}{2020}]{Eide2020}
Eide M.~B.,  Ciardi B.,  Graziani L.,  Busch P.,  Feng Y.,   {Di Matteo} T.,
  2020, \mn@doi [Mon. Not. R. Astron. Soc.] {10.1093/mnras/staa2774}, 498, 6083

\bibitem[\protect\citeauthoryear{Eldridge \& Stanway}{Eldridge \&
  Stanway}{2009}]{Eldridge2009}
Eldridge J.~J.,  Stanway E.~R.,  2009, \mn@doi [Mon. Not. R. Astron. Soc.]
  {10.1111/j.1365-2966.2009.15514.x}, 400, 1019

\bibitem[\protect\citeauthoryear{Endsley, Stark, Whitler, Topping, Chen, Plat,
  Chisholm  \& Charlot}{Endsley et~al.}{2023}]{Endsley2023a}
Endsley R.,  Stark D.~P.,  Whitler L.,  Topping M.~W.,  Chen Z.,  Plat A.,
  Chisholm J.,   Charlot S.,  2023, \mn@doi [Mon. Not. R. Astron. Soc.]
  {10.1093/mnras/stad1919}, 524, 2312

\bibitem[\protect\citeauthoryear{Endsley et~al.,}{Endsley
  et~al.}{2024}]{Endsley2023}
Endsley R.,  et~al., 2024, \mn@doi [Mon. Not. R. Astron. Soc.]
  {10.1093/mnras/stae1857}, 533, 1111

\bibitem[\protect\citeauthoryear{Fan, Carilli  \& Keating}{Fan
  et~al.}{2006}]{Fan2006a}
Fan X.,  Carilli C.~L.,   Keating B.,  2006, \mn@doi [Annu. Rev. Astron.
  Astrophys.] {10.1146/annurev.astro.44.051905.092514}, 44, 415

\bibitem[\protect\citeauthoryear{Faucher-Gigu{\`{e}}re}{Faucher-Gigu{\`{e}}re}{2018}]{Faucher-Giguere2018a}
Faucher-Gigu{\`{e}}re C.-A.,  2018, \mn@doi [Nat. Astron.]
  {10.1038/s41550-018-0427-y}, 2, 368

\bibitem[\protect\citeauthoryear{Faucher-Gigu{\`{e}}re, Lidz, Zaldarriaga  \&
  Hernquist}{Faucher-Gigu{\`{e}}re et~al.}{2009}]{Faucher-Giguere2009}
Faucher-Gigu{\`{e}}re C.-A.,  Lidz A.,  Zaldarriaga M.,   Hernquist L.,  2009,
  \mn@doi [Astrophys. J.] {10.1088/0004-637X/703/2/1416}, 703, 1416

\bibitem[\protect\citeauthoryear{Faucher-Gigu{\`{e}}re, Kere{\v{s}}, Dijkstra,
  Hernquist  \& Zaldarriaga}{Faucher-Gigu{\`{e}}re
  et~al.}{2010}]{Faucher-Giguere2010}
Faucher-Gigu{\`{e}}re C.~A.,  Kere{\v{s}} D.,  Dijkstra M.,  Hernquist L.,
  Zaldarriaga M.,  2010, \mn@doi [Astrophys. J.] {10.1088/0004-637X/725/1/633},
  725, 633

\bibitem[\protect\citeauthoryear{Feldmann}{Feldmann}{2015}]{Feldmann2015a}
Feldmann R.,  2015, \mn@doi [Mon. Not. R. Astron. Soc.] {10.1093/mnras/stv552},
  449, 3274

\bibitem[\protect\citeauthoryear{Feldmann, Gnedin  \& Kravtsov}{Feldmann
  et~al.}{2012}]{Feldmann2012c}
Feldmann R.,  Gnedin N.~Y.,   Kravtsov A.~V.,  2012, \mn@doi [Astrophys. J.]
  {10.1088/0004-637X/758/2/127}, 758, 127

\bibitem[\protect\citeauthoryear{Feldmann, Quataert, Hopkins,
  Faucher-Gigu{\`{e}}re  \& Kere{\v{s}}}{Feldmann et~al.}{2017}]{Feldmann2017a}
Feldmann R.,  Quataert E.,  Hopkins P.~F.,  Faucher-Gigu{\`{e}}re C.-A.,
  Kere{\v{s}} D.,  2017, \mn@doi [Mon. Not. R. Astron. Soc.]
  {10.1093/mnras/stx1120}, 470, 1050

\bibitem[\protect\citeauthoryear{Feldmann et~al.,}{Feldmann
  et~al.}{2023}]{Feldmann2023}
Feldmann R.,  et~al., 2023, \mn@doi [Mon. Not. R. Astron. Soc.]
  {10.1093/mnras/stad1205}, 522, 3831

\bibitem[\protect\citeauthoryear{Ferland, Korista, Verner, Ferguson, Kingdon
  \& Verner}{Ferland et~al.}{1998}]{Ferland1998}
Ferland G.~J.,  Korista K.~T.,  Verner D.~A.,  Ferguson J.~W.,  Kingdon J.~B.,
   Verner E.~M.,  1998, \mn@doi [Publ. Astron. Soc. Pacific] {10.1086/316190},
  110, 761

\bibitem[\protect\citeauthoryear{Ferrara}{Ferrara}{2024}]{Ferrara2024}
Ferrara A.,  2024, \mn@doi [Astron. Astrophys.] {10.1051/0004-6361/202348321},
  684, 1

\bibitem[\protect\citeauthoryear{Ferrara, Pallottini  \& Dayal}{Ferrara
  et~al.}{2023}]{Ferrara2023}
Ferrara A.,  Pallottini A.,   Dayal P.,  2023, \mn@doi [Mon. Not. R. Astron.
  Soc.] {10.1093/mnras/stad1095}, 522, 3986

\bibitem[\protect\citeauthoryear{Field}{Field}{1958}]{Field1958}
Field G.,  1958, \mn@doi [Proc. IRE] {10.1109/JRPROC.1958.286741}, 46, 240

\bibitem[\protect\citeauthoryear{Finkelstein et~al.,}{Finkelstein
  et~al.}{2015}]{Finkelstein2015}
Finkelstein S.~L.,  et~al., 2015, \mn@doi [Astrophys. J.]
  {10.1088/0004-637X/810/1/71}, 810, 71

\bibitem[\protect\citeauthoryear{Finkelstein et~al.,}{Finkelstein
  et~al.}{2022}]{Finkelstein2022}
Finkelstein S.~L.,  et~al., 2022, \mn@doi [Astrophys. J. Lett.]
  {10.3847/2041-8213/ac966e}, 940, L55

\bibitem[\protect\citeauthoryear{Finkelstein et~al.,}{Finkelstein
  et~al.}{2023}]{Finkelstein2023}
Finkelstein S.~L.,  et~al., 2023, \mn@doi [Astrophys. J. Lett.]
  {10.3847/2041-8213/acade4}, 946, L13

\bibitem[\protect\citeauthoryear{Finkelstein et~al.,}{Finkelstein
  et~al.}{2024}]{Finkelstein2024}
Finkelstein S.~L.,  et~al., 2024, \mn@doi [Astrophys. J. Lett.]
  {10.3847/2041-8213/ad4495}, 969, L2

\bibitem[\protect\citeauthoryear{{Flores Vel{\'{a}}zquez} et~al.,}{{Flores
  Vel{\'{a}}zquez} et~al.}{2021}]{FloresVelazquez2021}
{Flores Vel{\'{a}}zquez} J.~A.,  et~al., 2021, \mn@doi [Mon. Not. R. Astron.
  Soc.] {10.1093/mnras/staa3893}, 501, 4812

\bibitem[\protect\citeauthoryear{Fudamoto et~al.,}{Fudamoto
  et~al.}{2020}]{Fudamoto2020}
Fudamoto Y.,  et~al., 2020, \mn@doi [Astron. Astrophys.]
  {10.1051/0004-6361/202038163}, 643, 1

\bibitem[\protect\citeauthoryear{Fujimoto et~al.,}{Fujimoto
  et~al.}{2023}]{Fujimoto2023}
Fujimoto S.,  et~al., 2023, \mn@doi [Astrophys. J. Lett.]
  {10.3847/2041-8213/acd2d9}, 949, L25

\bibitem[\protect\citeauthoryear{Furlanetto, Mirocha, Mebane  \&
  Sun}{Furlanetto et~al.}{2017}]{Furlanetto2017}
Furlanetto S.~R.,  Mirocha J.,  Mebane R.~H.,   Sun G.,  2017, \mn@doi [Mon.
  Not. R. Astron. Soc.] {10.1093/MNRAS/STX2132}, 472, 1576

\bibitem[\protect\citeauthoryear{Gaikwad et~al.,}{Gaikwad
  et~al.}{2023}]{Gaikwad2023}
Gaikwad P.,  et~al., 2023, \mn@doi [Mon. Not. R. Astron. Soc.]
  {10.1093/mnras/stad2566}, 525, 4093

\bibitem[\protect\citeauthoryear{Gallerani, Ferrara, Fan  \&
  Choudhury}{Gallerani et~al.}{2007}]{Gallerani2008}
Gallerani S.,  Ferrara A.,  Fan X.,   Choudhury T.~R.,  2007, \mn@doi [Mon.
  Not. R. Astron. Soc.] {10.1111/j.1365-2966.2008.13029.x}, 386, 359

\bibitem[\protect\citeauthoryear{Garrison-Kimmel et~al.,}{Garrison-Kimmel
  et~al.}{2019}]{Garrison-Kimmel2019}
Garrison-Kimmel S.,  et~al., 2019, \mn@doi [Mon. Not. R. Astron. Soc.]
  {10.1093/mnras/stz1317}, 487, 1380

\bibitem[\protect\citeauthoryear{Gelli, Mason  \& Hayward}{Gelli
  et~al.}{2024}]{Gelli2024}
Gelli V.,  Mason C.,   Hayward C.~C.,  2024, \mn@doi [Astrophys. J.]
  {10.3847/1538-4357/ad7b36}, 975, 192

\bibitem[\protect\citeauthoryear{Gensior, Feldmann, Mayer, Wetzel, Hopkins  \&
  Faucher-Gigu{\`{e}}re}{Gensior et~al.}{2023}]{Gensior2023}
Gensior J.,  Feldmann R.,  Mayer L.,  Wetzel A.,  Hopkins P.~F.,
  Faucher-Gigu{\`{e}}re C.~A.,  2023, \mn@doi [Mon. Not. R. Astron. Soc. Lett.]
  {10.1093/mnrasl/slac138}, 518, L63

\bibitem[\protect\citeauthoryear{Ghara, Mellema, Giri, Choudhury, Datta  \&
  Majumdar}{Ghara et~al.}{2018}]{Ghara2018}
Ghara R.,  Mellema G.,  Giri S.~K.,  Choudhury T.~R.,  Datta K.~K.,   Majumdar
  S.,  2018, \mn@doi [Mon. Not. R. Astron. Soc.] {10.1093/mnras/sty314}, 476,
  1741

\bibitem[\protect\citeauthoryear{Gill, Knebe  \& Gibson}{Gill
  et~al.}{2004}]{Gill2004}
Gill S. P.~D.,  Knebe A.,   Gibson B.~K.,  2004, \mn@doi [Mon. Not. R. Astron.
  Soc.] {10.1111/j.1365-2966.2004.07786.x}, 351, 399

\bibitem[\protect\citeauthoryear{Glazebrook et~al.,}{Glazebrook
  et~al.}{2024}]{Glazebrook2023}
Glazebrook K.,  et~al., 2024, \mn@doi [Nature] {10.1038/s41586-024-07191-9},
  628, 277

\bibitem[\protect\citeauthoryear{Gnedin}{Gnedin}{2000}]{Gnedin2000a}
Gnedin N.~Y.,  2000, \mn@doi [Astrophys. J.] {10.1086/308876}, 535, 530

\bibitem[\protect\citeauthoryear{Gnedin}{Gnedin}{2014}]{Gnedin2014a}
Gnedin N.~Y.,  2014, \mn@doi [Astrophys. J.] {10.1088/0004-637X/793/1/29}, 793,
  1

\bibitem[\protect\citeauthoryear{Gnedin \& Madau}{Gnedin \&
  Madau}{2022}]{Gnedin2022}
Gnedin N.~Y.,  Madau P.,  2022, \mn@doi [Living Rev. Comput. Astrophys.]
  {10.1007/s41115-022-00015-5}, 8, 3

\bibitem[\protect\citeauthoryear{Gnedin, Tassis  \& Kravtsov}{Gnedin
  et~al.}{2009}]{Gnedin2009a}
Gnedin N.~Y.,  Tassis K.,   Kravtsov A.~V.,  2009, \mn@doi [Astrophys. J.]
  {10.1088/0004-637X/697/1/55}, 697, 55

\bibitem[\protect\citeauthoryear{Gorce, Douspis  \& Salvati}{Gorce
  et~al.}{2022}]{Gorce2022}
Gorce A.,  Douspis M.,   Salvati L.,  2022, \mn@doi [Astron. Astrophys.]
  {10.1051/0004-6361/202243351}, 662

\bibitem[\protect\citeauthoryear{Greig, Mesinger, Haiman  \& Simcoe}{Greig
  et~al.}{2017}]{Greig2017}
Greig B.,  Mesinger A.,  Haiman Z.,   Simcoe R.~A.,  2017, \mn@doi [Mon. Not.
  R. Astron. Soc.] {10.1093/mnras/stw3351}, 466, 4239

\bibitem[\protect\citeauthoryear{Greig, Mesinger, Davies, Wang, Yang  \&
  Hennawi}{Greig et~al.}{2022}]{Greig2022}
Greig B.,  Mesinger A.,  Davies F.~B.,  Wang F.,  Yang J.,   Hennawi J.~F.,
  2022, \mn@doi [Mon. Not. R. Astron. Soc.] {10.1093/mnras/stac825}, 512, 5390

\bibitem[\protect\citeauthoryear{Hahn \& Abel}{Hahn \& Abel}{2011}]{Hahn2011}
Hahn O.,  Abel T.,  2011, \mn@doi [Mon. Not. R. Astron. Soc.]
  {10.1111/j.1365-2966.2011.18820.x}, 415, 2101

\bibitem[\protect\citeauthoryear{Harikane et~al.,}{Harikane
  et~al.}{2016}]{Harikane2016}
Harikane Y.,  et~al., 2016, \mn@doi [Astrophys. J.]
  {10.3847/0004-637x/821/2/123}, 821, 123

\bibitem[\protect\citeauthoryear{Harikane et~al.,}{Harikane
  et~al.}{2022}]{Harikane2022}
Harikane Y.,  et~al., 2022, \mn@doi [Astrophys. J. Suppl. Ser.]
  {10.3847/1538-4365/ac3dfc}, 259, 20

\bibitem[\protect\citeauthoryear{Harikane et~al.,}{Harikane
  et~al.}{2023}]{Harikane2023}
Harikane Y.,  et~al., 2023, \mn@doi [Astrophys. J. Suppl. Ser.]
  {10.3847/1538-4365/acaaa9}, 265, 5

\bibitem[\protect\citeauthoryear{Harikane, Nakajima, Ouchi, Umeda, Isobe, Ono,
  Xu  \& Zhang}{Harikane et~al.}{2024}]{Harikane2023a}
Harikane Y.,  Nakajima K.,  Ouchi M.,  Umeda H.,  Isobe Y.,  Ono Y.,  Xu Y.,
  Zhang Y.,  2024, \mn@doi [Astrophys. J.] {10.3847/1538-4357/ad0b7e}, 960, 56

\bibitem[\protect\citeauthoryear{Hassan et~al.,}{Hassan
  et~al.}{2023}]{Hassan2023}
Hassan S.,  et~al., 2023, \mn@doi [Astrophys. J. Lett.]
  {10.3847/2041-8213/ad0239}, 958, L3

\bibitem[\protect\citeauthoryear{Hegde, Wyatt  \& Furlanetto}{Hegde
  et~al.}{2024}]{Hegde2024}
Hegde S.,  Wyatt M.~M.,   Furlanetto S.~R.,  2024, \mn@doi [J. Cosmol.
  Astropart. Phys.] {10.1088/1475-7516/2024/08/025}, 2024, 025

\bibitem[\protect\citeauthoryear{Hills, Kulkarni, Meerburg  \& Puchwein}{Hills
  et~al.}{2018}]{Hills2018}
Hills R.,  Kulkarni G.,  Meerburg P.~D.,   Puchwein E.,  2018, \mn@doi [Nature]
  {10.1038/s41586-018-0796-5}, 564, E32

\bibitem[\protect\citeauthoryear{Hopkins}{Hopkins}{2015}]{Hopkins2015a}
Hopkins P.~F.,  2015, \mn@doi [Mon. Not. R. Astron. Soc.]
  {10.1093/mnras/stv195}, 450, 53

\bibitem[\protect\citeauthoryear{Hopkins, Keres, Onorbe, Faucher-Giguere,
  Quataert, Murray  \& Bullock}{Hopkins et~al.}{2014}]{Hopkins2014}
Hopkins P.~F.,  Keres D.,  Onorbe J.,  Faucher-Giguere C.-A.,  Quataert E.,
  Murray N.,   Bullock J.~S.,  2014, \mn@doi [Mon. Not. R. Astron. Soc.]
  {10.1093/mnras/stu1738}, 445, 581

\bibitem[\protect\citeauthoryear{Hopkins et~al.,}{Hopkins
  et~al.}{2018}]{Hopkins2018}
Hopkins P.~F.,  et~al., 2018, \mn@doi [Mon. Not. R. Astron. Soc.]
  {10.1093/mnras/sty1690}, 480, 800

\bibitem[\protect\citeauthoryear{Hsiao et~al.,}{Hsiao et~al.}{2024}]{Hsiao2023}
Hsiao T. Y.-Y.,  et~al., 2024, \mn@doi [Astrophys. J.]
  {10.3847/1538-4357/ad5da8}, 973, 8

\bibitem[\protect\citeauthoryear{Hunter}{Hunter}{2007}]{Hunter2007}
Hunter J.~D.,  2007, \mn@doi [Comput. Sci. Eng.] {10.1109/MCSE.2007.55}, 9, 90

\bibitem[\protect\citeauthoryear{Inayoshi, Visbal  \& Haiman}{Inayoshi
  et~al.}{2019}]{Inayoshi2020}
Inayoshi K.,  Visbal E.,   Haiman Z.,  2019, \mn@doi [Annu. Rev. Astron.
  Astrophys.] {10.1146/annurev-astro-120419-014455}, 58, 27

\bibitem[\protect\citeauthoryear{Inoue, Shimizu, Iwata  \& Tanaka}{Inoue
  et~al.}{2014}]{Inoue2014}
Inoue A.~K.,  Shimizu I.,  Iwata I.,   Tanaka M.,  2014, \mn@doi [Mon. Not. R.
  Astron. Soc.] {10.1093/mnras/stu936}, 442, 1805

\bibitem[\protect\citeauthoryear{Kannan, Garaldi, Smith, Pakmor, Springel,
  Vogelsberger  \& Hernquist}{Kannan et~al.}{2022}]{Kannan2022}
Kannan R.,  Garaldi E.,  Smith A.,  Pakmor R.,  Springel V.,  Vogelsberger M.,
   Hernquist L.,  2022, \mn@doi [Mon. Not. R. Astron. Soc.]
  {10.1093/mnras/stab3710}, 511, 4005

\bibitem[\protect\citeauthoryear{Kannan et~al.,}{Kannan
  et~al.}{2023}]{Kannan2023}
Kannan R.,  et~al., 2023, \mn@doi [Mon. Not. R. Astron. Soc.]
  {10.1093/mnras/stac3743}, 524, 2594

\bibitem[\protect\citeauthoryear{Katz, Weinberg  \& Hernquist}{Katz
  et~al.}{1996}]{Katz1996a}
Katz N.,  Weinberg D.~H.,   Hernquist L.,  1996, \mn@doi [Astrophys. J. Suppl.
  Ser.] {10.1086/192305}, 105, 19

\bibitem[\protect\citeauthoryear{Katz, Kimm, Haehnelt, Sijacki, Rosdahl  \&
  Blaizot}{Katz et~al.}{2018}]{Katz2018}
Katz H.,  Kimm T.,  Haehnelt M.,  Sijacki D.,  Rosdahl J.,   Blaizot J.,  2018,
  \mn@doi [Mon. Not. R. Astron. Soc.] {10.1093/mnras/sty1225}, 478, 4986

\bibitem[\protect\citeauthoryear{Katz, Kimm, Haehnelt, Sijacki, Rosdahl  \&
  Blaizot}{Katz et~al.}{2019}]{Katz2019}
Katz H.,  Kimm T.,  Haehnelt M.~G.,  Sijacki D.,  Rosdahl J.,   Blaizot J.,
  2019, \mn@doi [Mon. Not. R. Astron. Soc.] {10.1093/mnras/sty3154}, 483, 1029

\bibitem[\protect\citeauthoryear{Kimm \& Cen}{Kimm \& Cen}{2014}]{Kimm2014a}
Kimm T.,  Cen R.,  2014, \mn@doi [Astrophys. J.] {10.1088/0004-637X/788/2/121},
  788, 121

\bibitem[\protect\citeauthoryear{Knollmann \& Knebe}{Knollmann \&
  Knebe}{2009}]{Knollmann2009}
Knollmann S.~R.,  Knebe A.,  2009, \mn@doi [Astrophys. J. Suppl. Ser.]
  {10.1088/0067-0049/182/2/608}, 182, 608

\bibitem[\protect\citeauthoryear{Koopmans et~al.,}{Koopmans
  et~al.}{2015}]{Koopmans2015}
Koopmans L.,  et~al., 2015, in Proc. Adv. Astrophys. with Sq. Km. Array —
  PoS(AASKA14). Sissa Medialab, Trieste, Italy, p.~001 (\mn@eprint {arXiv}
  {1505.07568}), \mn@doi{10.22323/1.215.0001}, \url
  {https://pos.sissa.it/215/001}

\bibitem[\protect\citeauthoryear{Kravtsov \& Belokurov}{Kravtsov \&
  Belokurov}{2024}]{Kravtsov2024}
Kravtsov A.,  Belokurov V.,  2024, 000 (\mn@eprint {arXiv} {2405.04578})

\bibitem[\protect\citeauthoryear{Kravtsov, Berlind, Wechsler, Klypin,
  Gottlober, Allgood  \& Primack}{Kravtsov et~al.}{2004}]{Kravtsov2004}
Kravtsov A.~V.,  Berlind A.~A.,  Wechsler R.~H.,  Klypin A.~A.,  Gottlober S.,
  Allgood B.,   Primack J.~R.,  2004, \mn@doi [Astrophys. J.] {10.1086/420959},
  609, 35

\bibitem[\protect\citeauthoryear{Krumholz \& Dekel}{Krumholz \&
  Dekel}{2012}]{Krumholz2012}
Krumholz M.~R.,  Dekel A.,  2012, \mn@doi [Astrophys. J.]
  {10.1088/0004-637X/753/1/16}, 753, 16

\bibitem[\protect\citeauthoryear{Krumholz \& Gnedin}{Krumholz \&
  Gnedin}{2011}]{Krumholz2011c}
Krumholz M.~R.,  Gnedin N.~Y.,  2011, \mn@doi [Astrophys. J.]
  {10.1088/0004-637X/729/1/36}, 729, 36

\bibitem[\protect\citeauthoryear{Kuhlen \& Faucher-Gigu{\`{e}}re}{Kuhlen \&
  Faucher-Gigu{\`{e}}re}{2012}]{Kuhlen2012a}
Kuhlen M.,  Faucher-Gigu{\`{e}}re C.~A.,  2012, \mn@doi [Mon. Not. R. Astron.
  Soc.] {10.1111/j.1365-2966.2012.20924.x}, 423, 862

\bibitem[\protect\citeauthoryear{Labb{\'{e}} et~al.,}{Labb{\'{e}}
  et~al.}{2023}]{Labbe2023a}
Labb{\'{e}} I.,  et~al., 2023, \mn@doi [Nature] {10.1038/s41586-023-05786-2},
  616, 266

\bibitem[\protect\citeauthoryear{Leethochawalit, Jones, Ellis, Stark  \&
  Zitrin}{Leethochawalit et~al.}{2016}]{Leethochawalit2016}
Leethochawalit N.,  Jones T.~A.,  Ellis R.~S.,  Stark D.~P.,   Zitrin A.,
  2016, \mn@doi [Astrophys. J.] {10.3847/0004-637X/831/2/152}, 831, 152

\bibitem[\protect\citeauthoryear{Leroy et~al.,}{Leroy
  et~al.}{2012}]{Leroy2012b}
Leroy A.~K.,  et~al., 2012, \mn@doi [Flux] {10.1088/0004-6256/144/1/3}, 99, 34

\bibitem[\protect\citeauthoryear{Lewis, Challinor  \& Lasenby}{Lewis
  et~al.}{2000}]{Lewis2000}
Lewis A.,  Challinor A.,   Lasenby A.,  2000, \mn@doi [Astrophys. J.]
  {10.1086/309179}, 538, 473

\bibitem[\protect\citeauthoryear{Lewis et~al.,}{Lewis et~al.}{2020}]{Lewis2020}
Lewis J.~S.,  et~al., 2020, \mn@doi [Mon. Not. R. Astron. Soc.]
  {10.1093/mnras/staa1748}, 496, 4342

\bibitem[\protect\citeauthoryear{Lewis et~al.,}{Lewis et~al.}{2022}]{Lewis2022}
Lewis J.~S.,  et~al., 2022, \mn@doi [Mon. Not. R. Astron. Soc.]
  {10.1093/mnras/stac2383}, 516, 3389

\bibitem[\protect\citeauthoryear{Li, Narayanan  \& Dav{\'{e}}}{Li
  et~al.}{2019}]{Li2019}
Li Q.,  Narayanan D.,   Dav{\'{e}} R.,  2019, \mn@doi [Mon. Not. R. Astron.
  Soc.] {10.1093/mnras/stz2684}, 490, 1425

\bibitem[\protect\citeauthoryear{Li, Dekel, Sarkar, Aung, Giavalisco, Mandelker
   \& Tacchella}{Li et~al.}{2024}]{Li2023a}
Li Z.,  Dekel A.,  Sarkar K.~C.,  Aung H.,  Giavalisco M.,  Mandelker N.,
  Tacchella S.,  2024, \mn@doi [Astron. Astrophys.]
  {10.1051/0004-6361/202348727}, 690, A108

\bibitem[\protect\citeauthoryear{Liang, Feldmann, Hayward, Narayanan,
  {\c{C}}atmabacak, Kere{\v{s}}, Faucher-Gigu{\`{e}}re  \& Hopkins}{Liang
  et~al.}{2021}]{Liang2021a}
Liang L.,  Feldmann R.,  Hayward C.~C.,  Narayanan D.,  {\c{C}}atmabacak O.,
  Kere{\v{s}} D.,  Faucher-Gigu{\`{e}}re C.-A.,   Hopkins P.~F.,  2021, \mn@doi
  [Mon. Not. R. Astron. Soc.] {10.1093/mnras/stab096}, 502, 3210

\bibitem[\protect\citeauthoryear{Liang et~al.,}{Liang et~al.}{2024}]{Liang2023}
Liang L.,  et~al., 2024, \mn@doi [Mon. Not. R. Astron. Soc.]
  {10.1093/mnras/stad3792}, 528, 499

\bibitem[\protect\citeauthoryear{Lilly, Carollo, Pipino, Renzini  \&
  Peng}{Lilly et~al.}{2013}]{Lilly2013c}
Lilly S.~J.,  Carollo C.~M.,  Pipino A.,  Renzini A.,   Peng Y.,  2013, \mn@doi
  [Astrophys. J.] {10.1088/0004-637X/772/2/119}, 772, 119

\bibitem[\protect\citeauthoryear{Lin et~al.,}{Lin et~al.}{2023}]{Lin2023}
Lin Y.-h.,  et~al., 2023, \mn@doi [Mon. Not. R. Astron. Soc.]
  {10.1093/mnras/stad3483}, 527, 4173

\bibitem[\protect\citeauthoryear{Lovell, Vijayan, Thomas, Wilkins, Barnes,
  Irodotou  \& Roper}{Lovell et~al.}{2020}]{Lovell2020}
Lovell C.~C.,  Vijayan A.~P.,  Thomas P.~A.,  Wilkins S.~M.,  Barnes D.~J.,
  Irodotou D.,   Roper W.,  2020, \mn@doi [Mon. Not. R. Astron. Soc.]
  {10.1093/mnras/staa3360}, 500, 2127

\bibitem[\protect\citeauthoryear{Lovell, Harrison, Harikane, Tacchella  \&
  Wilkins}{Lovell et~al.}{2023}]{Lovell2023a}
Lovell C.~C.,  Harrison I.,  Harikane Y.,  Tacchella S.,   Wilkins S.~M.,
  2023, \mn@doi [Mon. Not. R. Astron. Soc.] {10.1093/mnras/stac3224}, 518, 2511

\bibitem[\protect\citeauthoryear{Ma et~al.,}{Ma et~al.}{2018}]{Ma2018b}
Ma X.,  et~al., 2018, \mn@doi [Mon. Not. R. Astron. Soc.]
  {10.1093/mnras/sty1024}, 478, 1694

\bibitem[\protect\citeauthoryear{Ma et~al.,}{Ma et~al.}{2019}]{Ma2019}
Ma X.,  et~al., 2019, \mn@doi [Mon. Not. R. Astron. Soc.]
  {10.1093/mnras/stz1324}, 487, 1844

\bibitem[\protect\citeauthoryear{Ma, Quataert, Wetzel, Hopkins,
  Faucher-Gigu{\`{e}}re  \& Kere{\v{s}}}{Ma et~al.}{2020}]{Ma2020}
Ma X.,  Quataert E.,  Wetzel A.,  Hopkins P.~F.,  Faucher-Gigu{\`{e}}re C.~A.,
   Kere{\v{s}} D.,  2020, \mn@doi [Mon. Not. R. Astron. Soc.]
  {10.1093/mnras/staa2404}, 498, 2001

\bibitem[\protect\citeauthoryear{Madau}{Madau}{2018}]{Madau2018}
Madau P.,  2018, \mn@doi [Mon. Not. R. Astron. Soc. Lett.]
  {10.1093/mnrasl/sly125}, 480, L43

\bibitem[\protect\citeauthoryear{Madau \& Dickinson}{Madau \&
  Dickinson}{2014}]{Madau2014}
Madau P.,  Dickinson M.,  2014, \mn@doi [Annu. Rev. Astron. Astrophys.]
  {10.1146/annurev-astro-081811-125615}, 52, 415

\bibitem[\protect\citeauthoryear{Madau, Haardt  \& Rees}{Madau
  et~al.}{1999}]{Madau1999}
Madau P.,  Haardt F.,   Rees M.~J.,  1999, \mn@doi [Astrophys. J.]
  {10.1086/306975}, 514, 648

\bibitem[\protect\citeauthoryear{Maity \& Choudhury}{Maity \&
  Choudhury}{2022}]{Maity2022}
Maity B.,  Choudhury T.~R.,  2022, \mn@doi [Mon. Not. R. Astron. Soc.]
  {10.1093/mnras/stac1847}, 515, 617

\bibitem[\protect\citeauthoryear{Marszewski, Sun, Faucher-Gigu{\`{e}}re,
  Hayward  \& Feldmann}{Marszewski et~al.}{2024}]{Marszewski2024}
Marszewski A.,  Sun G.,  Faucher-Gigu{\`{e}}re C.-A.,  Hayward C.~C.,
  Feldmann R.,  2024, \mn@doi [Astrophys. J. Lett.] {10.3847/2041-8213/ad4cee},
  967, L41

\bibitem[\protect\citeauthoryear{Mascia et~al.,}{Mascia
  et~al.}{2023}]{Mascia2023}
Mascia S.,  et~al., 2023, \mn@doi [Astron. Astrophys.]
  {10.1051/0004-6361/202345866}, 672, A155

\bibitem[\protect\citeauthoryear{Mascia et~al.,}{Mascia
  et~al.}{2024}]{Mascia2024}
Mascia S.,  et~al., 2024, \mn@doi [Astron. Astrophys.]
  {10.1051/0004-6361/202347884}, 3, 1

\bibitem[\protect\citeauthoryear{Mason, Trenti  \& Treu}{Mason
  et~al.}{2015}]{Mason2015}
Mason C.~A.,  Trenti M.,   Treu T.,  2015, \mn@doi [Astrophys. J.]
  {10.1088/0004-637X/813/1/21}, 813, 21

\bibitem[\protect\citeauthoryear{Mason et~al.,}{Mason
  et~al.}{2019}]{Mason2019a}
Mason C.~A.,  et~al., 2019, \mn@doi [Mon. Not. R. Astron. Soc.]
  {10.1093/mnras/stz632}, 485, 3947

\bibitem[\protect\citeauthoryear{Matthee, Sobral, Best, Khostovan, Oteo,
  Bouwens  \& R{\"{o}}ttgering}{Matthee et~al.}{2017}]{Matthee2017a}
Matthee J.,  Sobral D.,  Best P.,  Khostovan A.~A.,  Oteo I.,  Bouwens R.,
  R{\"{o}}ttgering H.,  2017, \mn@doi [Mon. Not. R. Astron. Soc.]
  {10.1093/mnras/stw2973}, 465, 3637

\bibitem[\protect\citeauthoryear{McGreer, Mesinger  \& D'Odorico}{McGreer
  et~al.}{2015}]{McGreer2015}
McGreer I.~D.,  Mesinger A.,   D'Odorico V.,  2015, \mn@doi [Mon. Not. R.
  Astron. Soc.] {10.1093/mnras/stu2449}, 447, 499

\bibitem[\protect\citeauthoryear{McLure et~al.,}{McLure
  et~al.}{2013}]{McLure2013}
McLure R.~J.,  et~al., 2013, \mn@doi [Mon. Not. R. Astron. Soc.]
  {10.1093/mnras/stt627}, 432, 2696

\bibitem[\protect\citeauthoryear{Mellema et~al.,}{Mellema
  et~al.}{2013}]{Mellema2013}
Mellema G.,  et~al., 2013, \mn@doi [Exp. Astron.] {10.1007/s10686-013-9334-5},
  36, 235

\bibitem[\protect\citeauthoryear{Mertens et~al.,}{Mertens
  et~al.}{2020}]{Mertens2020}
Mertens F.~G.,  et~al., 2020, \mn@doi [Mon. Not. R. Astron. Soc.]
  {10.1093/mnras/staa327}, 493, 1662

\bibitem[\protect\citeauthoryear{Meurer, Heckman  \& Calzetti}{Meurer
  et~al.}{1999}]{Meurer1999}
Meurer G.~R.,  Heckman T.~M.,   Calzetti D.,  1999, \mn@doi [Astrophys. J.]
  {10.1086/307523}, 521, 64

\bibitem[\protect\citeauthoryear{Mirocha \& Furlanetto}{Mirocha \&
  Furlanetto}{2023}]{Mirocha2023a}
Mirocha J.,  Furlanetto S.~R.,  2023, \mn@doi [Mon. Not. R. Astron. Soc.]
  {10.1093/mnras/stac3578}, 519, 843

\bibitem[\protect\citeauthoryear{Moster, Naab  \& White}{Moster
  et~al.}{2013}]{Moster2013}
Moster B.~P.,  Naab T.,   White S. D.~M.,  2013, \mn@doi [Mon. Not. R. Astron.
  Soc.] {10.1093/mnras/sts261}, 428, 3121

\bibitem[\protect\citeauthoryear{Moster, Naab  \& White}{Moster
  et~al.}{2018}]{Moster2018}
Moster B.~P.,  Naab T.,   White S. D.~M.,  2018, \mn@doi [Mon. Not. R. Astron.
  Soc.] {10.1093/mnras/sty655}, 477, 1822

\bibitem[\protect\citeauthoryear{Mu{\~{n}}oz, Qin, Mesinger, Murray, Greig  \&
  Mason}{Mu{\~{n}}oz et~al.}{2022}]{Munoz2022}
Mu{\~{n}}oz J.~B.,  Qin Y.,  Mesinger A.,  Murray S.~G.,  Greig B.,   Mason C.,
   2022, \mn@doi [Mon. Not. R. Astron. Soc.] {10.1093/mnras/stac185}, 511, 3657

\bibitem[\protect\citeauthoryear{Mu{\~{n}}oz, Mirocha, Chisholm, Furlanetto  \&
  Mason}{Mu{\~{n}}oz et~al.}{2024}]{Munoz2024}
Mu{\~{n}}oz J.~B.,  Mirocha J.,  Chisholm J.,  Furlanetto S.~R.,   Mason C.,
  2024, \mn@doi [Mon. Not. R. Astron. Soc. Lett.] {10.1093/mnrasl/slae086},
  535, L37

\bibitem[\protect\citeauthoryear{Munshi et~al.,}{Munshi
  et~al.}{2024}]{Munshi2024}
Munshi S.,  et~al., 2024, \mn@doi [Astron. Astrophys.]
  {10.1051/0004-6361/202348329}, 681, 1

\bibitem[\protect\citeauthoryear{Murray}{Murray}{2014}]{Murray2014}
Murray S.,  2014, {HMF: Halo Mass Function calculator}, \url
  {https://ui.adsabs.harvard.edu/abs/2014ascl.soft12006M}

\bibitem[\protect\citeauthoryear{Murray, Power  \& Robotham}{Murray
  et~al.}{2013}]{Murray2013a}
Murray S.,  Power C.,   Robotham A.,  2013, \mn@doi [Astron. Comput.]
  {10.1016/j.ascom.2013.11.001}, 3-4, 23

\bibitem[\protect\citeauthoryear{Naidu et~al.,}{Naidu et~al.}{2017}]{Naidu2017}
Naidu R.~P.,  et~al., 2017, \mn@doi [Astrophys. J.] {10.3847/1538-4357/aa8863},
  847, 12

\bibitem[\protect\citeauthoryear{Naidu, Tacchella, Mason, Bose, Oesch  \&
  Conroy}{Naidu et~al.}{2020}]{Naidu2020}
Naidu R.~P.,  Tacchella S.,  Mason C.~A.,  Bose S.,  Oesch P.~A.,   Conroy C.,
  2020, \mn@doi [Astrophys. J.] {10.3847/1538-4357/ab7cc9}, 892, 109

\bibitem[\protect\citeauthoryear{Naidu et~al.,}{Naidu et~al.}{2022}]{Naidu2022}
Naidu R.~P.,  et~al., 2022, \mn@doi [Astrophys. J. Lett.]
  {10.3847/2041-8213/ac9b22}, 940, L14

\bibitem[\protect\citeauthoryear{Ocvirk et~al.,}{Ocvirk
  et~al.}{2020}]{Ocvirk2020}
Ocvirk P.,  et~al., 2020, \mn@doi [Mon. Not. R. Astron. Soc.]
  {10.1093/mnras/staa1266}, 496, 4087

\bibitem[\protect\citeauthoryear{Oesch, Bouwens, Illingworth, Labb{\'{e}}  \&
  Stefanon}{Oesch et~al.}{2018}]{Oesch2018}
Oesch P.~A.,  Bouwens R.~J.,  Illingworth G.~D.,  Labb{\'{e}} I.,   Stefanon
  M.,  2018, \mn@doi [Astrophys. J.] {10.3847/1538-4357/aab03f}, 855, 105

\bibitem[\protect\citeauthoryear{Oke \& Gunn}{Oke \& Gunn}{1983}]{Oke1983}
Oke J.~B.,  Gunn J.~E.,  1983, \mn@doi [Astrophys. J.] {10.1086/160817}, 266,
  713

\bibitem[\protect\citeauthoryear{Ouchi et~al.,}{Ouchi et~al.}{2010}]{Ouchi2010}
Ouchi M.,  et~al., 2010, \mn@doi [Astrophys. J.] {10.1088/0004-637X/723/1/869},
  723, 869

\bibitem[\protect\citeauthoryear{Pahl, Shapley, Steidel, Chen  \& Reddy}{Pahl
  et~al.}{2021}]{Pahl2021}
Pahl A.~J.,  Shapley A.,  Steidel C.~C.,  Chen Y.,   Reddy N.~A.,  2021,
  \mn@doi [Mon. Not. R. Astron. Soc.] {10.1093/mnras/stab1374}, 505, 2447

\bibitem[\protect\citeauthoryear{Pallottini \& Ferrara}{Pallottini \&
  Ferrara}{2023}]{Pallottini2023}
Pallottini A.,  Ferrara A.,  2023, \mn@doi [Astron. Astrophys.]
  {10.1051/0004-6361/202347384}, 677, L4

\bibitem[\protect\citeauthoryear{Pandya et~al.,}{Pandya
  et~al.}{2021}]{Pandya2021}
Pandya V.,  et~al., 2021, \mn@doi [Mon. Not. R. Astron. Soc.]
  {10.1093/mnras/stab2714}, 508, 2979

\bibitem[\protect\citeauthoryear{Park, Mesinger, Greig  \& Gillet}{Park
  et~al.}{2019}]{Park2019}
Park J.,  Mesinger A.,  Greig B.,   Gillet N.,  2019, \mn@doi [Mon. Not. R.
  Astron. Soc.] {10.1093/mnras/stz032}, 484, 933

\bibitem[\protect\citeauthoryear{Patil et~al.,}{Patil et~al.}{2017}]{Patil2017}
Patil A.~H.,  et~al., 2017, \mn@doi [Astrophys. J.] {10.3847/1538-4357/aa63e7},
  838, 65

\bibitem[\protect\citeauthoryear{Peimbert, Peimbert  \& Luridiana}{Peimbert
  et~al.}{2016}]{Peimbert2016}
Peimbert A.,  Peimbert M.,   Luridiana V.,  2016, Rev. Mex. Astron. y
  Astrofis., 52, 419

\bibitem[\protect\citeauthoryear{P{\'{e}}rez-Gonz{\'{a}}lez
  et~al.,}{P{\'{e}}rez-Gonz{\'{a}}lez et~al.}{2023}]{Perez-Gonzalez2023}
P{\'{e}}rez-Gonz{\'{a}}lez P.~G.,  et~al., 2023, \mn@doi [Astrophys. J. Lett.]
  {10.3847/2041-8213/acd9d0}, 951, L1

\bibitem[\protect\citeauthoryear{Qin, Balu  \& Wyithe}{Qin
  et~al.}{2023}]{Qin2023}
Qin Y.,  Balu S.,   Wyithe J. S.~B.,  2023, \mn@doi [Mon. Not. R. Astron. Soc.]
  {10.1093/mnras/stad2448}, 526, 1324

\bibitem[\protect\citeauthoryear{Rahmati, Schaye, Pawlik  \&
  Rai{\v{c}}evic}{Rahmati et~al.}{2013}]{Rahmati2013a}
Rahmati A.,  Schaye J.,  Pawlik A.~H.,   Rai{\v{c}}evic M.,  2013, \mn@doi
  [Mon. Not. R. Astron. Soc.] {10.1093/mnras/stt324}, 431, 2261

\bibitem[\protect\citeauthoryear{Reichardt et~al.,}{Reichardt
  et~al.}{2021}]{Reichardt2021}
Reichardt C.~L.,  et~al., 2021, \mn@doi [Astrophys. J.]
  {10.3847/1538-4357/abd407}, 908, 199

\bibitem[\protect\citeauthoryear{Robertson}{Robertson}{2022}]{Robertson2022}
Robertson B.~E.,  2022, \mn@doi [Annu. Rev. Astron. Astrophys.]
  {10.1146/annurev-astro-120221-044656}, 60, 121

\bibitem[\protect\citeauthoryear{Robertson, Ellis, Dunlop, McLure  \&
  Stark}{Robertson et~al.}{2010}]{Robertson2010}
Robertson B.~E.,  Ellis R.~S.,  Dunlop J.~S.,  McLure R.~J.,   Stark D.~P.,
  2010, \mn@doi [Nature] {10.1038/nature09527}, 468, 49

\bibitem[\protect\citeauthoryear{Robertson et~al.,}{Robertson
  et~al.}{2013}]{Robertson2013}
Robertson B.~E.,  et~al., 2013, \mn@doi [Astrophys. J.]
  {10.1088/0004-637X/768/1/71}, 768, 71

\bibitem[\protect\citeauthoryear{Robertson, Ellis, Furlanetto  \&
  Dunlop}{Robertson et~al.}{2015}]{Robertson2015}
Robertson B.~E.,  Ellis R.~S.,  Furlanetto S.~R.,   Dunlop J.~S.,  2015,
  \mn@doi [Astrophys. J.] {10.1088/2041-8205/802/2/L19}, 802, L19

\bibitem[\protect\citeauthoryear{Robertson et~al.,}{Robertson
  et~al.}{2023}]{Robertson2023}
Robertson B.~E.,  et~al., 2023, \mn@doi [Nat. Astron.]
  {10.1038/s41550-023-01921-1}, 7, 611

\bibitem[\protect\citeauthoryear{Robertson et~al.,}{Robertson
  et~al.}{2024}]{Robertson2023b}
Robertson B.,  et~al., 2024, \mn@doi [Astrophys. J.]
  {10.3847/1538-4357/ad463d}, 970, 31

\bibitem[\protect\citeauthoryear{Rodr{\'{i}}guez-Puebla, Behroozi, Primack,
  Klypin, Lee  \& Hellinger}{Rodr{\'{i}}guez-Puebla
  et~al.}{2016}]{Rodriguez-Puebla2016a}
Rodr{\'{i}}guez-Puebla A.,  Behroozi P.,  Primack J.,  Klypin A.,  Lee C.,
  Hellinger D.,  2016, \mn@doi [Mon. Not. R. Astron. Soc.]
  {10.1093/mnras/stw1705}, 462, 893

\bibitem[\protect\citeauthoryear{Rodr{\'{i}}guez-Puebla, Primack, Avila-Reese
  \& Faber}{Rodr{\'{i}}guez-Puebla et~al.}{2017}]{Rodriguez-Puebla2017}
Rodr{\'{i}}guez-Puebla A.,  Primack J.~R.,  Avila-Reese V.,   Faber S.~M.,
  2017, \mn@doi [Mon. Not. R. Astron. Soc.] {10.1093/mnras/stx1172}, 470, 651

\bibitem[\protect\citeauthoryear{Rosdahl et~al.,}{Rosdahl
  et~al.}{2018}]{Rosdahl2018a}
Rosdahl J.,  et~al., 2018, \mn@doi [Mon. Not. R. Astron. Soc.]
  {10.1093/mnras/sty1655}, 479, 994

\bibitem[\protect\citeauthoryear{Rosdahl et~al.,}{Rosdahl
  et~al.}{2022}]{Rosdahl2022}
Rosdahl J.,  et~al., 2022, \mn@doi [Mon. Not. R. Astron. Soc.]
  {10.1093/mnras/stac1942}, 515, 2386

\bibitem[\protect\citeauthoryear{Sabti, Mu{\~{n}}oz  \& Blas}{Sabti
  et~al.}{2022}]{Sabti2022a}
Sabti N.,  Mu{\~{n}}oz J.~B.,   Blas D.,  2022, \mn@doi [Phys. Rev. D]
  {10.1103/PhysRevD.105.043518}, 105, 043518

\bibitem[\protect\citeauthoryear{Saldana-Lopez et~al.,}{Saldana-Lopez
  et~al.}{2023}]{Saldana-Lopez2023}
Saldana-Lopez A.,  et~al., 2023, \mn@doi [Mon. Not. R. Astron. Soc.]
  {10.1093/mnras/stad1283}, 522, 6295

\bibitem[\protect\citeauthoryear{Salpeter}{Salpeter}{1955}]{Salpeter1955}
Salpeter E.~E.,  1955, Astrophys. J., 121, 161

\bibitem[\protect\citeauthoryear{Sawala, Frenk, Crain, Jenkins, Schaye, Theuns
  \& Zavala}{Sawala et~al.}{2013}]{Sawala2013}
Sawala T.,  Frenk C.~S.,  Crain R.~A.,  Jenkins A.,  Schaye J.,  Theuns T.,
  Zavala J.,  2013, \mn@doi [Mon. Not. R. Astron. Soc.] {10.1093/mnras/stt259},
  431, 1366

\bibitem[\protect\citeauthoryear{Schaeffer, Giri  \& Schneider}{Schaeffer
  et~al.}{2023}]{Schaeffer2023}
Schaeffer T.,  Giri S.~K.,   Schneider A.,  2023, \mn@doi [Mon. Not. R. Astron.
  Soc.] {10.1093/mnras/stad2937}, 526, 2942

\bibitem[\protect\citeauthoryear{Schaller et~al.,}{Schaller
  et~al.}{2015}]{Schaller2015}
Schaller M.,  et~al., 2015, \mn@doi [Mon. Not. R. Astron. Soc.]
  {10.1093/mnras/stv1067}, 451, 1247

\bibitem[\protect\citeauthoryear{Schenker, Ellis, Konidaris  \& Stark}{Schenker
  et~al.}{2014}]{Schenker2014}
Schenker M.~A.,  Ellis R.~S.,  Konidaris N.~P.,   Stark D.~P.,  2014, \mn@doi
  [Astrophys. J.] {10.1088/0004-637X/795/1/20}, 795, 20

\bibitem[\protect\citeauthoryear{Schroeder, Mesinger  \& Haiman}{Schroeder
  et~al.}{2013}]{Schroeder2013}
Schroeder J.,  Mesinger A.,   Haiman Z.,  2013, \mn@doi [Mon. Not. R. Astron.
  Soc.] {10.1093/mnras/sts253}, 428, 3058

\bibitem[\protect\citeauthoryear{Seeyave et~al.,}{Seeyave
  et~al.}{2023}]{Seeyave2023}
Seeyave L. T.~C.,  et~al., 2023, \mn@doi [Mon. Not. R. Astron. Soc.]
  {10.1093/mnras/stad2487}, 525, 2422

\bibitem[\protect\citeauthoryear{Shapley, Steidel, Strom, Bogosavljevi{\'{c}},
  Reddy, Siana, Mostardi  \& Rudie}{Shapley et~al.}{2016}]{Shapley2016a}
Shapley A.~E.,  Steidel C.~C.,  Strom A.~L.,  Bogosavljevi{\'{c}} M.,  Reddy
  N.~A.,  Siana B.,  Mostardi R.~E.,   Rudie G.~C.,  2016, \mn@doi [Astrophys.
  J. Lett.] {10.3847/2041-8205/826/2/L24}, 826, L24

\bibitem[\protect\citeauthoryear{Sharma, Theuns, Frenk, Bower, Crain, Schaller
  \& Schaye}{Sharma et~al.}{2016}]{Sharma2016}
Sharma M.,  Theuns T.,  Frenk C.,  Bower R.,  Crain R.,  Schaller M.,   Schaye
  J.,  2016, \mn@doi [Mon. Not. R. Astron. Soc. Lett.] {10.1093/mnrasl/slw021},
  458, L94

\bibitem[\protect\citeauthoryear{Shen, Vogelsberger, Boylan-Kolchin, Tacchella
  \& Kannan}{Shen et~al.}{2023a}]{Shen2023b}
Shen X.,  Vogelsberger M.,  Boylan-Kolchin M.,  Tacchella S.,   Kannan R.,
  2023a, \mn@doi [Mon. Not. R. Astron. Soc.] {10.1093/mnras/stad2508}, 525,
  3254

\bibitem[\protect\citeauthoryear{Shen, Vogelsberger, Boylan-Kolchin, Tacchella
  \& Kannan}{Shen et~al.}{2023b}]{Shen2023a}
Shen X.,  Vogelsberger M.,  Boylan-Kolchin M.,  Tacchella S.,   Kannan R.,
  2023b, \mn@doi [Mon. Not. R. Astron. Soc.] {10.1093/mnras/stad2508}, 525,
  3254

\bibitem[\protect\citeauthoryear{Simmonds et~al.,}{Simmonds
  et~al.}{2023a}]{Simmonds2023}
Simmonds C.,  et~al., 2023a, \mn@doi [Mon. Not. R. Astron. Soc.]
  {10.1093/mnras/stad1749}, 523, 5468

\bibitem[\protect\citeauthoryear{Simmonds et~al.,}{Simmonds
  et~al.}{2023b}]{Simmonds2024}
Simmonds C.,  et~al., 2023b, \mn@doi [Mon. Not. R. Astron. Soc.]
  {10.1093/mnras/stad3605}, 527, 6139

\bibitem[\protect\citeauthoryear{Sims \& Pober}{Sims \& Pober}{2020}]{Sims2020}
Sims P.~H.,  Pober J.~C.,  2020, \mn@doi [Mon. Not. R. Astron. Soc.]
  {10.1093/mnras/stz3388}, 492, 22

\bibitem[\protect\citeauthoryear{Singh \& Subrahmanyan}{Singh \&
  Subrahmanyan}{2019}]{Singh2019}
Singh S.,  Subrahmanyan R.,  2019, \mn@doi [Astrophys. J.]
  {10.3847/1538-4357/ab2879}, 880, 26

\bibitem[\protect\citeauthoryear{Singh et~al.,}{Singh et~al.}{2022}]{Singh2022}
Singh S.,  et~al., 2022, \mn@doi [Nat. Astron.] {10.1038/s41550-022-01610-5},
  6, 607

\bibitem[\protect\citeauthoryear{Sipple \& Lidz}{Sipple \&
  Lidz}{2024}]{Sipple2023}
Sipple J.,  Lidz A.,  2024, \mn@doi [Astrophys. J.] {10.3847/1538-4357/ad06a7},
  961, 50

\bibitem[\protect\citeauthoryear{Smit, Bouwens, Franx, Illingworth,
  Labb{\'{e}}, Oesch  \& {Van Dokkum}}{Smit et~al.}{2012}]{Smit2012}
Smit R.,  Bouwens R.~J.,  Franx M.,  Illingworth G.~D.,  Labb{\'{e}} I.,  Oesch
  P.~A.,   {Van Dokkum} P.~G.,  2012, \mn@doi [Astrophys. J.]
  {10.1088/0004-637X/756/1/14}, 756

\bibitem[\protect\citeauthoryear{So, Norman, Reynolds  \& Wise}{So
  et~al.}{2014}]{So2014}
So G.~C.,  Norman M.~L.,  Reynolds D.~R.,   Wise J.~H.,  2014, \mn@doi
  [Astrophys. J.] {10.1088/0004-637X/789/2/149}, 789

\bibitem[\protect\citeauthoryear{Sobacchi \& Mesinger}{Sobacchi \&
  Mesinger}{2015}]{Sobacchi2015}
Sobacchi E.,  Mesinger A.,  2015, \mn@doi [Mon. Not. R. Astron. Soc.]
  {10.1093/mnras/stv1751}, 453, 1843

\bibitem[\protect\citeauthoryear{Song et~al.,}{Song et~al.}{2016}]{Song2015}
Song M.,  et~al., 2016, \mn@doi [Astrophys. J.] {10.3847/0004-637X/825/1/5},
  825, 5

\bibitem[\protect\citeauthoryear{Springel}{Springel}{2005}]{Springel2005a}
Springel V.,  2005, \mn@doi [Mon. Not. R. Astron. Soc.]
  {10.1111/j.1365-2966.2005.09655.x}, 364, 1105

\bibitem[\protect\citeauthoryear{Springel et~al.,}{Springel
  et~al.}{2008}]{Springel2008}
Springel V.,  et~al., 2008, \mn@doi [Mon. Not. R. Astron. Soc.]
  {10.1111/j.1365-2966.2008.14066.x}, 391, 1685

\bibitem[\protect\citeauthoryear{Stanway \& Eldridge}{Stanway \&
  Eldridge}{2018}]{Stanway2018}
Stanway E.~R.,  Eldridge J.~J.,  2018, \mn@doi [Mon. Not. R. Astron. Soc.]
  {10.1093/mnras/sty1353}, 479, 75

\bibitem[\protect\citeauthoryear{Stark}{Stark}{2016}]{Stark2016}
Stark D.~P.,  2016, \mn@doi [Annu. Rev. Astron. Astrophys.]
  {10.1146/annurev-astro-081915-023417}, 54, 761

\bibitem[\protect\citeauthoryear{Stark, Schenker, Ellis, Robertson, McLure  \&
  Dunlop}{Stark et~al.}{2013}]{Stark2013}
Stark D.~P.,  Schenker M.~a.,  Ellis R.,  Robertson B.,  McLure R.,   Dunlop
  J.,  2013, \mn@doi [Astrophys. J.] {10.1088/0004-637X/763/2/129}, 763, 129

\bibitem[\protect\citeauthoryear{Stefanon, Bouwens, Labb{\'{e}}, Illingworth,
  Gonzalez  \& Oesch}{Stefanon et~al.}{2021}]{Stefanon2021}
Stefanon M.,  Bouwens R.~J.,  Labb{\'{e}} I.,  Illingworth G.~D.,  Gonzalez V.,
    Oesch P.~A.,  2021, \mn@doi [Astrophys. J.] {10.3847/1538-4357/ac1bb6},
  922, 29

\bibitem[\protect\citeauthoryear{Su, Hopkins, Hayward, Faucher-Giguere, Keres,
  Ma  \& Robles}{Su et~al.}{2016}]{Su2017a}
Su K.-Y.,  Hopkins P.~F.,  Hayward C.~C.,  Faucher-Giguere C.-A.,  Keres D.,
  Ma X.,   Robles V.~H.,  2016, \mn@doi [Mon. Not. R. Astron. Soc.]
  {10.1093/mnras/stx1463}, 471, 144

\bibitem[\protect\citeauthoryear{Sun \& Furlanetto}{Sun \&
  Furlanetto}{2016}]{Sun2016}
Sun G.,  Furlanetto S.~R.,  2016, \mn@doi [Mon. Not. R. Astron. Soc.]
  {10.1093/mnras/stw980}, 460, 417

\bibitem[\protect\citeauthoryear{Sun, Faucher-Gigu{\`{e}}re, Hayward  \&
  Shen}{Sun et~al.}{2023a}]{Sun2023}
Sun G.,  Faucher-Gigu{\`{e}}re C.-A.,  Hayward C.~C.,   Shen X.,  2023a,
  \mn@doi [Mon. Not. R. Astron. Soc.] {10.1093/mnras/stad2902}, 526, 2665

\bibitem[\protect\citeauthoryear{Sun, Mas-Ribas, Chang, Furlanetto, Mebane,
  Gonzalez, Parsons  \& Trapp}{Sun et~al.}{2023b}]{Sun2023c}
Sun G.,  Mas-Ribas L.,  Chang T.-C.,  Furlanetto S.~R.,  Mebane R.~H.,
  Gonzalez M.~O.,  Parsons J.,   Trapp A.~C.,  2023b, \mn@doi [Astrophys. J.]
  {10.3847/1538-4357/acc9b3}, 950, 40

\bibitem[\protect\citeauthoryear{Sun, Faucher-Gigu{\`{e}}re, Hayward, Shen,
  Wetzel  \& Cochrane}{Sun et~al.}{2023c}]{Sun2023a}
Sun G.,  Faucher-Gigu{\`{e}}re C.-A.,  Hayward C.~C.,  Shen X.,  Wetzel A.,
  Cochrane R.~K.,  2023c, \mn@doi [Astrophys. J. Lett.]
  {10.3847/2041-8213/acf85a}, 955, L35

\bibitem[\protect\citeauthoryear{Tacchella, Trenti  \& Carollo}{Tacchella
  et~al.}{2013}]{Tacchella2013}
Tacchella S.,  Trenti M.,   Carollo C.~M.,  2013, \mn@doi [Astrophys. J. Lett.]
  {10.1088/2041-8205/768/2/L37}, 768

\bibitem[\protect\citeauthoryear{Tacchella, Bose, Conroy, Eisenstein  \&
  Johnson}{Tacchella et~al.}{2018}]{Tacchella2018}
Tacchella S.,  Bose S.,  Conroy C.,  Eisenstein D.~J.,   Johnson B.~D.,  2018,
  \mn@doi [Astrophys. J.] {10.3847/1538-4357/aae8e0}, 868, 92

\bibitem[\protect\citeauthoryear{Tacchella et~al.,}{Tacchella
  et~al.}{2023}]{Tacchella2023}
Tacchella S.,  et~al., 2023, \mn@doi [Astrophys. J.]
  {10.3847/1538-4357/acdbc6}, 952, 74

\bibitem[\protect\citeauthoryear{Tang et~al.,}{Tang et~al.}{2023}]{Tang2023}
Tang M.,  et~al., 2023, \mn@doi [Mon. Not. R. Astron. Soc.]
  {10.1093/mnras/stad2763}, 526, 1657

\bibitem[\protect\citeauthoryear{Tinker, Kravtsov, Klypin, Abazajian, Warren,
  Yepes, Gottl{\"{o}}ber  \& Holz}{Tinker et~al.}{2008}]{Tinker2008}
Tinker J.,  Kravtsov A.~V.,  Klypin A.,  Abazajian K.,  Warren M.,  Yepes G.,
  Gottl{\"{o}}ber S.,   Holz D.~E.,  2008, \mn@doi [Astrophys. J.]
  {10.1086/591439}, 688, 709

\bibitem[\protect\citeauthoryear{Trenti, Stiavelli, Bouwens, Oesch, Shull,
  Illingworth, Bradley  \& Carollo}{Trenti et~al.}{2010}]{Trenti2010}
Trenti M.,  Stiavelli M.,  Bouwens R.~J.,  Oesch P.,  Shull J.~M.,  Illingworth
  G.~D.,  Bradley L.~D.,   Carollo C.~M.,  2010, \mn@doi [Astrophys. J.]
  {10.1088/2041-8205/714/2/L202}, 714, L202

\bibitem[\protect\citeauthoryear{Umeda, Ouchi, Nakajima, Harikane, Ono, Xu,
  Isobe  \& Zhang}{Umeda et~al.}{2024}]{Umeda2023}
Umeda H.,  Ouchi M.,  Nakajima K.,  Harikane Y.,  Ono Y.,  Xu Y.,  Isobe Y.,
  Zhang Y.,  2024, \mn@doi [Astrophys. J.] {10.3847/1538-4357/ad554e}, 971, 124

\bibitem[\protect\citeauthoryear{Vale \& Ostriker}{Vale \&
  Ostriker}{2004}]{Vale2004}
Vale A.,  Ostriker J.~P.,  2004, \mn@doi [Mon. Not. R. Astron. Soc.]
  {10.1111/j.1365-2966.2004.08059.x}, 353, 189

\bibitem[\protect\citeauthoryear{Velliscig, van Daalen, Schaye, McCarthy,
  Cacciato, {Le Brun}  \& Vecchia}{Velliscig et~al.}{2014}]{Velliscig2014}
Velliscig M.,  van Daalen M.~P.,  Schaye J.,  McCarthy I.~G.,  Cacciato M.,
  {Le Brun} A.~M.,   Vecchia C.~D.,  2014, \mn@doi [Mon. Not. R. Astron. Soc.]
  {10.1093/mnras/stu1044}, 442, 2641

\bibitem[\protect\citeauthoryear{Verner \& Ferland}{Verner \&
  Ferland}{1996}]{Verner1996}
Verner D.~A.,  Ferland G.~J.,  1996, \mn@doi [Astrophys. J. Suppl. Ser.]
  {10.1086/192284}, 103, 467

\bibitem[\protect\citeauthoryear{Villasenor, Robertson, Madau  \&
  Schneider}{Villasenor et~al.}{2022}]{Villasenor2021}
Villasenor B.,  Robertson B.,  Madau P.,   Schneider E.,  2022, \mn@doi
  [Astrophys. J.] {10.3847/1538-4357/ac704e}, 933, 59

\bibitem[\protect\citeauthoryear{Vogelsberger et~al.,}{Vogelsberger
  et~al.}{2020}]{Vogelsberger2020a}
Vogelsberger M.,  et~al., 2020, \mn@doi [Mon. Not. R. Astron. Soc.]
  {10.1093/MNRAS/STAA137}, 492, 5167

\bibitem[\protect\citeauthoryear{Wang, Lei, Tang, Yuan  \& Fan}{Wang
  et~al.}{2024}]{Wang2024a}
Wang Y.-Y.,  Lei L.,  Tang S.-P.,  Yuan G.-W.,   Fan Y.-Z.,  2024, \mn@doi
  [Astrophys. J.] {10.3847/1538-4357/ad8080}, 975, 285

\bibitem[\protect\citeauthoryear{Watson}{Watson}{2011}]{Watson2011}
Watson D.,  2011, \mn@doi [Astron. Astrophys.] {10.1051/0004-6361/201117120},
  533, A16

\bibitem[\protect\citeauthoryear{Weingartner \& Draine}{Weingartner \&
  Draine}{2001}]{Weingartner2001b}
Weingartner J.~C.,  Draine B.~T.,  2001, \mn@doi [Astrophys. J.]
  {10.1086/318651}, 548, 296

\bibitem[\protect\citeauthoryear{Wetzel, Hopkins, Kim, Faucher-Gigu{\`{e}}re,
  Kere{\v{s}}  \& Quataert}{Wetzel et~al.}{2016}]{Wetzel2016}
Wetzel A.~R.,  Hopkins P.~F.,  Kim J.-h.,  Faucher-Gigu{\`{e}}re C.-A.,
  Kere{\v{s}} D.,   Quataert E.,  2016, \mn@doi [Astrophys. J.]
  {10.3847/2041-8205/827/2/L23}, 827, L23

\bibitem[\protect\citeauthoryear{Wiersma, Schaye  \& Smith}{Wiersma
  et~al.}{2009}]{Wiersma2009b}
Wiersma R. P.~C.,  Schaye J.,   Smith B.~D.,  2009, \mn@doi [Mon. Not. R.
  Astron. Soc.] {10.1111/j.1365-2966.2008.14191.x}, 393, 99

\bibitem[\protect\citeauthoryear{Wilkins, Feng, Di-Matteo, Croft, Stanway,
  Bouwens  \& Thomas}{Wilkins et~al.}{2016}]{Wilkins2016c}
Wilkins S.~M.,  Feng Y.,  Di-Matteo T.,  Croft R.,  Stanway E.~R.,  Bouwens
  R.~J.,   Thomas P.,  2016, \mn@doi [Mon. Not. R. Astron. Soc. Lett.]
  {10.1093/mnrasl/slw007}, 458, L6

\bibitem[\protect\citeauthoryear{Wilkins et~al.,}{Wilkins
  et~al.}{2022}]{Wilkins2022}
Wilkins S.~M.,  et~al., 2022, \mn@doi [Mon. Not. R. Astron. Soc.]
  {10.1093/mnras/stac3280}, 519, 3118

\bibitem[\protect\citeauthoryear{Wise, Demchenko, Halicek, Norman, Turk, Abel
  \& Smith}{Wise et~al.}{2014}]{Wise2014}
Wise J.~H.,  Demchenko V.~G.,  Halicek M.~T.,  Norman M.~L.,  Turk M.~J.,  Abel
  T.,   Smith B.~D.,  2014, \mn@doi [Mon. Not. R. Astron. Soc.]
  {10.1093/mnras/stu979}, 442, 2560

\bibitem[\protect\citeauthoryear{Wouthuysen}{Wouthuysen}{1952}]{Wouthuysen1952}
Wouthuysen S.~A.,  1952, \mn@doi [Astron. J.] {10.1086/106661}, 57, 31

\bibitem[\protect\citeauthoryear{Wu \& Kravtsov}{Wu \& Kravtsov}{2024}]{Wu2024}
Wu Z.,  Kravtsov A.,  2024, \mn@doi [Open J. Astrophys.]
  {10.33232/001c.121193}, 7, 56

\bibitem[\protect\citeauthoryear{Yeh et~al.,}{Yeh et~al.}{2023}]{Yeh2023}
Yeh J.~Y.,  et~al., 2023, \mn@doi [Mon. Not. R. Astron. Soc.]
  {10.1093/mnras/stad210}, 520, 2757

\bibitem[\protect\citeauthoryear{Yung, Somerville, Finkelstein, Wilkins  \&
  Gardner}{Yung et~al.}{2023}]{Yung2023}
Yung L. Y.~A.,  Somerville R.~S.,  Finkelstein S.~L.,  Wilkins S.~M.,   Gardner
  J.~P.,  2023, \mn@doi [Mon. Not. R. Astron. Soc.] {10.1093/mnras/stad3484},
  527, 5929

\bibitem[\protect\citeauthoryear{Zackrisson, Rydberg, Schaerer, Stlin  \&
  Tuli}{Zackrisson et~al.}{2011}]{Zackrisson2011}
Zackrisson E.,  Rydberg C.~E.,  Schaerer D.,  Stlin G.,   Tuli M.,  2011,
  \mn@doi [Astrophys. J.] {10.1088/0004-637X/740/1/13}, 740

\makeatother
\end{thebibliography}

%%%%%%%%%%%%%%%%%%%%%%%%%%%%%%%%%%%%%%%%%%%%%%%%%%

%%%%%%%%%%%%%%%%% APPENDICES %%%%%%%%%%%%%%%%%%%%%

\appendix

\section{Calculation of UV magnitudes}

\begin{figure}
\begin{tabular}{c}
\includegraphics[width=75mm]{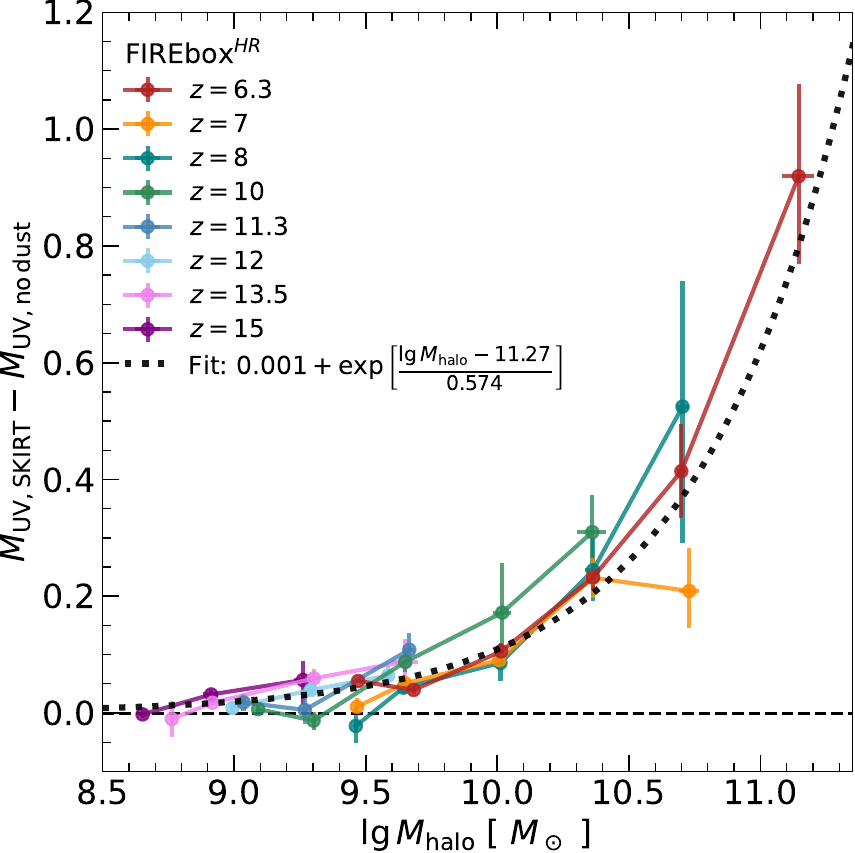}
\end{tabular}
\caption{Dependence on halo mass of the average difference between the rest-frame UV magnitude obtained from a dust-radiative transfer computation with SKIRT and from a dust-free calculation obtained by adding the UV luminosities of all star particles in \hr{} halos. Error bars show confidence intervals (16th--84th percentiles) of the average difference obtained via bootstrapping. The dotted curve is the result of fitting an exponential function with variable offset to all shown data points and taking their uncertainties into account. The best fit parameters are provided in the legend. On average, the much faster, dust-free calculation is accurate to better than 0.05 mag for halos with $M_{\rm halo}<3\times{}10^9\,M_\odot$. The importance of dust attenuation increases with halo mass but does not appear to strongly evolve with redshift over $z\sim{}6-15$.}
\label{fig:MUV_SKIRT_vs_nodust}
\end{figure}

As detailed in Section \ref{sect:postprocessing}, this study uses two methods to determine the UV magnitudes of galaxies. The first method, which accounts for dust attenuation, involves running radiative transfer calculations with SKIRT and measuring fluxes within circular apertures centered on the respective galaxy. The resulting UV magnitude is thus in principle viewing-angle dependent. The second method, which is computationally cheaper and viewing angle-independent, ignores dust and calculates UV fluxes by summing contributions from all star particles within a given three-dimensional distance from the given galaxy. Fig.~\ref{fig:MUV_SKIRT_vs_nodust} compares the average difference in UV magnitude between the two methods for central galaxies residing in halos of a given mass. As the figure demonstrates, the second method approximates the results of the full radiative transfer calculation in low mass halos, with a difference of less than 0.05 magnitude on average in halos with masses below $\sim{}3\times{}10^9\,M_\odot$. The figure also shows that UV magnitudes are increasingly affected by dust absorption and scattering in more massive halos, e.g., by $\sim{}0.2-0.3$ mag on average in $\sim{}3\times{}10^{10}\,M_\odot$ halos.

\section{Masses of halos and galaxies}
\label{sect:MUV_vs_MhaloMstar}

\begin{figure*}
\begin{tabular}{cc}
\includegraphics[width=80mm]{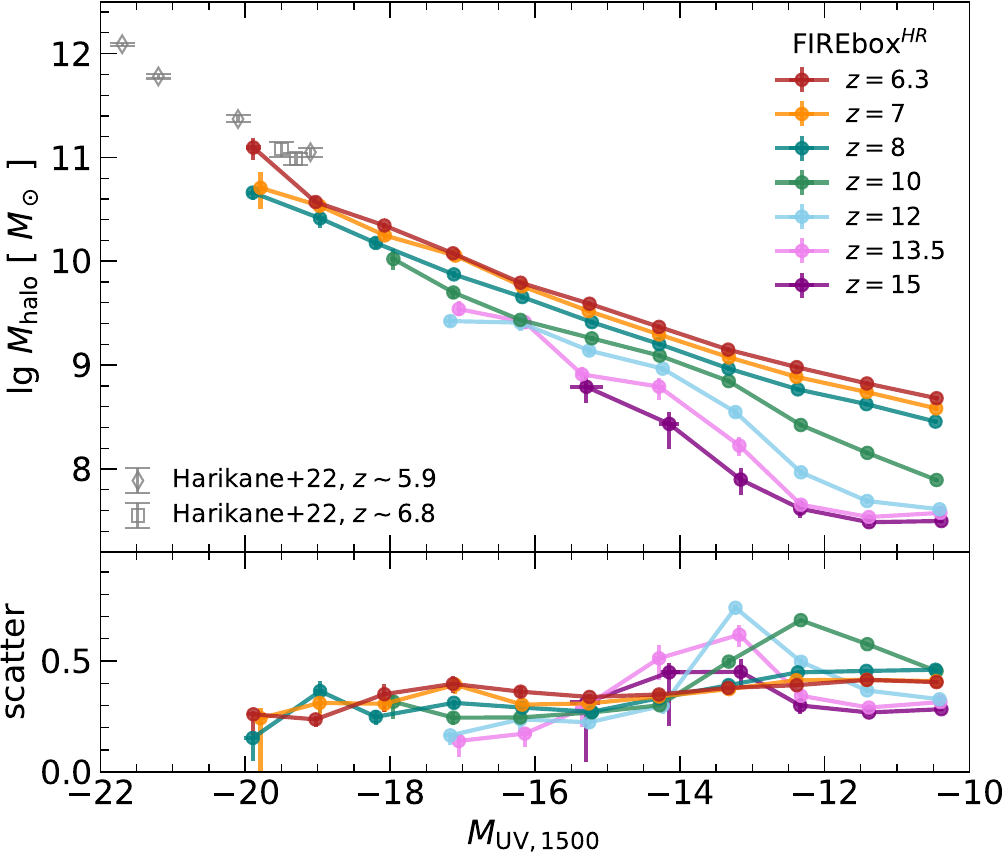} &
\includegraphics[width=80mm]{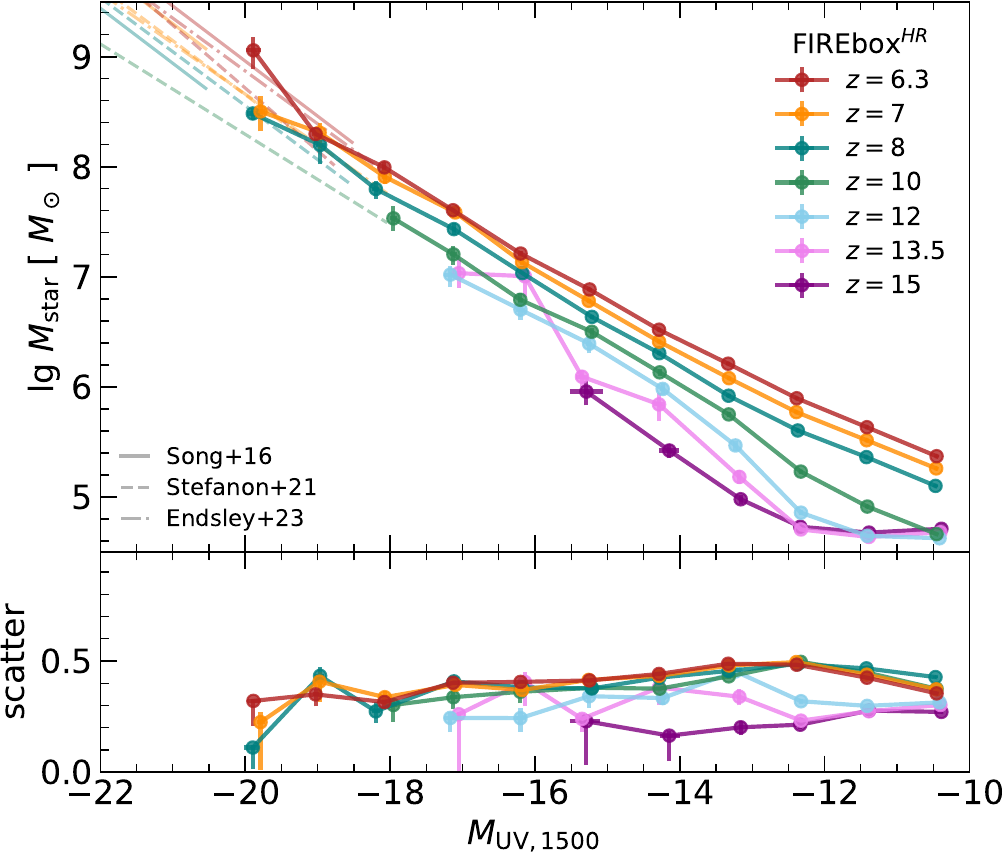}
\end{tabular}
\caption{Halo mass (left panels) and stellar mass (right panels) in bins of UV magnitude for galaxies at $z\sim{}6-15$ in \hr{}. The larger panels at the top show the logarithm of the average mass in each UV magnitude bin while the smaller panels at the bottom show the corresponding scatter of the mass in dex.
 Filled symbols and thick lines show results for \hr{} for different redshifts (see legend). Bins with less than 3 data points are not shown. The figure only includes halos with at least 300 dark matter particles ($M_{\rm halo}\gtrsim{}10^7\,M_\odot$). 
UV magnitudes shown on the x-axis include the effects of dust attenuation. Uncertainties (16th--84th percentiles) are computed via bootstrapping. Empty squares and diamonds in the top left panel are observational estimates at $z\sim{}6-7$ based on galaxy clustering by \protect\cite{Harikane2022}. Thin lines show the observed $M_{\rm UV}-M_{\rm star}$ relation at $z\sim{}6-10$ \citep{Song2015, Stefanon2021, Endsley2023}. The normalization of this relation is sensitive to the adopted star formation history, see, e.g., \protect\cite{Endsley2023}. On average, galaxies that are more UV luminous are more massive and reside in more massive halos. Stellar and halo masses at fixed UV magnitude increase with decreasing redshift. The scatter ($\sim{}0.3-0.4$ dex) in either relation is significant.}
\label{fig:MUV_Mhalo_Mstar}
\end{figure*}

Fig.~\ref{fig:MUV_Mhalo_Mstar} reports the average halo mass and the average stellar mass of \hr{} galaxies in bins of their (dust-attenuated) UV magnitude at $z\sim{}6-15$. On average, more UV luminous galaxies tend to have higher stellar masses and reside in more massive halos, as expected.  Both relations show a substantial scatter ($\sim{}0.3-0.4$ dex) that does not strongly depend on luminosity and redshift over $z=6-8$. As a comparison, \cite{Song2015} report a scatter of 0.36, 0.40, 0.30 dex for $M_{\rm UV}$ -- $M_{\rm star}$ relation at $z=6, 7, 8$ with no noticeable correlation of the scatter with redshift or UV luminosity.

Furthermore, the stellar and halo mass of galaxies of a given UV luminosity decreases with increasing redshift. The decrease in galaxy stellar mass with increasing redshift at fixed UV luminosity is consistent with observational findings (e.g., \citealt{Stark2013, Song2015, Stefanon2021}) and usually understood to reflect an evolution in the specific SFR. At $z>10$, halo and stellar masses of UV faint galaxies with $M_{\rm UV}>-16$ decrease quickly with increasing redshift. We interpret this result as the lack of suppression of star formation in low mass halos during the pre-reionization era caused by the lack of a cosmic UV background.

The evolution of the average halo and stellar mass at fixed UV luminosity can be understood in the context of a non-evolving SFE -- halo mass relation. A galaxy in a halo of any given mass will, on average, have a higher UV luminosity at a higher $z$ because the specific halo growth rate increases with $z$ at fixed mass \citep{Behroozi2015, Rodriguez-Puebla2016a} and the UV luminosity is proportional to the halo growth rate and the SFE (Eq. \ref{eq:lum_to_SFE}). This observation combined with a comparably low and approximately mass and redshift independent scatter implies that the average halo mass hosting galaxies of a given UV luminosity will decrease with increasing redshift. Given that the stellar mass -- halo mass relation has low scatter and is approximately redshift independent at high $z$ in FIRE-2 zoom-in simulation \citep{Ma2018b, Ma2019}, the stellar mass will also decrease with increasing redshift at fixed UV magnitude.

\section{SFE for different averaging times of star formation}
\label{sect:SFEavgTimeVariation}

\begin{figure*}
\begin{tabular}{cc}
\includegraphics[width=80mm]{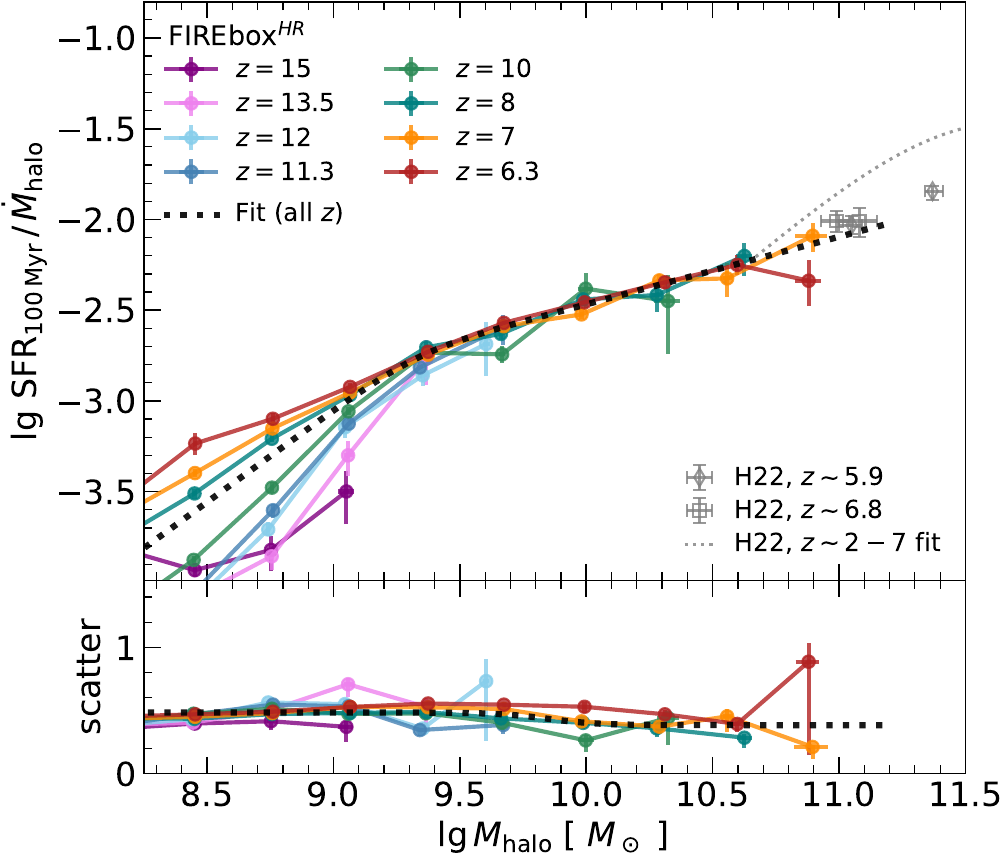} &
\includegraphics[width=80mm]{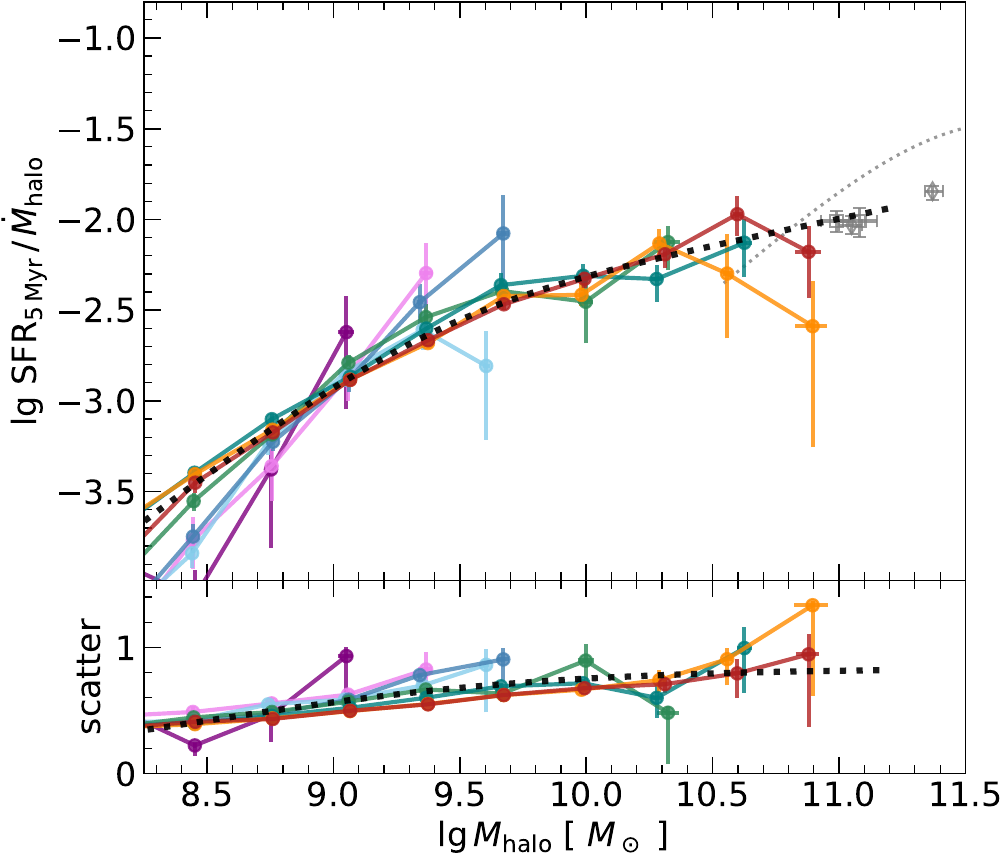}
\end{tabular}
\caption{
Same as the bottom left panel in Fig.~\ref{fig:efficiency} except for different star formation rate (SFR) averaging times: 100 Myr (left panel) and 5 Myr (right panel). 
}
\label{fig:efficiencySupp}
\end{figure*}

Fig.~\ref{fig:efficiencySupp} shows the SFE -- halo mass relation for two alternative choices of the SFR averaging time (100 Myr and 5 Myr). A longer averaging time of 100 Myr reduces the scatter of the relation but introduces a bias as many galaxies are growing quickly at high redshift and the SFRs are thus underestimated compared to those measured on shorter timescales.

\section{The faint end slope of the UV luminosity function}
\label{sect:FaintEndUVLF} 

Eq. (\ref{eq:Model_UV_LF}) expresses the UV luminosity function (UV LF) in terms of the halo mass function (HMF) $dn/dM$ and the probability distribution $p(\mathcal{M}=m\vert{}M, z)$ of the UV magnitude $\mathcal{M}$ as function of halo mass $M$ and redshift $z$. Following the discussion in Section \ref{sect:TheoreticalModel}, we model $p(\mathcal{M}=m\vert{}M, z)$ as a normal distribution with the mean $\langle{}\mathcal{M}\rangle{}$ and standard deviation $\sigma_{\mathcal{M}}$ given by equation (\ref{eq:mag_pdf_params}). The UV luminosity function can thus be written as
\begin{equation}
\phi(m, z) = \int_0^{\infty} dM \frac{d n}{d M}\frac{1}{\sqrt{2\pi}\sigma_{\mathcal{M}}}\exp\left[-\frac{(\mathscr{M} + \frac{c}{2}\sigma^2_{\mathcal{M}} -m)^2}{2\sigma_{\mathcal{M}}^2}\right],
\label{eq:phi}
\end{equation}
with $\mathscr{M}={\rm Mag}(\langle{}\mathcal{S}\rangle\dot{M}/\kappa)$ and $c=\ln(10)/2.5$.

We can solve Eq. (\ref{eq:phi}) analytically if the halo mass function, the SFE, and the halo growth rate are all power-law functions of the halo mass, and $\sigma_\mathcal{M}$ is a constant, i.e., if
\begin{align}
\frac{dn}{dM} &= \left(\frac{dn}{dM}\right)_0\,\tilde{M}^{\alpha_M},
&\langle{}\mathcal{S}\rangle\ &= \langle{}\mathcal{S}\rangle_0\,\tilde{M}^{\beta},
&\dot{M} &= \dot{M}_0\,\tilde{M}^{\gamma},
\label{eq:PL_scalings}
\end{align} 
where $\tilde{M}=M/M_0$, $M_0$ is a chosen reference halo mass, and pre-factors with the subscript $0$ do not depend on halo mass but potentially on other quantities such as redshift. Inserting Eq.~(\ref{eq:PL_scalings}) into Eq. (\ref{eq:phi}) and integrating over $M$ results in
\begin{equation}
\phi(m, z) = c\phi_* 10^{-0.4(1 + \alpha)(m - m_*)},
\label{eq:phi_PL}
\end{equation}
with
\begin{align}
\alpha &= \frac{1 + \alpha_M}{\beta + \gamma} - 1\label{eq:alpha2}\\ 
m_* &= {\rm Mag}(\langle{}\mathcal{S}\rangle_0\dot{M}_0/\kappa)\\
\phi_* &= \left(\frac{dn}{dM}\right)_0 M_0 \frac{1}{\beta + \gamma} \exp\left( \frac{c^2}{2} (1+\alpha) (2 + \alpha) \sigma_\mathcal{M}^2 \right)\label{eq:phi_star}.
\end{align}

According to Eq.~(\ref{eq:phi_PL}), the UV LF is then a power law with respect to UV luminosities. Moreover, since Eq.~(\ref{eq:phi_PL}) has the same functional form as the faint magnitude limit of a traditional Schechter or double power law function, we can identify the following parameters: $\alpha$ represents the faint-end slope, $m_*$ corresponds to the characteristic magnitude, and $\phi_*$ is the normalization of the luminosity function. Due to the scale-free nature of a power law, either $m_*$ or $\phi_*$ (but not both independently) can be freely adjusted by selecting an appropriate value for $M_0$. Multiplying the normalization $\langle{}\mathcal{S}\rangle_0$ of the SFE -- halo mass relation by a factor $f$ increases $m_*$ by $-2.5\lg{}f$, i.e., it amounts to a horizontal shift by this amount. Alternatively, such a shift may be interpreted as a vertical increase of $\lg{}\phi$ by $-(1+\alpha)\lg{}f$.

Eq.~(\ref{eq:alpha2}) shows that the faint-end slope $\alpha$ of the UV LF is directly related to the combination of $\alpha_M$ and $\beta + \gamma$. Provided the scatter $\sigma_M$ is mass independent as assumed in the derivation above or small enough to be ignored, we can directly infer the low-mass slope of the SFE–halo mass relation (denoted by $\beta$ here and by $\alpha_1$ in the main text) from measurements of the faint-end slope of the UV LF for a given cosmological model with well constrained $\alpha_M$ and $\gamma$. Alternatively, if $\beta$ can be measured independently by other means, the faint end of the UV LF may be used to constrain the growth properties of halos ($\alpha_M$ and $\gamma$) and perhaps even test alternative dark matter models.

The faint-end slope also determines how a larger scatter in the SFE, given by ${\rm std}(\lg{}\mathcal{S}) = \sigma_\mathcal{M}/2.5$, affects the normalization of the faint end of the UV LF. According to Eq.~(\ref{eq:phi_star}), an increase in the scatter of the SFE will raise the normalization of the UV LF when $\alpha<-2$ or $\alpha>-1$ and  decrease $\phi_*$ when $-2 < \alpha < -1$. 
These considerations clarify how the UV LFs shown in the right panel of Fig.~\ref{fig:UVLD_Model_highz} are affected by the scatter in the SFE. At sufficiently faint magnitudes, a power-law approximation of the UV LF results in a slope $\alpha$ that is shallower than $-2$ because of the modest turn-over at faint magnitudes seen in \hr. As a result, a larger scatter reduces the UV LF at the faintest magnitudes. In contrast, at brighter UV magnitudes, the UV LF has a local slope steeper than $-2$, leading to an increase in number density with increasing scatter. The scatter has little impact on the UV LF when the local slope is near $-2$. This condition is met around $M_{\rm UV}\sim{}-18$ at $z=6$, and $M_{\rm UV}\sim{}-15$ at $z=14$ for the model with a fixed scatter of 0.4 dex shown in Fig.~\ref{fig:UVLD_Model_highz}. This result explains why the UV luminosity density integrated down to $M_{\rm UV}=-17$ is only weakly dependent on the scatter of the SFE as shown in the left panel of Fig.~\ref{fig:UVLD_ModelVariations}.

\vspace{1 cm}

\section{Fitting parameters for the UV luminosity functions}
\label{sect:FitParamsUVLF}

Table~\ref{tab:ModUVLFfit} presents the parameters derived from fitting a modified Schechter function (as described in \citealt{Bouwens2017}) to the UV LFs predicted by the theoretical model of Section~\ref{sect:TheoreticalModel} and shown in Fig.\ref{fig:UVLF_Model_highz}. Additionally, Table~\ref{tab:ModUVLFfitDPL} displays the parameters obtained when employing a double power law to fit the UV LFs (e.g., \citealt{Harikane2022}). The scatter in the SFE -- halo mass relation is modeled in two ways: its mass dependence is either inferred directly from \hr{} (denoted as 'fiducial') or it is assumed to be a constant value between 0.2 and 0.6 dex. All fits were performed using the \textsc{scipy.optimize.curve\_fit} routine in Python, spanning UV magnitudes from -21.5 to -13 with a uniform spacing of 0.36 mag. The double power law generally provides as superior fit to the UV LFs of the theoretical model, see Section~\ref{sect:ModelImplicationsUV}.

\begin{table}
\begin{tabular}{cccccc}
$z$ & $M_{\rm UV}^*$ & $\lg{}\phi^*$ & $\alpha$ & $\delta{}$ & $M_{\rm UV, T}$ \\
\hline \multicolumn{6}{c}{mass-dependent scatter (fiducial)} \\ \hline
6 & -20.52 & -3.17 & -1.95 & 0.17 & -13.24 \\
8 & -20.11 & -3.36 & -1.99 & 0.16 & -12.90 \\
10 & -19.82 & -3.74 & -2.04 & 0.14 & -12.28 \\
12 & -19.63 & -4.27 & -2.09 & 0.11 & -11.24 \\
14 & -19.59 & -4.98 & -2.16 & 0.09 & -9.79 \\
16 & -19.71 & -5.91 & -2.27 & 0.09 & -8.61 \\
18 & -19.91 & -7.01 & -2.40 & 0.09 & -7.99 \\
20 & -20.13 & -8.21 & -2.54 & 0.09 & -7.61 \\
\hline \multicolumn{6}{c}{constant scatter (0.2 dex)} \\ \hline
6 & -20.54 & -3.23 & -1.98 & 0.14 & -12.39 \\
8 & -20.20 & -3.53 & -2.08 & 0.15 & -12.41 \\
10 & -19.96 & -4.04 & -2.20 & 0.16 & -12.29 \\
12 & -19.78 & -4.72 & -2.33 & 0.17 & -12.12 \\
14 & -19.43 & -5.32 & -2.42 & 0.15 & -11.40 \\
16 & -19.33 & -6.25 & -2.59 & 0.16 & -11.15 \\
18 & -19.05 & -7.09 & -2.71 & 0.15 & -10.21 \\
20 & -18.81 & -8.03 & -2.84 & 0.13 & -9.14 \\
\hline \multicolumn{6}{c}{constant scatter (0.4 dex)} \\ \hline
6 & -20.73 & -3.26 & -1.93 & 0.12 & -11.97 \\
8 & -20.48 & -3.58 & -2.01 & 0.12 & -11.90 \\
10 & -20.32 & -4.10 & -2.11 & 0.13 & -11.66 \\
12 & -20.22 & -4.79 & -2.23 & 0.13 & -11.36 \\
14 & -20.16 & -5.62 & -2.35 & 0.14 & -11.02 \\
16 & -20.13 & -6.57 & -2.50 & 0.14 & -10.68 \\
18 & -20.11 & -7.64 & -2.65 & 0.15 & -10.35 \\
20 & -19.88 & -8.59 & -2.76 & 0.13 & -9.34 \\
\hline \multicolumn{6}{c}{constant scatter (0.6 dex)} \\ \hline
6 & -20.99 & -3.32 & -1.88 & 0.09 & -11.27 \\
8 & -20.83 & -3.65 & -1.94 & 0.10 & -11.07 \\
10 & -20.75 & -4.18 & -2.02 & 0.10 & -10.68 \\
12 & -20.71 & -4.86 & -2.12 & 0.10 & -10.21 \\
14 & -20.71 & -5.67 & -2.22 & 0.10 & -9.71 \\
16 & -20.71 & -6.59 & -2.34 & 0.10 & -9.20 \\
18 & -20.73 & -7.61 & -2.46 & 0.10 & -8.70 \\
20 & -20.74 & -8.71 & -2.59 & 0.10 & -8.23 \\
\end{tabular}
\caption{Parameters derived from fitting a modified Schechter function to the ultraviolet luminosity function (UV LF) predicted by the theoretical model for a non-evolving star formation efficiency (SFE) -- halo mass relation with different choices of the scatter. Columns 1-5 list the redshift $z$, the characteristic UV magnitude $M_{\rm UV}^*$, the common logarithm of the LF normalization $\phi^*$ in units of ${\rm mag}^{-1}\,{\rm cMpc}^{-3}$, the faint-end slope $\alpha$, and the roll-over parameter $\delta$ of the modified Schechter function parametrization. The roll-over parameter equals zero for the traditional Schechter function. The final column provides the UV magnitude $M_{\rm UV, T}$ at which the LF is expected to reach its maximum based on the modified Schechter parametrization. The theoretical model predicts a decrease in $\alpha$ (steepening) with increasing redshift and a slight deviation from a traditional Schechter function at the faintest magnitudes.
}
\label{tab:ModUVLFfit}
\end{table}

\begin{table}
\begin{tabular}{ccccc}
$z$ & $M_{\rm UV}^*$ & $\lg{}\phi^*$ & $\alpha$ & $\beta$ \\
\hline \multicolumn{5}{c}{mass-dependent scatter (fiducial)} \\ \hline
6 & -17.25 & -1.55 & -1.42 & -2.61 \\
8 & -17.86 & -2.11 & -1.56 & -2.99 \\
10 & -18.29 & -2.81 & -1.71 & -3.41 \\
12 & -18.63 & -3.62 & -1.86 & -3.82 \\
14 & -18.88 & -4.52 & -2.00 & -4.09 \\
16 & -18.98 & -5.42 & -2.13 & -4.08 \\
18 & -18.97 & -6.33 & -2.25 & -3.94 \\
20 & -18.86 & -7.24 & -2.36 & -3.81 \\
\hline \multicolumn{5}{c}{constant scatter (0.2 dex)} \\ \hline
6 & -17.75 & -1.80 & -1.59 & -2.69 \\
8 & -18.00 & -2.26 & -1.70 & -3.04 \\
10 & -18.14 & -2.86 & -1.82 & -3.38 \\
12 & -18.23 & -3.58 & -1.96 & -3.72 \\
14 & -18.11 & -4.31 & -2.09 & -3.90 \\
16 & -18.12 & -5.23 & -2.25 & -4.19 \\
18 & -17.93 & -6.09 & -2.39 & -4.28 \\
20 & -17.70 & -7.00 & -2.53 & -4.35 \\
\hline \multicolumn{5}{c}{constant scatter (0.4 dex)} \\ \hline
6 & -17.79 & -1.84 & -1.58 & -2.57 \\
8 & -18.02 & -2.27 & -1.68 & -2.81 \\
10 & -18.15 & -2.83 & -1.78 & -3.02 \\
12 & -18.21 & -3.50 & -1.90 & -3.22 \\
14 & -18.23 & -4.27 & -2.03 & -3.39 \\
16 & -18.22 & -5.13 & -2.16 & -3.56 \\
18 & -18.20 & -6.08 & -2.31 & -3.72 \\
20 & -17.97 & -6.95 & -2.43 & -3.78 \\
\hline \multicolumn{5}{c}{constant scatter (0.6 dex)} \\ \hline
6 & -17.87 & -1.91 & -1.58 & -2.43 \\
8 & -18.07 & -2.30 & -1.65 & -2.58 \\
10 & -18.17 & -2.82 & -1.74 & -2.71 \\
12 & -18.21 & -3.45 & -1.84 & -2.83 \\
14 & -18.22 & -4.16 & -1.95 & -2.93 \\
16 & -18.19 & -4.94 & -2.06 & -3.04 \\
18 & -18.14 & -5.80 & -2.17 & -3.15 \\
20 & -18.08 & -6.71 & -2.29 & -3.26 \\
\end{tabular}
\caption{Parameters derived from fitting a double power law function to the ultraviolet luminosity function (UV LF) predicted by the theoretical model for a non-evolving star formation efficiency (SFE) -- halo mass relation with different choices of the scatter. Columns 1-5 list the redshift $z$, the characteristic UV magnitude $M_{\rm UV}^*$, the common logarithm of the LF normalization $\phi^*$ in units of ${\rm mag}^{-1}\,{\rm cMpc}^{-3}$, the faint-end slope $\alpha$, and the bright-end slope $\beta$. The theoretical model predicts a decrease (i.e., steepening) of both the faint-end and the bright-end slope of the UV LFs with increasing redshift.
}
\label{tab:ModUVLFfitDPL}
\end{table}

%%%%%%%%%%%%%%%%%%%%%%%%%%%%%%%%%%%%%%%%%%%%%%%%%%

% Don't change these lines
\bsp	% typesetting comment
\label{lastpage}
\end{document}